\PassOptionsToPackage{unicode}{hyperref}
\PassOptionsToPackage{hyphens}{url}
\PassOptionsToPackage{dvipsnames,svgnames,x11names}{xcolor}
\documentclass[
  12pt,
  letterpaper,
]{article}

\usepackage{amsmath,amssymb}
\usepackage{setspace}
\usepackage{iftex}
\ifPDFTeX
  \usepackage[T1]{fontenc}
  \usepackage[utf8]{inputenc}
  \usepackage{textcomp} 
\else 
  \usepackage{unicode-math}
  \defaultfontfeatures{Scale=MatchLowercase}
  \defaultfontfeatures[\rmfamily]{Ligatures=TeX,Scale=1}
\fi
\usepackage{lmodern}
\ifPDFTeX\else  
\fi
\IfFileExists{upquote.sty}{\usepackage{upquote}}{}
\IfFileExists{microtype.sty}{
  \usepackage[]{microtype}
  \UseMicrotypeSet[protrusion]{basicmath} 
}{}
\makeatletter
\@ifundefined{KOMAClassName}{
  \IfFileExists{parskip.sty}{%
    \usepackage{parskip}
  }{
    \setlength{\parindent}{0pt}
    \setlength{\parskip}{6pt plus 2pt minus 1pt}}
}{
  \KOMAoptions{parskip=half}}
\makeatother
\usepackage{xcolor}
\usepackage[margin=1in]{geometry}
\makeatletter
\ifx\paragraph\undefined\else
  \let\oldparagraph\paragraph
  \renewcommand{\paragraph}{
    \@ifstar
      \xxxParagraphStar
      \xxxParagraphNoStar
  }
  \newcommand{\xxxParagraphStar}[1]{\oldparagraph*{#1}\mbox{}}
  \newcommand{\xxxParagraphNoStar}[1]{\oldparagraph{#1}\mbox{}}
\fi
\ifx\subparagraph\undefined\else
  \let\oldsubparagraph\subparagraph
  \renewcommand{\subparagraph}{
    \@ifstar
      \xxxSubParagraphStar
      \xxxSubParagraphNoStar
  }
  \newcommand{\xxxSubParagraphStar}[1]{\oldsubparagraph*{#1}\mbox{}}
  \newcommand{\xxxSubParagraphNoStar}[1]{\oldsubparagraph{#1}\mbox{}}
\fi
\makeatother

\usepackage{color}
\usepackage{fancyvrb}

\DefineVerbatimEnvironment{Highlighting}{Verbatim}{commandchars=\\\{\}}
\usepackage{framed}
\definecolor{shadecolor}{RGB}{241,243,245}
\newenvironment{Shaded}{\begin{snugshade}}{\end{snugshade}}

\newcommand{\AttributeTok}[1]{\textcolor[rgb]{0.40,0.45,0.13}{#1}}

\newcommand{\FunctionTok}[1]{\textcolor[rgb]{0.28,0.35,0.67}{#1}}

\newcommand{\NormalTok}[1]{\textcolor[rgb]{0.00,0.23,0.31}{#1}}

\newcommand{\OtherTok}[1]{\textcolor[rgb]{0.00,0.23,0.31}{#1}}

\newcommand{\SpecialCharTok}[1]{\textcolor[rgb]{0.37,0.37,0.37}{#1}}

\newcommand{\StringTok}[1]{\textcolor[rgb]{0.13,0.47,0.30}{#1}}

\providecommand{\tightlist}{%
  \setlength{\itemsep}{0pt}\setlength{\parskip}{0pt}}\usepackage{longtable,booktabs,array}
\usepackage{calc} 
\usepackage{etoolbox}
\makeatletter
\patchcmd\longtable{\par}{\if@noskipsec\mbox{}\fi\par}{}{}
\makeatother
\IfFileExists{footnotehyper.sty}{\usepackage{footnotehyper}}{\usepackage{footnote}}
\makesavenoteenv{longtable}
\usepackage{graphicx}
\makeatletter
\newsavebox\pandoc@box
\newcommand*\pandocbounded[1]{
  \sbox\pandoc@box{#1}%
  \Gscale@div\@tempa{\textheight}{\dimexpr\ht\pandoc@box+\dp\pandoc@box\relax}%
  \Gscale@div\@tempb{\linewidth}{\wd\pandoc@box}%
  \ifdim\@tempb\p@<\@tempa\p@\let\@tempa\@tempb\fi
  \ifdim\@tempa\p@<\p@\scalebox{\@tempa}{\usebox\pandoc@box}%
  \else\usebox{\pandoc@box}%
  \fi%
}
\def\fps@figure{htbp}
\makeatother
\NewDocumentCommand\citeproctext{}{}
\NewDocumentCommand\citeproc{mm}{%
  \begingroup\def\citeproctext{#2}\cite{#1}\endgroup}
\makeatletter
 \let\@cite@ofmt\@firstofone
 \def\@biblabel#1{}
 \def\@cite#1#2{{#1\if@tempswa , #2\fi}}
\makeatother
\newlength{\cslhangindent}
\newlength{\csllabelwidth}
\newenvironment{CSLReferences}[2] 
 {\begin{list}{}{%
  \setlength{\itemindent}{0pt}
  \setlength{\leftmargin}{0pt}
  \setlength{\parsep}{0pt}
  \ifodd #1
   \setlength{\leftmargin}{\cslhangindent}
   \setlength{\itemindent}{-1\cslhangindent}
  \fi
  \setlength{\itemsep}{#2\baselineskip}}}
 {\end{list}}
\usepackage{calc}

\usepackage{amsmath,amssymb,booktabs,etoolbox}
\AtBeginEnvironment{longtable}{\small\setstretch{1.0}}
\ifPDFTeX\DeclareUnicodeCharacter{2212}{\ensuremath{-}}\DeclareUnicodeCharacter{0394}{\ensuremath{\Delta}}\fi
\makeatletter
\@ifpackageloaded{caption}{}{\usepackage{caption}}
\AtBeginDocument{%
\ifdefined\contentsname
  \renewcommand*\contentsname{Table of contents}
\else
  \newcommand\contentsname{Table of contents}
\fi
\ifdefined\listfigurename
  \renewcommand*\listfigurename{List of Figures}
\else
  \newcommand\listfigurename{List of Figures}
\fi
\ifdefined\listtablename
  \renewcommand*\listtablename{List of Tables}
\else
  \newcommand\listtablename{List of Tables}
\fi
\ifdefined\figurename
  \renewcommand*\figurename{Figure}
\else
  \newcommand\figurename{Figure}
\fi
\ifdefined\tablename
  \renewcommand*\tablename{Table}
\else
  \newcommand\tablename{Table}
\fi
}
\@ifpackageloaded{float}{}{\usepackage{float}}
\floatstyle{ruled}
\@ifundefined{c@chapter}{\newfloat{codelisting}{h}{lop}}{\newfloat{codelisting}{h}{lop}[chapter]}
\floatname{codelisting}{Listing}

\makeatother
\makeatletter
\@ifpackageloaded{caption}{}{\usepackage{caption}}
\@ifpackageloaded{subcaption}{}{\usepackage{subcaption}}
\makeatother

\usepackage{bookmark}

\IfFileExists{xurl.sty}{\usepackage{xurl}}{} 
\hypersetup{
  pdftitle={A Multiverse of Good and Bad Controls: Candidate Causal Graphs for Interpreting Model Robustness Analysis},
  pdfauthor={Shoki Okubo},
  pdfkeywords={model uncertainty, multiverse
analysis, robustness, causal inference, directed acyclic
graphs, covariate selection},
  colorlinks=true,
  linkcolor={blue},
  filecolor={Maroon},
  citecolor={Blue},
  urlcolor={Blue},
  pdfcreator={LaTeX via pandoc}}

\title{A Multiverse of Good and Bad Controls: Candidate Causal Graphs
for Interpreting Model Robustness
Analysis\thanks{This research benefited from helpful discussions and feedback in presentations at the Institute of Social Science, University of Tokyo, the Japan Sociological Society, and the Japanese Association for Mathematical Sociology. The R package dagmv that implements the framework is available at https://github.com/sokubo/dagmv, and the replication archive at https://github.com/sokubo/paper-multiverse-dag-replication. This work was supported by JSPS KAKENHI Grant Number 26K05332.}}
\author{Shoki
Okubo\thanks{Department of Sociology, Toyo University, Tokyo, Japan. Email: okubo080@toyo.jp. Website: sokubo.github.io.}}
\date{September 15, 2026}

\begin{document}
\maketitle
\begin{abstract}
Model robustness analysis estimates an effect across a multiverse of
specifications that pools control sets identifying the declared estimand
with sets that condition on mediators or colliders. We propose stating
rival assumptions about contested controls as a small set of candidate
causal graphs, enumerating the adjustment sets each graph licenses, and
reporting robustness metrics conditional on each graph. A finite-mixture
identity splits the licensed multiverse's dispersion into within-graph
and between-graph components; the between-graph share is a conditional
descriptive summary whose reading depends on the candidate set, the
weights, and a common estimand. Simulations examine misleading pooled
robustness assessments and the limits of the decomposition. Applications
to hurricane fatalities, job training, and union wages show fragility
that survives every graph, instability produced by unlicensed
specifications, and a fragility verdict concealing a significant premium
in each adjustment-identified candidate world. An R package implements
the workflow.
\end{abstract}


\setstretch{1.5}
\section{Introduction}\label{sec-intro}

How should researchers respond to the fact that published estimates
depend on modeling choices no theory fully pins down? One influential
answer is computational: estimate the model in every plausible
specification and report the whole distribution. In sociology this
program runs from Young (\citeproc{ref-young2009}{2009}) through the
multimodel framework of Young and Holsteen
(\citeproc{ref-youngholsteen2017}{2017}) to Muñoz and Young
(\citeproc{ref-munozyoung2018}{2018}), who ran nine billion regressions
to show that model-robustness screening eliminates most of the false
positives manufactured by significance-based selection; Young and
Cumberworth (\citeproc{ref-youngcumberworth2025}{2025}) consolidate the
program in book form, and many-analyst studies
(\citeproc{ref-breznau2022}{Breznau et al. 2022}) and large-scale
replications in political science
(\citeproc{ref-ganslmeier2025}{Ganslmeier and Vlandas 2025a}) have made
the underlying problem --- a hidden universe of model uncertainty ---
impossible to ignore.

Yet the multiverse has a foundational problem that its own architects
now acknowledge: not all specifications are created equal. The
distribution of estimates across all \(2^J\) combinations of \(J\)
candidate controls pools specifications that identify the declared
causal effect with specifications that do not --- models that condition
on mediators, colliders, or other variables downstream of the causal
path from treatment to outcome (\citeproc{ref-elwertwinship2014}{Elwert
and Winship 2014}; \citeproc{ref-montgomery2018}{Montgomery, Nyhan, and
Torres 2018}; \citeproc{ref-cinelli2024}{Cinelli, Forney, and Pearl
2024}). Such a model may still estimate \emph{some} well-defined
quantity --- a mediator-adjusted coefficient can, under conditions of
its own, target a controlled direct effect
(\citeproc{ref-cinelli2024}{Cinelli, Forney, and Pearl 2024}) --- but it
does not estimate the total effect the multiverse declares, and
averaging it with models that do produces a distribution with no single
referent. Slez (\citeproc{ref-slez2019}{2019}) objected, from a
different angle, that the pooled modeling distribution treats every
model as equally credible and so overstates uncertainty; the graphical
literature objects that it treats every model as answering the same
question. That second point generalizes far beyond sociology: the
graphical literature on covariate selection shows that which controls
belong in a regression is a function of the estimand and the causal
structure, not a dimension of free analytic choice
(\citeproc{ref-pearl2009}{Pearl 2009};
\citeproc{ref-shpitser2010}{Shpitser, VanderWeele, and Robins 2010};
\citeproc{ref-vanderweele2019}{VanderWeele 2019};
\citeproc{ref-keele2020}{Keele, Stevenson, and Elwert 2020};
\citeproc{ref-lundberg2021}{Lundberg, Johnson, and Stewart 2021}). Muñoz
and Young (\citeproc{ref-munozyoung2018}{2018, 18}) themselves caution
against controls that are endogenous or intermediate outcomes, and Young
and Cumberworth (\citeproc{ref-youngcumberworth2025}{2025}, ch.~7) now
counsel skepticism toward every control and describe the conditions
under which a control creates bias --- advice that still leaves the
analyst to judge each case. The latest round of the debate --- whether
robustness is better assessed ``with a few thoughtful models than with
billions of regressions'' (\citeproc{ref-auspurg2025}{Auspurg 2025};
\citeproc{ref-ganslmeier2025reply}{Ganslmeier and Vlandas 2025b}) ---
turns on exactly this judgment. Auspurg
(\citeproc{ref-auspurg2025}{2025}) states explicit rules for a justified
model set (no inferior estimators, no non-equivalent samples or
measures, and no models that ``omit confounders or include posttreatment
variables''); Ganslmeier and Vlandas
(\citeproc{ref-ganslmeier2025reply}{2025b}) answer that ``what counts as
justified is itself heavily contested,'' since omitting a variable risks
bias as surely as including a bad one. Both are right: the rules are
written down, but the one rule that matters here --- which controls are
confounders and which are post-treatment --- cannot be applied without a
causal claim that the data do not supply, and the two sides have no
shared way of stating that claim and letting readers audit it.

This paper proposes a way of stating that claim. Our starting point is
that the researcher does not know the true causal graph --- if she did,
the graph would determine which adjustment sets are admissible, although
choices among admissible specifications could remain. What she can do is
state the rival causal assumptions about the contested controls as a
small set of \textbf{candidate DAGs}, each a transparent, criticizable
claim about how the contested variables relate to treatment and outcome.
Given each candidate graph, the valid specification set follows
mechanically from the generalized adjustment criterion
(\citeproc{ref-shpitser2010}{Shpitser, VanderWeele, and Robins 2010};
\citeproc{ref-perkovic2018}{Perković et al. 2018}); the multiverse
within a graph contains only specifications that identify the declared
estimand under that graph's assumptions. The dispersion of the
disciplined multiverse --- a finite mixture of world-specific
specification distributions on the data at hand --- then admits an exact
decomposition into a \textbf{within-world} component --- specification
dispersion, the thing robustness analysis was designed to measure ---
and a \textbf{between-world} component --- structural disagreement, the
thing robustness analysis silently mixes in. The between-world share of
that dispersion is a conditional descriptive summary, not a test
statistic, and it depends on the candidate set and the weights attached
to it; read beside the world-specific estimates and their uncertainty,
it supplies something no pooled metric does: a statement of whether the
spread of estimates is evidence against a finding or evidence that the
discipline has not settled a causal question.

The idea of running separate multiverses under rival causal structures
is not ours. Del Giudice and Gangestad
(\citeproc{ref-delgiudice2021}{2021, 11--13}) build two
six-specification multiverses for a simulated example in which a
variable is a collider under one graph and a mediator under the other,
display them side by side, and read within-world homogeneity against
between-world disagreement that ``would require additional empirical
evidence''; Wysocki, Lawson, and Rhemtulla
(\citeproc{ref-wysocki2022}{2022, 12--13}, Table 1) lay out the
procedure in words --- state the plausible causal structures, derive the
appropriate control set under each, present the estimates as competing
when the sets differ, and say so when a structure admits no appropriate
set; and Lundberg, Johnson, and Stewart
(\citeproc{ref-lundberg2021}{2021, sec. 4.2}) state the principle that
robustness across conditioning sets ``only matters for those sets which
credibly identify the causal effect.'' The recent multidisciplinary
guide of Short et al. (\citeproc{ref-short2026}{2026, 15--16})
recommends separate multiverses by estimand. This paper automates and
extends the steps that these treatments leave to manual analysis: the
enumeration of every admissible control set under every candidate graph
over a full specification grid (rather than one fixed set per world),
the bookkeeping of cells licensed by several worlds and of worlds that
license none, an exact finite-mixture decomposition of the licensed
dispersion with explicit world weights and their sensitivity, and joint
resampling uncertainty for the between-world contrasts.
Table~\ref{tbl-predecessors} distinguishes the existing contributions
from the additions made here; the union application of
Section~\ref{sec-reanalysis} exercises every one of the additions.

\begin{longtable}[]{@{}
  >{\raggedright\arraybackslash}p{(\linewidth - 4\tabcolsep) * \real{0.3333}}
  >{\raggedright\arraybackslash}p{(\linewidth - 4\tabcolsep) * \real{0.3333}}
  >{\raggedright\arraybackslash}p{(\linewidth - 4\tabcolsep) * \real{0.3333}}@{}}
\caption{What the framework inherits and what it
adds.}\label{tbl-predecessors}\tabularnewline
\toprule\noalign{}
\begin{minipage}[b]{\linewidth}\raggedright
Element
\end{minipage} & \begin{minipage}[b]{\linewidth}\raggedright
Present in
\end{minipage} & \begin{minipage}[b]{\linewidth}\raggedright
This paper
\end{minipage} \\
\midrule\noalign{}
\endfirsthead
\toprule\noalign{}
\begin{minipage}[b]{\linewidth}\raggedright
Element
\end{minipage} & \begin{minipage}[b]{\linewidth}\raggedright
Present in
\end{minipage} & \begin{minipage}[b]{\linewidth}\raggedright
This paper
\end{minipage} \\
\midrule\noalign{}
\endhead
\bottomrule\noalign{}
\endlastfoot
Separate multiverses under rival causal structures & Del Giudice and
Gangestad (\citeproc{ref-delgiudice2021}{2021}) (two hand-built six-cell
multiverses; direct effect); Wysocki, Lawson, and Rhemtulla
(\citeproc{ref-wysocki2022}{2022}) (procedure in words); Short et al.
(\citeproc{ref-short2026}{2026}) (recommendation) & Automated over a
full specification grid; total effect; any number of named worlds \\
Admissible control sets from a graph & Shpitser, VanderWeele, and Robins
(\citeproc{ref-shpitser2010}{2010}); Perković et al.
(\citeproc{ref-perkovic2018}{2018}); DAGassist for one graph
(\citeproc{ref-dagassist2026}{Goff and Denly 2026}) & Enumerated under
every candidate graph; overlap of licensed cells across worlds
recorded \\
Worlds with no admissible set & Wysocki, Lawson, and Rhemtulla
(\citeproc{ref-wysocki2022}{2022}) (row 3 of their Table 1) & Retained
as named, non-identified candidates (W5 in the union application) \\
Decomposition of multiverse dispersion & specr's variance components by
analytic choice (\citeproc{ref-specr2020}{Masur and Scharkow 2023}) &
Exact finite-mixture split by graph membership tied to validity
screening, with world weights, restricted-simplex bounds, and a
Dirichlet sweep \\
Uncertainty of between-world contrasts & --- & Paired bootstrap
refitting every cell on each resample; contrasts against an external
benchmark \\
\end{longtable}

Three further contributions follow. First, we show by simulation that
the standard robustness metrics --- the robustness ratio, sign
stability, and significance rate (\citeproc{ref-youngholsteen2017}{Young
and Holsteen 2017}) --- can simultaneously certify a finding as robust
and average over causally incompatible worlds whose implied effects
differ by a factor of five (Section~\ref{sec-simulation}). Second, we
sketch how the framework extends to panel data, where two-way fixed
effects and modern difference-in-differences estimators add an estimator
dimension to the multiverse and where time-varying controls raise the
post-treatment problem in acute form (\citeproc{ref-imaikim2019}{Imai
and Kim 2019}; \citeproc{ref-callawaysantanna2021}{Callaway and
Sant'Anna 2021}; \citeproc{ref-dechaisemartin2020}{de Chaisemartin and
D'Haultfœuille 2020}) (Section~\ref{sec-panel}). Third, we provide an R
package, dagmv, whose core --- DAG parsing, d-separation, the
generalized adjustment criterion, valid-specification enumeration, and
the variance decomposition --- has no heavy dependencies, and which
implements the Young--Holsteen robustness metrics of the Stata modules
mrobust and MULTIVRS (\citeproc{ref-youngholsteen2017}{Young and
Holsteen 2017}, \citeproc{ref-multivrs2021}{2021}) (with the two
departures noted in Section~\ref{sec-background}) while adding the
causal layer that existing multiverse software lacks
(Section~\ref{sec-software}).

We position this paper as extending the model-robustness program with
the causal layer whose absence its architects have acknowledged, not as
refuting it. Our reanalyses of Muñoz and Young's own applications make
the point concretely (Section~\ref{sec-reanalysis}). For female-named
hurricanes, the decomposition leaves their fragility verdict intact at
the average-severity contrast we examine: the effect is weak and rarely
significant within every candidate graph, so the fragility cannot be
blamed on pooling bad controls; standardizing the fitted models to a
common one-point contrast also shows that the cells which separate the
candidate worlds are the cells whose fitted means are least credible, so
that the between-world spread in that application is coefficient
sensitivity rather than a measured structural disagreement. For job
training, the decomposition \emph{traces} their headline pathology to
its source: the naive multiverse's sign instability --- estimates
swinging from about −\$8,500 to +\$1,700 --- is produced by
specifications that omit lagged earnings, which no causal account of
program selection we can cite licenses; within the candidate graphs that
treat pre-program earnings as confounders, the licensed regression
coefficients sit near \(+\$1{,}000\) and the licensed treated-population
(AIPW) estimates near \(+\$1{,}200\) to \(+\$1{,}400\), and an
experimental benchmark rejects the specifications that only the
remaining, undefended world licenses and is compatible with the two
defended worlds once the estimand is matched to it. A third application,
new to this paper, takes the framework to a canonical sociological
quantity: the union wage premium. There the pooled multiverse
\emph{fails} conventional robustness standards --- a robustness ratio of
1.05, with one in five specifications negative --- even though each of
the six candidate worlds that identify the premium by adjustment
delivers a positive, statistically significant coefficient; the pooled
metric mistakes an unresolved structural question (whether occupation
and industry are confounders or mediators of union wages) for fragility.
A seventh candidate, in which occupation is a confounder before union
entry and a mediator after it, is not identified by any control set in
the data, so the reversal is conditional on the candidate set and not a
verdict robust to every structure we considered. Robustness analysis
without causal discipline misreads all three cases; with it, one verdict
survives, one is traced to unlicensed specifications, and one is
reversed as a reading of the same estimates.

\section{Background: Two Programs on a Collision
Course}\label{sec-background}

\subsection{The Computational Model Robustness
Program}\label{the-computational-model-robustness-program}

The model-robustness program begins from an asymmetry of information:
analysts know how sensitive their estimates are to specification
choices; readers do not. Young (\citeproc{ref-young2009}{2009}) brought
model uncertainty to the attention of applied sociologists --- building
on the sensitivity-analysis and model-averaging literature he cites
(\citeproc{ref-young2009}{Young 2009, 380}) --- by showing that
celebrated findings on religion and economic growth dissolve across
plausible specifications. Young and Holsteen
(\citeproc{ref-youngholsteen2017}{2017}) systematized the response:
enumerate the model space implied by \(J\) plausible controls (all
\(2^J\) combinations, times functional-form and estimation variants),
estimate every model, and report the \textbf{modeling distribution}
alongside the preferred estimate, summarized by the \textbf{robustness
ratio} (the preferred estimate divided by the total standard error
combining sampling and modeling variance), \textbf{sign stability}, and
the \textbf{significance rate} --- implemented in the Stata module
mrobust. Muñoz and Young (\citeproc{ref-munozyoung2018}{2018}) scaled
the program to nine billion regressions and gave it its sharpest payoff:
significance-driven two-stage selection manufactures false positives,
and robustness screening eliminates most of them (in their headline
simulation with 50 candidate variables and 100 observations, cutting the
false-positive rate from 11.3 to 4.4 percent). Parallel movements arose
in psychology --- the multiverse program of Steegen et al.
(\citeproc{ref-steegen2016}{2016}) and specification-curve analysis of
Simonsohn, Simmons, and Nelson (\citeproc{ref-simonsohn2020}{2020}) ---
and the scale of the underlying problem is no longer in doubt:
seventy-three teams given the same data and hypothesis produced a
``hidden universe'' of diverging conclusions
(\citeproc{ref-breznau2022}{Breznau et al. 2022}), and 3.6 billion
estimates across four political-science topics reveal model uncertainty
in the sign as well as the significance of many well-studied effects ---
driven, in that study, more by sample selection and outcome measurement
than by the control set (\citeproc{ref-ganslmeier2025}{Ganslmeier and
Vlandas 2025a}). The program's consolidation in book form
(\citeproc{ref-youngcumberworth2025}{Young and Cumberworth 2025})
confirms its status as a leading answer to the credibility crisis.

The program's metrics require explicit definition, because we report
each of them conditional on a candidate graph. Let \(\hat\tau_k\) be the
estimate in specification \(k\), with standard error \(\text{SE}_k\),
and \(\bar\tau\) the mean estimate across the space. The \emph{modeling
SD} is the standard deviation of \(\hat\tau_k\) across the space; the
\emph{robustness ratio} is
\(\bar\tau / \sqrt{\overline{\text{SE}^2} + \text{SD}_{\text{model}}^2}\)
(the mean estimate over the total standard error). Two conventions are
ours to state. Young and Holsteen
(\citeproc{ref-youngholsteen2017}{2017, 13}) define the ratio with the
preferred estimate in the numerator but report the mean-estimate version
in their tables {[}pp.~14, 18, 27 and note 8{]}, as does the MULTIVRS
module; we use the mean-estimate version throughout, because no
specification is privileged in a multiverse. For the sampling part of
the total standard error we use the root mean squared standard error,
where MULTIVRS uses the arithmetic mean of the standard errors (the two
coincide when the standard errors are equal; ours is the form of Slez's
equation 2). The ratio is read against the conventional threshold of 2;
the \emph{significance rate} is the share with \(p < .05\). For
\emph{sign stability} we use the share of estimates that share the sign
of the mean estimate \(\bar\tau\) --- of the world mean, when the metric
is computed per world. Young and Holsteen
(\citeproc{ref-youngholsteen2017}{2017, 13}) define it as the percentage
of estimates that have the same sign, without naming a referent; the
mean-sign and modal-sign readings coincide whenever the two agree, and
we flag the one application (job training, Section~\ref{sec-reanalysis})
in which they do not.\footnote{The mean-sign definition is what the
  accompanying software computes and what every sign-stability figure in
  this paper reports. When a few large estimates of one sign pull the
  mean across zero while most estimates carry the other sign, the
  mean-sign share falls below 50 percent and the modal-sign share is its
  complement; we report both values wherever this occurs.} Each is a
summary of a distribution whose referent is the entire model space ---
which is exactly where the critique below bites: a distribution over
models that answer different questions has no single referent, so the
metrics inherit whatever incoherence the space contains.

Throughout, the program treats the control set as a dimension of
\emph{analytic freedom}: since theory rarely pins down the exact
controls, all plausible combinations are estimated and the distribution
is reported transparently.

\subsection{The Causal Selection
Critique}\label{the-causal-selection-critique}

An independent literature reaches the opposite conclusion about that
same dimension: which controls belong in a regression is not free at all
--- it is fixed, up to a set of admissible choices, by the estimand and
the causal structure. Graphical identification theory gives the complete
answer: a control set is admissible if and only if it avoids the
forbidden set --- descendants of the non-treatment nodes on the causal
path from treatment to outcome --- and blocks all proper non-causal
paths (\citeproc{ref-pearl2009}{Pearl 2009};
\citeproc{ref-shpitser2010}{Shpitser, VanderWeele, and Robins 2010};
\citeproc{ref-perkovic2018}{Perković et al. 2018}), and practical
translations of this theory now exist for neighboring fields
(\citeproc{ref-vanderweele2019}{VanderWeele 2019};
\citeproc{ref-keele2020}{Keele, Stevenson, and Elwert 2020}). Sociology
has largely absorbed the lesson: conditioning on colliders manufactures
bias (\citeproc{ref-elwertwinship2014}{Elwert and Winship 2014}),
conditioning on post-treatment variables can ruin even experiments
(\citeproc{ref-montgomery2018}{Montgomery, Nyhan, and Torres 2018}),
research questions must be stated as estimands before models are fit
(\citeproc{ref-lundberg2021}{Lundberg, Johnson, and Stewart 2021}),
statistical control requires a causal justification
(\citeproc{ref-wysocki2022}{Wysocki, Lawson, and Rhemtulla 2022}), and
the crash course of Cinelli, Forney, and Pearl
(\citeproc{ref-cinelli2024}{2024}) supplies the working taxonomy of good
and bad controls. A further wing of the literature shows that even among
admissible sets some are more efficient than others
(\citeproc{ref-henckel2022}{Henckel, Perković, and Maathuis 2022};
\citeproc{ref-rotnitzkysmucler2020}{Rotnitzky and Smucler 2020}):
control choice is doubly non-arbitrary. Both admissibility and
efficiency depend on the estimand: the graphical optimality theory is
proved for the average treatment effect, and it has been known since
White and Lu (\citeproc{ref-whitelu2011}{2011, 1454, 1456}) that an
efficiency ordering of adjustment sets for the average effect need not
carry over to the effect on the treated. A companion paper separates two
further points: for the effect on the treated, graph structure alone
need not determine an efficiency-optimal adjustment set
(\citeproc{ref-okubo2026att}{Okubo 2026}, Theorem 2), and for
overlap-weighted estimands changing the adjustment set can additionally
change the target population itself, which complicates any efficiency
comparison (\citeproc{ref-okubo2026att}{Okubo 2026, sec. 7}). So ``which
controls?'' cannot even be posed, let alone averaged over, before
``which effect, in which structure?'' is fixed.

Slez (\citeproc{ref-slez2019}{2019}) criticized the robustness program
from a different direction. His objection {[}pp.~401--402, 405--411{]}
is that the unweighted modeling distribution reports \emph{numerical
variation} across models of unequal credibility and so overstates
statistical uncertainty; his alternative is to weight models by fit,
through BIC-approximate or fully Bayesian model averaging. Young
(\citeproc{ref-young2019}{2019}) replied that fit-based weighting is
itself a form of model selection --- in Slez's own example one model
receives 94 percent of the weight {[}pp.~440--441{]} --- and that the
framework presumes a theoretically informed space of plausible controls,
which is where bad controls are to be excluded {[}pp.~437--438{]}.
Neither side of that exchange framed the problem as one of estimands:
the models in the space were assumed to target one coefficient, and the
dispute was over how to weight them. The graphical critique enters at
the earlier step --- whether the models target the same quantity at all
--- and it is that step this paper formalizes.

\subsection{An Unresolved Debate}\label{an-unresolved-debate}

The standoff is now playing out publicly. Auspurg
(\citeproc{ref-auspurg2025}{2025}) argues that robustness is better
assessed ``with a few thoughtful models than with billions of
regressions'' and lists the rules that separate justified from
unjustified models --- no inferior estimators, no non-equivalent samples
or outcome measures, and no models that omit confounders or include
post-treatment variables; in her re-specification of one of Ganslmeier
and Vlandas (\citeproc{ref-ganslmeier2025}{2025a})'s applications, 1,152
of 92,160 specifications pass. The reply by Ganslmeier and Vlandas
(\citeproc{ref-ganslmeier2025reply}{2025b}) counters that ``what counts
as justified is itself heavily contested,'' since omitting a variable
risks bias as surely as including a bad one, and that narrowing the
space ``risks assuming away what sensitivity analyses are meant to
explore.'' Critics of the multiverse in psychology make a related point:
in the simulated example of Del Giudice and Gangestad
(\citeproc{ref-delgiudice2021}{2021, 11--12}) about 99 percent of a
mechanically generated multiverse consists of specifications that are
not equivalent alternatives for the question asked. Both sides of the
exchange are right about the other's weakness. Rules for statistical
adequacy --- estimator, sample, measurement --- can be written down and
audited, and Auspurg writes them down. The one rule that cannot be
applied by inspection is the causal one: whether a given control is a
confounder or a post-treatment variable is a claim about the world, not
about the data, and two careful analysts can disagree about it. The
framework of Section~\ref{sec-framework} addresses that rule and only
that rule: enumerate the \emph{structural disagreement itself},
transparently, and let the data display what follows from each side of
it.

Three features of the debate explain why it has not converged, and each
maps onto a component of the framework below. The first is a
\emph{conflation of two kinds of dispersion}. The pooled modeling
distribution mixes the sensitivity of an estimate to defensible choices
within a fixed causal question with the disagreement between estimates
that answer different causal questions; Del Giudice and Gangestad
(\citeproc{ref-delgiudice2021}{2021, 13}) read exactly this contrast off
their two hand-built multiverses --- homogeneous within each world,
different between them --- and Lundberg, Johnson, and Stewart
(\citeproc{ref-lundberg2021}{2021, sec. 4.2}) state the principle that
robustness across conditioning sets matters only among sets that
identify the effect. The within/between decomposition of
Section~\ref{sec-framework} is our formalization of that contrast: it
distinguishes specification dispersion within each candidate graph from
dispersion between the graph-conditioned means for the same declared
estimand under rival identifying assumptions, and the structural share
is the between-world component divided by the total dispersion. It is a
different distinction from Slez's, which concerns the weighting of
models assumed to target one coefficient. The second feature is an
\emph{asymmetry of auditability}. The brute-force side can point to a
fully enumerated model space; the few-thoughtful-models side states its
rules, but the causal rule among them rests on judgments about
confounding and mediation that readers cannot check from the rule alone.
A candidate-graph set makes those judgments auditable: the curation is
still there, but it is a small set of explicit, citable, falsifiable
structures rather than a verdict about which controls are
``post-treatment.'' The third feature is \emph{the missing adjudication
step}. Both sides agree that some specifications are wrong, and both
have procedures for parts of the problem --- the statistical rules on
one side, the causal reasoning of Wysocki, Lawson, and Rhemtulla
(\citeproc{ref-wysocki2022}{2022}) and Del Giudice and Gangestad
(\citeproc{ref-delgiudice2021}{2021}) on the other --- but no procedure
ties a stated structural claim to the full specification space
mechanically, so the disagreement is relitigated paper by paper. The
generalized adjustment criterion is that procedure --- mechanical once a
graph is stated, silent until it is --- and it converts an argument
about taste into an argument about edges, which is an argument the
substantive literature already knows how to have. The framework does not
settle the debate so much as relocate it to where it can be settled:
from the robustness table to the causal diagram.

\subsection{Adjacent Tools and the Missing
Piece}\label{adjacent-tools-and-the-missing-piece}

The tooling landscape has responded to both programs, and mapping it
clarifies what is missing. On the robustness side, mrobust and MULTIVRS
implement the Young-Holsteen metrics and multiverse distributions in
Stata (\citeproc{ref-youngholsteen2017}{Young and Holsteen 2017},
\citeproc{ref-multivrs2021}{2021}); specr supports specification-curve
analysis in R and already decomposes the variance of the estimates
across analytic choices with a multilevel model
(\citeproc{ref-specr2020}{Masur and Scharkow 2023},
\texttt{icc\_specs()}); multiverse declares branches inside arbitrary R
code, with conditional branch exclusions and a table that records each
universe's parameter assignments and code
(\citeproc{ref-multiverse2024}{Sarma and Kay 2024}); and RobustiPy
brings the program to Python with bootstrap and joint inference,
out-of-sample metrics, Bayesian model averaging over the control
coefficients, and predictive feature attribution for the full
specification (\citeproc{ref-robustipy2025}{Valdenegro Ibarra et al.
2026, v4}). All accept a user-curated specification space, but none
screens that space for causal validity: variance components and
influence diagnostics can report \emph{which} choice moves the estimates
--- the model-influence analysis of Young and Cumberworth
(\citeproc{ref-youngcumberworth2025}{2025}, ch.~6) --- but not whether
either side of an influential choice answers the declared question. On
the causal side, DAGassist automates covariate-role classification and
validity screening for a user-supplied graph and a single model call
(\citeproc{ref-dagassist2026}{Goff and Denly 2026}), operationalizing
the crash-course taxonomy (\citeproc{ref-cinelli2024}{Cinelli, Forney,
and Pearl 2024}); its current release (0.3.0) also enumerates every
acyclic orientation of a set of edges marked as uncertain --- up to
1,024 orientations by default --- and evaluates each added hypothesized
edge as its own branch, reporting whether the roles and the minimal
adjustment sets change across them, a local form of graph uncertainty
around one maintained graph. It does not fit the model under each
orientation, combine the screening with a user-specified specification
multiverse, treat rival graphs as named worlds, or decompose the
dispersion of estimates across them. Sensitivity analysis relaxes one
maintained structure toward unobserved confounding
(\citeproc{ref-oster2019}{Oster 2019};
\citeproc{ref-cinellihazlett2020}{Cinelli and Hazlett 2020}) without
representing rival structures.

Closest to this paper on the statistical side, Hu and van der Pas
(\citeproc{ref-huvdpas2025}{2025}) address graph uncertainty by
learning: taking the skeleton of the graph as known (in practice
estimated), they sample orientations of it by Markov chain Monte Carlo,
test subsets of each sampled graph's optimal adjustment set for
validity, and return the list of adjustment sets whose validity
frequency exceeds a threshold. The contrast is instructive. A
frequency-ranked list can guide adjustment-set choice when the
uncertainty is represented by a distribution over graphs. When competing
theoretical accounts imply different orientations of the contested
edges, and a skeptical reader grants none of them, separately reporting
the implications of each account makes that disagreement explicit. Our
framework keeps the disagreement on the table as a small set of named
worlds, reports what each implies, and summarizes the between-world
contribution to dispersion with the structural share \(\rho\) of
Section~\ref{sec-framework} instead of integrating it away; where a
defensible distribution over graphs exists, it can of course supply the
world weights \(w_g\) defined there. Among the tools we reviewed, none
combines a user-specified specification multiverse with a set of named
candidate graphs, and none decomposes multiverse dispersion by graph
membership tied to validity screening (Table~\ref{tbl-software} in
Section~\ref{sec-software}).

\section{The Framework}\label{sec-framework}

\subsection{Setup}\label{setup}

A researcher declares a causal estimand --- here, the total effect
\(\tau\) of a point exposure \(X\) on an outcome \(Y\)
(\citeproc{ref-lundberg2021}{Lundberg, Johnson, and Stewart 2021}) ---
and faces \(J\) candidate controls \(Z_1, \dots, Z_J\): variables that
some defensible analysis would include. Core covariates that every
account requires may be fixed in all models. The Muñoz-Young model space
is the set of \(2^J\) control subsets, optionally crossed with
functional-form, sample, and estimator variants.

The estimand declaration is not decorative, because admissibility is
estimand-relative: a control forbidden for the total effect (a mediator)
may be required --- together with additional adjustment for
mediator--outcome confounding --- for a controlled direct effect
(\citeproc{ref-cinelli2024}{Cinelli, Forney, and Pearl 2024}, Model 11
and its variation), so a multiverse that pools both kinds of
specification answers two questions at once. We call this, by analogy, a
multiverse-scale Table 2 fallacy: Westreich and Greenland
(\citeproc{ref-westreich2013}{2013, 293}) describe one model whose
several coefficients are read as if all were total effects; here the
several models of one multiverse are read as if all estimated the same
one. Conditioning on a mediator does not by itself identify a direct
effect either --- absent adjustment for mediator--outcome confounding it
opens a collider path (\citeproc{ref-cinelli2024}{Cinelli, Forney, and
Pearl 2024}, variation of Model 11;
\citeproc{ref-cinelli2022preprint}{Cinelli, Forney, and Pearl 2022, 8});
what it reliably does is fail to identify the declared total effect. We
fix the total effect throughout and flag where the framework
generalizes.

Three conditions are distinct and the inventory of a specification space
should keep them apart (\citeproc{ref-wysocki2022}{Wysocki, Lawson, and
Rhemtulla 2022, 3}; \citeproc{ref-westreich2013}{Westreich and Greenland
2013, 294--96}; \citeproc{ref-keele2020}{Keele, Stevenson, and Elwert
2020}). \emph{Graph eligibility} is what the adjustment criterion
certifies: under the candidate graph, the adjustment functional
\(E_Z\,E[Y \mid X, Z]\) over the licensed set identifies the total
effect. \emph{Estimation-model adequacy} is what a fitted regression
adds: a coefficient equals that functional's contrast only if the
model's functional form is right, and a coefficient at a reference value
of an interacted moderator is a conditional contrast, not the
population-averaged one. \emph{Target and sample compatibility} is what
the comparison of cells requires: the same exposure contrast, the same
outcome scale, the same target population, and a sample that has not
been selected on the outcome. Licensing by a graph settles only the
first. Where the applications compare cells across worlds we therefore
hold the contrast, scale and population fixed by construction --- an
ATT-targeted estimator over the treated population in the job-training
application, a standardized one-point contrast over a common population
in the hurricane application --- and display outcome-selected cells
separately as a selection-sensitivity family; where we do not (the union
application), we say that the comparison is of graph-conditioned
coefficients under an effect-homogeneity assumption. Throughout, the
candidate graphs are rival \emph{identifying assumptions for one
declared effect}, not rival questions. A specification licensed under an
incorrect graph need not identify that effect.

\subsection{Step 1: State the Disagreement as Candidate
Graphs}\label{step-1-state-the-disagreement-as-candidate-graphs}

The reason control choice is contested is rarely ignorance of
everything; it is disagreement about \emph{specific edges}. We therefore
encode rival positions as a set of candidate DAGs
\(\mathcal{G} = \{G_1, \dots, G_M\}\) built by a fixed protocol:

\begin{enumerate}
\def\labelenumi{\arabic{enumi}.}
\tightlist
\item
  \textbf{Fix the uncontested elements}: the exposure \(X\), the outcome
  \(Y\), the declared estimand, and all agreed parts of the causal
  structure (the target is the declared effect of \(X\) on \(Y\), which
  a candidate graph may carry through a direct edge, through mediators,
  or both).
\item
  \textbf{For each contested control}, list its plausible roles from the
  standard menu --- confounder, mediator, collider ancestor (via latent
  parents), instrument, or pure outcome cause --- including only roles
  supported by a mechanism citation or an explicit substantive
  justification for a hypothetical stress-test scenario.
\item
  \textbf{Take the product} of the contested alternatives, pruning
  combinations that are jointly incoherent, and (optionally) attach
  weights \(w_g\) reflecting prior plausibility (equal weights by
  default).
\end{enumerate}

Two contested variables with two plausible roles each yield \(M = 4\)
candidate graphs, as in our simulations; realistic applications rarely
need more than a handful. The protocol answers the natural objection
that candidate graphs merely relocate arbitrariness: the graph set does
not \emph{launder} the disagreement, it \emph{exposes} it. Nothing is
averaged away silently --- the between-graph spread is itself reported
(Step 3), which is precisely what a defensible treatment of
arbitrariness requires. The analogy is to multiple imputation or
Bayesian model averaging, except that our default output is a
\emph{display} of structural disagreement, not a pooled number.

Where do candidate graphs come from? Four sources cover practice. (i)
\emph{Published disputes.} When a literature has argued for decades
about whether occupation belongs in the union-premium equation
(Section~\ref{sec-reanalysis}), the two sides of that argument supply
candidate graphs, with citations attached to the edges the literature
actually supports; where the argument is conducted in terms of controls
rather than mechanisms, as in the union case, the analyst must still
draw the edges, and the drawing is then a stated hypothesis rather than
a documented mechanism. (ii) \emph{Temporal and institutional
structure.} Measurement timing, eligibility rules, and assignment
mechanisms rule roles in or out (a covariate whose value is
\emph{realized} after treatment cannot be a pre-treatment confounder ---
measurement timing is irrelevant, since a birth date collected
afterwards is still pre-treatment; a policy assigned by a predetermined
rule cannot depend on subsequently realized outcomes); pruning by such
constraints is usually what keeps \(M\) small. (iii) \emph{Adversarial
elicitation.} Before estimation, ask the colleague --- or anticipate the
referee --- most likely to dispute the finding to draw the graph under
which it fails; including that world is what makes the eventual
robustness claim credible to that reader. (iv) \emph{Prior taxonomies.}
The good-and-bad-controls inventory of Cinelli, Forney, and Pearl
(\citeproc{ref-cinelli2024}{2024}) enumerates the role menu; walking the
candidate pool through it is a mechanical first pass. Three constraints
keep the protocol honest. Every contested role must carry explicit
support of one of two kinds --- a mechanism citation that supports the
edge as drawn, or a substantive justification for a hypothetical
stress-test scenario --- and the analyst must say which kind is being
offered, so that a hypothetical mechanism does not silently acquire the
status of established evidence and no citation is attached to an edge it
does not support; ``someone might disagree'' is not a world. The set
should include the analyst's \emph{least favorable} defensible world,
for the same reason pre-registration exists --- and registering
\(\mathcal{G}\) itself before estimation is the natural extension. And
the stopping rule is \emph{validity-profile equivalence}: two worlds are
interchangeable for the analysis if and only if they license exactly the
same specifications --- their admissibility indicator vectors over all
\(2^J\) subsets of the candidate pool coincide (and, where these are
used, so do their within-world weights and estimand definitions). The
candidate set is complete when no defensible world outside it would
alter that vector. The required/forbidden/optional table of Step 2 is a
summary of the vector, not a substitute for it: two graphs can agree on
every role label yet license different specification sets, so
equivalence is never judged from the labels.\footnote{With candidate
  controls \(\{Z_1, Z_2\}\), compare \(G_0\): \(X \to Y\),
  \(Z_1 \to Y\), \(Z_2 \to Y\) with \(G_1\): \(Z_2 \to Z_1 \to X\),
  \(Z_2 \to Y\), \(X \to Y\). Under \(G_0\) all four subsets, the empty
  set included, are admissible; under \(G_1\) the empty set is not (the
  back-door path through \(Z_1\) and \(Z_2\) must be blocked) while the
  other three are. Both controls are \emph{optional} in both graphs, yet
  the licensed specification sets differ. We thank an early reader for
  the example.}

\subsection{Step 2: Validity and Roles within Each
Graph}\label{step-2-validity-and-roles-within-each-graph}

Under each \(G_m\), a control subset \(S\) is admissible for the total
effect if and only if it satisfies the generalized adjustment criterion
(\citeproc{ref-shpitser2010}{Shpitser, VanderWeele, and Robins 2010};
\citeproc{ref-perkovic2018}{Perković et al. 2018}, Definitions 3--4):
(i) \(S\) contains neither \(X\) itself nor any descendant of a node
\emph{other than \(X\)} that lies on a proper causal path from \(X\) to
\(Y\) --- the \emph{forbidden set},
\(\text{forb}(X, Y) = \text{de}(\text{cn}(X, Y)) \cup \{X\}\), where
\(\text{cn}\) collects the causal nodes so defined; and (ii) \(S\)
\(d\)-separates \(X\) and \(Y\) in the proper back-door graph, obtained
by deleting the first edge of every proper causal path out of \(X\). The
forbidden set is narrower than the conservative rule ``adjust for no
post-treatment variable'': in the graph \(X \to Y\), \(X \to Z\) with no
other edges, \(Z\) is a descendant of \(X\) but of no causal node, so
\(\{Z\}\) is admissible (if inefficient). The distinction does not bite
in our applications, where every forbidden control is a mediator, a
descendant of a mediator, or a descendant of the outcome, but the
criterion we implement is the complete one. Applied to the candidate
pool, the criterion classifies every contested control, per graph, as
\textbf{required} (in every admissible subset), \textbf{forbidden} (in
none), or \textbf{optional}; the classification is computed mechanically
by our software (Section~\ref{sec-software}) and typically fills a
one-page table that makes the structural stakes of each control visible
at a glance. A miniature example fixes ideas. Take candidate controls
\(\{Z_1, Z_2\}\) and two worlds that agree \(Z_1\) is a confounder but
disagree about \(Z_2\). In the \emph{mediator world}
(\(X \to Z_2 \to Y\)), \(Z_2\) is forbidden and the only admissible set
is \(\{Z_1\}\). In the \emph{outcome-cause world} (\(Z_2 \to Y\) only),
\(Z_2\) is optional and both \(\{Z_1\}\) and \(\{Z_1, Z_2\}\) are
admissible; were \(Z_2\) instead a second confounder, it would be
required and \(\{Z_1, Z_2\}\) alone would qualify. \(Z_1\) is required
in every world --- and any between-world gap in the multiverse now has
an interpretable address, the \(Z_2\) edge.

\subsection{Step 3: The Disciplined Multiverse and Its
Decomposition}\label{step-3-the-disciplined-multiverse-and-its-decomposition}

Within graph \(G_g\), the multiverse runs over admissible subsets only
(equally weighted), optionally crossed with the within-graph dimensions
--- functional form, outlier rules, estimation commands --- that do not
affect admissibility, provided estimates remain commensurable across
cells. All of the Young-Holsteen metrics are then reported
\emph{conditionally on the graph}.

The decomposition is an identity for a finite mixture defined on the
data at hand, and it is worth stating the mixture exactly. Fix the
observed dataset \(D\). Let \(\mathcal{S}_g\) be the set of
specifications licensed by world \(g\), \(b_k\) the estimate
specification \(k\) returns on \(D\), \(w_g\) the world weights (equal
by default), and \(q_{k \mid g}\) the within-world weights (uniform over
\(\mathcal{S}_g\) by default). Draw a world \(G \sim w\) and then a
specification \(K \mid G = g \sim q_{\cdot
\mid g}\). With world means and within-world variances

\[
\mu_g = \sum_{k \in \mathcal{S}_g} q_{k \mid g}\, b_k , \qquad
v_g = \sum_{k \in \mathcal{S}_g} q_{k \mid g}\, (b_k - \mu_g)^2 , \qquad
\bar\mu = \sum_g w_g\, \mu_g ,
\]

the law of total variance gives, exactly,

\[
\operatorname{Var}_{G,K}(b_K \mid D)
 = \underbrace{W(w) = \textstyle\sum_g w_g\, v_g}_{\text{within-world specification dispersion}}
 \; + \;
 \underbrace{B(w) = \textstyle\sum_g w_g\, (\mu_g - \bar\mu)^2}_{\text{between-world dispersion (structural disagreement)}} ,
\]

and the summary we report is the \textbf{structural share}
\(\rho(w) = B(w) / \{B(w) + W(w)\}\): the fraction of the licensed
multiverse's dispersion attributable to disagreement between worlds
rather than to specification choice within any one world. Five features
of this object govern its reading. First, both components are computed
on the fixed sample: \(W(w)\) is dispersion \emph{across
specifications}, not sampling variance, and \(\rho\) carries no
repeated-sampling uncertainty of its own --- the world-specific
robustness ratios (which include the mean squared standard error) and
resampling are the vehicles for that (the simulation designs of
Section~\ref{sec-simulation} pool Monte Carlo replications, so there the
within-world component also contains sampling variation, as we note
where it matters). Second, the identity uses population-weighted
variances, so \(v_g\) divides by the number of licensed specifications
rather than by that number minus one; a world with a single licensed
specification has \(v_g = 0\) mechanically, and if \(B(w) + W(w) = 0\)
the share is undefined, not zero. Third, the mixture is the
\emph{licensed} multiverse: unlicensed cells of the naive space are
absent from it, so it is not a decomposition of the naive multiverse's
variance. Fourth, a specification licensed by several worlds enters the
mixture with implicit weight \(\sum_g w_g\, q_{k \mid g}\) --- not with
equal weight among unique specifications --- and the degree of such
overlap should be reported with each application. Fifth, worlds with no
admissible set in the candidate pool are reported as \emph{not
identified by adjustment}, with \(|\mathcal{S}_g| = 0\), in the role and
decomposition tables; they are excluded from \(\rho\), the remaining
weights are renormalized, and the exclusion is stated rather than
allowed to disappear silently. The union application
(Section~\ref{sec-reanalysis}) supplies a worked example: a time-indexed
world in which occupation is a confounder before union entry and a
mediator after it has no admissible set in cross-sectional data.

Sampling uncertainty enters through a paired bootstrap. Units are
resampled, every cell of the multiverse is refitted on each resample,
and the world means, the world contrasts, and \(\rho\) are recomputed,
so that cells which share the data are resampled together and a contrast
between two worlds, or between a world and an external benchmark, is
assessed on the same resamples. We report 500 paired resamples (seed
20260906) for every application, as a bootstrap standard error and the
2.5th and 97.5th percentiles, and we report with each application the
number of licensed specifications that more than one world licenses. Two
conventions keep the resampled statistic equal to the stated one. First,
the set of positive-weight, adjustment-identified worlds is fixed before
resampling. A cell whose fit fails on a resample --- defined for
iterative estimators by the final convergence diagnostics, not only by
an error --- is not dropped and renormalized away, which would change
the specification mixture within the world; the world mean is left
missing for that resample. And a missing world mean does not lead to a
share recomputed over the surviving worlds with renormalized weights,
which would change the mixture between worlds; the share is missing for
that resample too, and each world contrast is evaluated on the resamples
where both of its terms are available. The number of contributing
resamples is reported with every statistic. This is different from a
world excluded at the design stage because no adjustment set exists (W5
in the union application), which is removed before the share is formed.
Second, the percentiles are conditional numerical diagnostics rather
than validated confidence limits: they condition on fitting success,
which need not be random across resamples, and for a ratio such as
\(\rho\), whose between-world numerator can be near zero, they are not
asserted to have nominal coverage. The world contrasts in effect units
are the quantities whose intervals we read.

\subsection{Reading Rules}\label{reading-rules}

A caution comes before the rules. The share \(\rho\) is computed on one
data set, and the world means it compares are estimates: two worlds
whose population means coincide will still differ by sampling noise, and
when the within-world cells are also tight that noise is the whole of
the between-world component. Scenario D of Section~\ref{sec-simulation}
makes the point with a design in which the contested edges carry zero
coefficients: the per-data-set share has a median of 49 percent and a
95th percentile of 93 percent although no structural disagreement
exists. A high fixed-data share therefore does not by itself identify a
structural source; it says that the world means differ on this sample by
more than the cells within worlds do. The reading rules below are to be
applied to the prespecified world contrasts in effect units, with their
resampling uncertainty (item 5 of the protocol below), and the share is
the summary of those contrasts, not a substitute for them.

With that caution, three configurations exhaust the useful cases.
\textbf{High \(\rho\) with world contrasts that exceed their sampling
uncertainty} (our simulations): the multiverse is a structural dispute
wearing the costume of specification noise; do not pool --- report
DAG-conditional estimates and resolve the dispute by design (benchmarks,
negative controls, sensitivity analysis), not by more specifications.
\textbf{Low \(\rho\), tight within-graph distributions}: a genuinely
robust finding --- the label ``robust'' now means something, because it
holds within every causal world on the table. \textbf{Low \(\rho\), wide
within-graph distributions} (the hurricane case of
Section~\ref{sec-reanalysis}): genuine fragility that cannot be blamed
on pooling bad controls.

Several properties qualify the interpretation. First, if the candidate
graphs agree on all admissibility judgments, \(\rho = 0\) and the
framework reduces exactly to Young-Holsteen analysis on the valid set
--- the disciplined multiverse is a strict generalization. Second, the
share depends on the specification inventory and on the weights, and not
monotonically. A world with exactly one licensed specification has
\(v_g = 0\) by identity and inflates the share; a world with two need
not --- two cells at \(-100\) and \(100\) have a population variance of
10,000. Adding a within-world dimension (an estimator, an outlier rule)
lowers \(\rho\) when it leaves the world means unchanged and raises
every within-world variance, but it can also raise \(\rho\): if the
added variant returns each world's mean, the between component is
unchanged while the within component halves (worlds \(\{-1, 1\}\) and
\(\{0, 2\}\) have \(\rho = 0.2\); crossed with such a variant they
become \(\{-1, 1, 0, 0\}\) and \(\{0, 2, 1, 1\}\) with \(\rho = 1/3\)).
The replication archive carries these cases as regression checks. The
dimension inventory, the per-world specification counts
\(|\mathcal{S}_g|\), and the between- and within-world variance
components \(B\) and \(W\) in squared effect units (with the
between-world range in effect units) must therefore be reported next to
the share whenever a space is enlarged, and a share built on worlds with
one specification is a description, not a finding. Third, \(\rho\) is a
\emph{share}, and should be read beside the between-world range in
effect units: a large share of a negligible total dispersion is
substantive stability, and a small share can coexist with a world
contrast that matters. Fourth, and most importantly, everything is
conditional on \(\mathcal{G}\): the decomposition cannot see structures
no one proposed, nor violations of sufficiency shared by all candidates.
Section~\ref{sec-simulation} quantifies exactly what this caveat costs.
Finally, \(\rho\) depends on the world weights, and not in a way that
any finite sweep can bound: as \(w\) approaches the vertex of a world
\(r\) with \(v_r > 0\), \(B(w) \to
0\) while \(W(w) \to v_r\), so \(\rho(w) \to 0\) along that path
whatever the between-world gaps. A single-weighting value is therefore
never reported alone. We report the equal-weight value beside two
further summaries, computed once and in the same way for every
application and simulation design. The first is a Dirichlet sweep:
\(N = 10{,}000\) weight vectors are drawn from a symmetric Dirichlet
distribution with concentration \(\alpha = 1\) (seed 20260906) and the
5th, 50th, and 95th percentiles of \(\rho\) are reported. These
quantiles describe where the sweep went, not a bound over the simplex.
The second supplies the bound on an explicitly restricted domain: the
infimum and supremum of \(\rho\) over the weightings that give no world
less than 5 percent of the weight, and over those that give no world
less than 10 percent, obtained by numerical optimization from the
per-world means and variances (constrained Nelder--Mead from the
centroid, the near-vertices and twenty random starts) and, for
decompositions with at most five worlds, checked against a fine grid on
the restricted simplex followed by local refinement (the two agree to
within \(10^{-6}\) in every such case; the six-world union
decompositions rest on the optimization alone; Table~\ref{tbl-simplex},
Supplemental Materials). Where the text below cites a Dirichlet quantile
or a restricted-simplex bound, it refers to this protocol.

\begin{quote}
\textbf{A one-page reporting protocol.}

\begin{enumerate}
\def\labelenumi{\arabic{enumi}.}
\tightlist
\item
  Declare the estimand --- treatment contrast, outcome scale, target
  population, estimating functional --- and the candidate-control pool;
  fix the uncontested roles. State separately which cells share the
  contrast, scale, population and sample (target and sample
  compatibility) and which estimator makes the fitted quantity equal the
  declared functional (estimation-model adequacy); graph eligibility
  settles neither.
\item
  Elicit candidate graphs from published disputes, institutional
  constraints, and adversarial review. For each contested role, provide
  either a supporting mechanism citation or an explicit substantive
  justification for a hypothetical stress-test scenario, and identify
  which kind of support is being offered; register the set.
\item
  Report the role-classification table (required / forbidden / optional,
  per world), flagging worlds not identified by adjustment.
\item
  Run the multiverse once; map admissibility per world; report the
  DAG-conditional robustness metrics beside the naive ones, with the
  number of licensed specifications per world and the overlap between
  worlds.
\item
  Report the structural share \(\rho\) at equal weights beside the
  Dirichlet quantiles (parameters, draws, and seed stated), the
  restricted-simplex bounds, the between-world range in effect units,
  and paired-bootstrap intervals for the world contrasts.
\item
  Read by the three-configuration rule, applied to the world contrasts
  and their intervals rather than to the share alone: high \(\rho\) with
  contrasts that exceed their uncertainty --- a structural dispute; seek
  design leverage, not more specifications. Low \(\rho\), tight
  within-world distributions --- robust in every world on the table. Low
  \(\rho\), wide within-world distributions --- genuinely fragile. In
  every case, check the claim world by world, and remember that a high
  share on one data set can arise from sampling noise alone when the
  within-world cells are tight.
\item
  Archive the graphs, the code, and the specification-to-validity map
  alongside the data.
\end{enumerate}
\end{quote}

\subsection{Scaling and Feasibility}\label{scaling-and-feasibility}

Two costs could threaten the framework at scale; neither computational
cost nor the number of candidate graphs was limiting in the applications
considered here. The \emph{computational} cost is trivial: admissibility
is checked by \(d\)-separation in the proper back-door graph via
moralization, the \(2^J\) subset checks run only over the contested
pool, and \texttt{mv\_run} fits each unique specification exactly once,
mapping validity per graph afterwards --- so adding candidate graphs
costs no additional estimation. The largest single run in this paper
(1,296 cells \(\times\) 3 graphs) completes in seconds on a laptop; the
paired bootstrap that refits every regression cell of every application,
and the 256 control-set AIPW-ATT cells of the job-training application,
500 times takes about 17 minutes on two cores of the laptop recorded in
the archive's README (31 minutes on the Linux machine used earlier;
bootstrapping all 1,296 AIPW-ATT cells is estimated to take about five
times the job-training stage, extrapolating from the per-cell cost, and
this runtime has not been measured separately); and the random-structure
battery of Section~\ref{sec-simulation} (up to 8 graphs \(\times\) 4,096
specifications per structure) takes seconds per structure.

The \emph{combinatorial} cost lives in \(M\), the product of contested
roles, and four devices keep it small. First, validity-profile
equivalence (Section~\ref{sec-framework}, Step 1): worlds that license
exactly the same specifications --- identical admissibility vectors over
the candidate pool, not merely identical role labels --- are
interchangeable, so the effective number of worlds is the number of
distinct admissibility vectors, typically far below the nominal product.
Second, incoherent combinations --- roles that jointly violate temporal
order or institutional constraints --- are pruned at elicitation. Third,
when the contested pool is genuinely large, sample worlds rather than
enumerate them and report the resulting range of \(\rho\), exactly as
one subsamples a specification space. Fourth, weights are reported
rather than argued over: the equal-weight share is accompanied by the
Dirichlet quantiles, the restricted-simplex bounds, and, where the
reading turns on it, by the world-by-world claim check, so that no
verdict rests on the default weights alone (in our applications the
readings survive that check, Section~\ref{sec-reanalysis}, although, as
noted above, no finite sweep bounds \(\rho\) over the whole simplex).
What deliberately does not scale is the demand that each world be
defended in print --- that constraint is the contribution, because it is
what converts an unbounded p-hacking surface into a finite, citable
disagreement.

\section{Simulation Evidence}\label{sec-simulation}

We present five designs: an illustrative design that shows what the
disciplined multiverse looks like (Design I), and four stress tests
(Designs II--V) that probe the framework --- including against itself.
One bookkeeping note applies throughout: the within-world distributions
and variances reported for Designs I to III pool estimates across Monte
Carlo replications, so there the within-world component contains
sampling variation as well as specification dispersion; in the empirical
applications of Section~\ref{sec-reanalysis}, which analyze one dataset
each, it is specification dispersion alone. Monte Carlo standard errors
for every design are collected in Table~\ref{tbl-simmcse} (Supplemental
Materials), with the replication as the unit: the MCSE of a bias is the
standard deviation of the per-replication world mean divided by
\(\sqrt{R}\), that of a coverage rate the same quantity for the
per-replication coverage share, and that of the pooled structural share
a batch-means estimate over ten batches of replications. No fit failed
in any design (0 of 204,800 cells in each scenario of Design II, 0 of
9,600 in Design I, 0 in Design III and in Scenario D). Because the
pooled share of Designs I to III carries sampling variation in its
within-world component, the table also reports the share computed one
replication at a time, as it is in the applications; that version is
higher (median 99.6 percent in Design I) because its within-world
component is specification dispersion alone.

\subsection{Design I: The Anatomy of a Misleading
Multiverse}\label{design-i-the-anatomy-of-a-misleading-multiverse}

\begin{figure}

\centering{

\includegraphics[width=0.85\linewidth,height=\textheight,keepaspectratio]{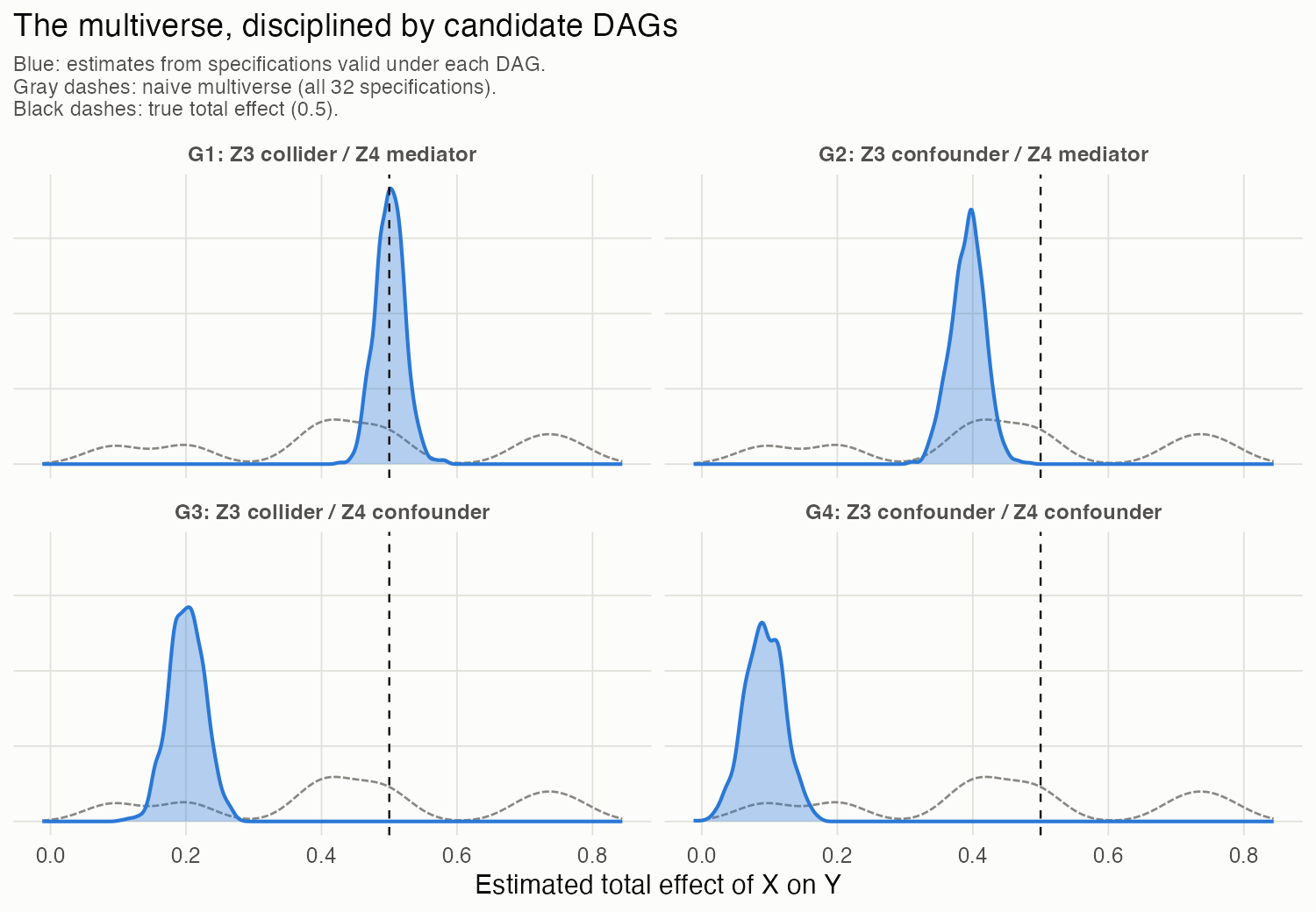}

}

\caption{\label{fig-bydag}The disciplined multiverse in Design I. Blue:
specifications valid under each candidate graph; gray: the naive
multiverse; dashed line: the true effect.}

\end{figure}

\begin{figure}

\centering{

\pandocbounded{\includegraphics[keepaspectratio]{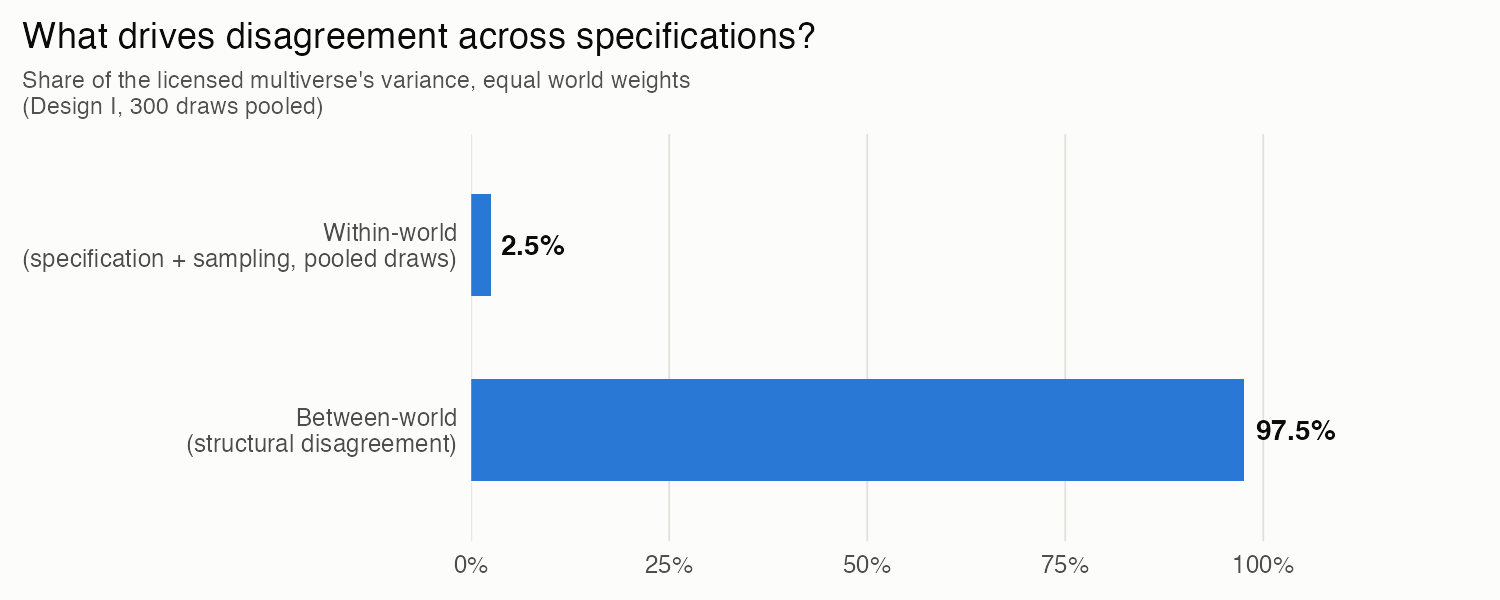}}

}

\caption{\label{fig-decomp}Decomposition of Design I multiverse
variance.}

\end{figure}

The illustrative design uses five candidate controls with one of each
role: a true confounder (\(Z_1\)), a pure outcome cause (\(Z_2\)), an
M-collider with latent parents (\(Z_3\)), a mediator (\(Z_4\)), and an
instrument (\(Z_5\)); the true total effect is \(\tau = 0.5\). The
analyst is uncertain whether \(Z_3\) is a collider or a confounder and
whether \(Z_4\) is a mediator or a confounder, giving four candidate
DAGs. Across 300 Monte Carlo draws (\(n = 2{,}000\); all \(2^5 = 32\)
subsets), the naive multiverse looks \emph{robust} by conventional
standards --- sign stability 100 percent, significance rate 98.7
percent, robustness ratio 2.04, above the customary bar of 2
(\citeproc{ref-youngholsteen2017}{Young and Holsteen 2017}) --- while
pooling causally incompatible answers: the DAG-conditional means are
0.501 (true graph), 0.393, 0.201, and 0.092, a five-fold range in which
each world is internally tight (within-graph SD 0.02--0.03) and
internally ``robust'' (Figure~\ref{fig-bydag}). The figure repays a slow
read: the gray cloud is what a referee currently sees --- a wide but
sign-stable distribution that passes every pooled robustness metric; the
four blue panels are the same estimates sorted by the causal world that
licenses them, and the instability collapses \emph{within} each panel
while the panels themselves disagree --- the cloud's width was never
specification noise but four tight estimates of the same declared effect
under four incompatible identifying assumptions, only one of which can
be right, standing side by side. The structural share is \(\rho = 97.5\)
percent (Monte Carlo standard error 0.1 percentage points;
Figure~\ref{fig-decomp}): almost none of the multiverse spread is
specification noise. A reader given only the naive summary learns
nothing about the one question that matters --- which causal world to
believe.

\subsection{Design II: What the Decomposition Can and Cannot
See}\label{design-ii-what-the-decomposition-can-and-cannot-see}

\begin{figure}

\centering{

\pandocbounded{\includegraphics[keepaspectratio]{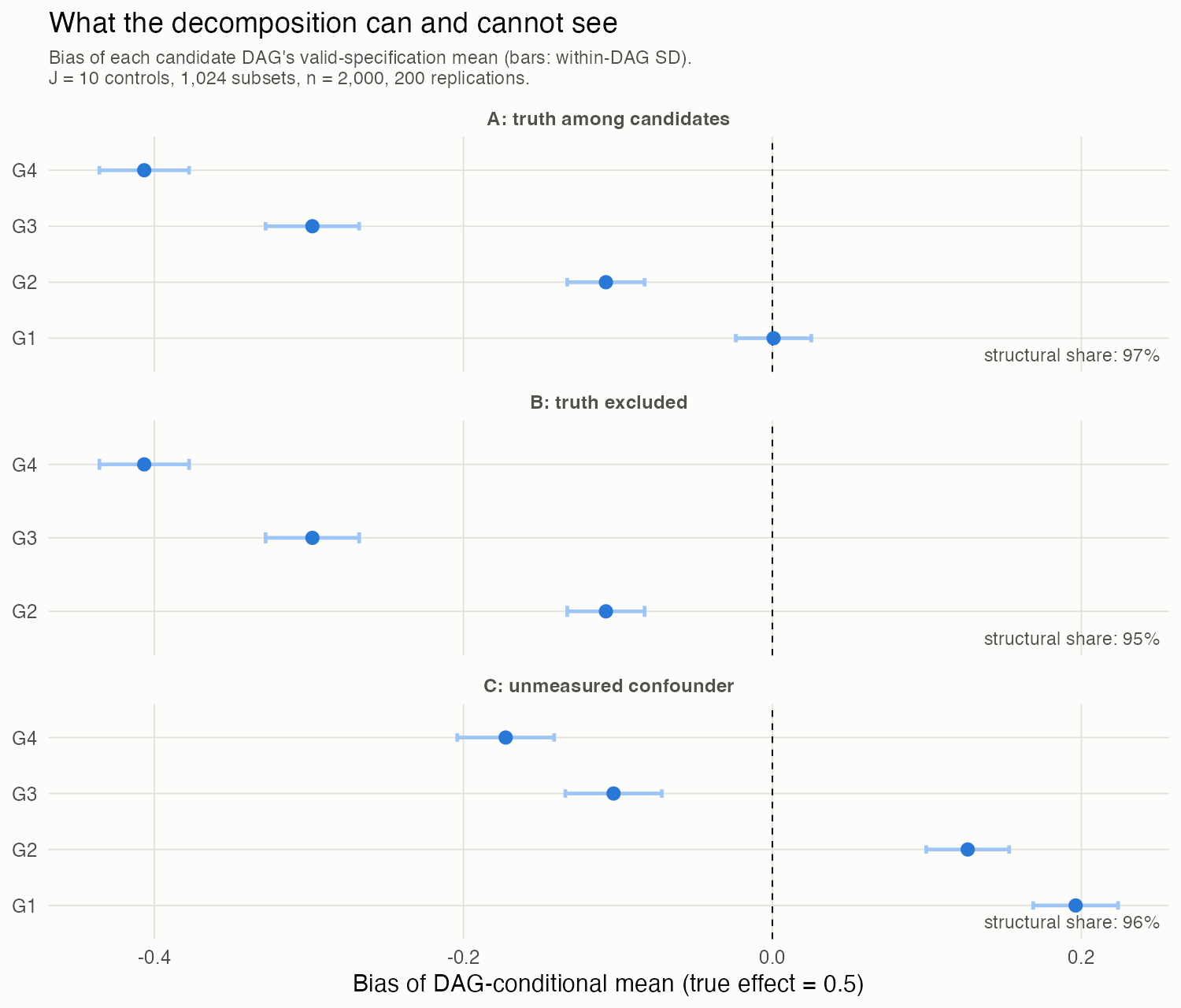}}

}

\caption{\label{fig-mainsim}Design II: bias of each candidate graph's
valid-specification mean across scenarios.}

\end{figure}

The main design enlarges the pool to \(J = 10\) (the five role variables
plus five pure-noise candidates), giving \(2^{10} = 1{,}024\) subsets
per draw (\(n = 2{,}000\), 200 replications), and evaluates three
scenarios (Figure~\ref{fig-mainsim}):

\textbf{Scenario A --- truth among the candidates.} With the true graph
in \(\mathcal{G}\), the framework behaves exactly as designed: the true
graph's valid-specification mean is unbiased (bias 0.001, MCSE 0.002),
its licensed cells' individual 95 percent intervals cover the truth 95.2
percent of the time on average (MCSE 1.3 percentage points; this is the
average coverage of the single-cell intervals, not the coverage of an
interval for the world mean, which would require the joint distribution
of the correlated cells), the wrong DAGs' distributions are tight but
wrong (average cell coverage 0.0--0.4 percent), and the structural share
is 97.1 percent (MCSE 0.3 percentage points). The naive multiverse's
specifications, by contrast, cover the truth only 23 percent of the time
(MCSE 0.4 percentage points). Noise candidates are correctly classified
optional everywhere and do essentially no harm.

\textbf{Scenario B --- truth excluded from the candidates.} When the
analyst never entertains the true graph (candidates are the three wrong
ones), every DAG-conditional distribution remains internally tight and
internally significant --- \emph{nothing inside the multiverse warns
that all candidates are wrong}. Two things still go right: the
structural share stays high (95 percent, MCSE 0.5 percentage points),
correctly reporting that the causal question is unresolved, and coverage
collapses only for the quantity no candidate can identify. The lesson is
the reading rule of Section~\ref{sec-framework}: a high structural share
is an instruction to resolve structure \emph{by design} --- benchmarks,
testable implications, sensitivity analysis --- never a license to
average across the candidates and move on.

\textbf{Scenario C --- a violation shared by all candidates.} Adding an
unmeasured confounder of \(X\) and \(Y\) (absent from every candidate
graph) shifts every DAG-conditional mean (bias +0.20, MCSE 0.002, under
the otherwise-true graph; coverage 0 percent) while leaving all internal
diagnostics untouched. Sufficiency violations are invisible to any
specification search --- the decomposition included. This is not a
defect of the decomposition so much as the boundary of what model
variation can ever detect; it is why the framework's output should be
paired with sensitivity analysis (\citeproc{ref-oster2019}{Oster 2019};
\citeproc{ref-cinellihazlett2020}{Cinelli and Hazlett 2020}), and why we
read the job-training reanalysis (Section~\ref{sec-reanalysis}) as the
empirical face of exactly this scenario.

\subsection{Designs III--V: Stress Tests of the Share, the Scale, and
the
Selectors}\label{designs-iiiv-stress-tests-of-the-share-the-scale-and-the-selectors}

\textbf{Design III --- is the structural share an artifact of the
design?} A grid over sample size and signal strength (all structural
coefficients scaled by \(s\); 200 replications per cell) shows that
\(\rho\) behaves as a \emph{diagnostic}, not a constant of the
framework. With strong signals it is high regardless of \(n\) (97.1
percent at \(n = 2{,}000\), 89.1 percent at \(n = 500\)); with halved
signals (\(s = 0.5\)) the between-DAG gaps shrink quadratically --- the
between-world range is 0.09 to 0.10 across the two sample sizes --- and
\(\rho\) falls to 39.7 percent at \(n = 500\) and 72.4 percent at
\(n = 2{,}000\) (MCSE 2.8 and 1.8 percentage points), correctly
reporting that the structural dispute matters little for the estimate at
that signal strength. The true graph's DAG-conditional mean is unbiased
in every cell (\textbar bias\textbar{} \textless{} 0.002). This
diagnostic behavior belongs to the pooled share, whose within-world
component contains sampling variation; computed one replication at a
time, the share has a median above 96 percent in every cell
(Table~\ref{tbl-simmcse}), because on one dataset the within-world
specification dispersion is small relative to the between-world gaps. On
a single dataset, therefore, the share must be read with the
between-world range in effect units, 0.09 to 0.10 in the halved-signal
cells against 0.41, which is the point Scenario D below makes in its
sharpest form.

(Full sensitivity grid: Table~\ref{tbl-designiii}, Supplemental
Materials.)

\textbf{Scenario D: world means that differ by sampling noise alone.}
The shares above are high partly by construction, because the contested
edges carry large coefficients. Scenario D (Supplemental Materials)
supplies the noise floor: the two contested coefficients, \(Z_4 \to Y\)
and \(U_2 \to Y\), are set to zero, so that adjusting for the mediator
or for the M-collider is harmless and all four worlds share the
population mean 0.2. Pooled across 200 replications the share is 0.1
percent (MCSE 0.1 percentage points) and the between-world range of the
pooled means is 0.002. Computed one replication at a time, however, the
share has a median of 49 percent with 5th and 95th percentiles of 5 and
93 percent, while the per-replication between-world range averages
0.017. On a single dataset the share alone cannot distinguish a
structural dispute from sampling noise when the within-world
specification dispersion is small, which is why the reading rule pairs
it with the between-world range in effect units and, in
Section~\ref{sec-reanalysis}, with paired-bootstrap intervals for that
range: in Scenario D the range is negligible, in Design I it is 0.41.

\textbf{Design IV --- random structures.} As a scaling and validity
check, we drew 100 random data-generating structures at \(J = 12\)
(roles assigned at random among confounder, precision, instrument,
mediator, M-collider, and noise; coefficients random; up to three
contested mediator/collider roles per structure, giving candidate sets
of up to eight DAGs; 4,096 specifications per multiverse; 20
replications each). Across all 100 structures the true DAG's conditional
mean is unbiased (maximum absolute bias 0.017, median signed bias
0.0007, median absolute bias 0.003 with a bootstrap standard error over
structures of 0.0005), and the structural share behaves as designed:
median 96.5 percent (IQR 94.9--97.4; bootstrap standard error of the
median over the 95 contested structures 0.2 percentage points) when
roles are genuinely contested, and exactly zero --- by construction ---
for the five structures whose candidate set contains no disagreement,
where the framework reduces to Young-Holsteen analysis on the valid set.
Enumeration and estimation for a full structure (up to eight graphs
\(\times\) 4,096 specifications) take seconds on a laptop, and no fit
failed.

\textbf{Design V --- can a selector replace the multiverse?} A natural
objection holds that modern variable selection makes enumeration
obsolete: let a data-driven selector choose the controls. Design V is a
selector-and-OLS stress test on the Design I data-generating process,
not a performance comparison of published estimators, whose eligibility
conditions the pool below violates. Two world-blind arms are fed the
full candidate pool. The first is post-double-selection in the sense of
Belloni, Chernozhukov, and Hansen (\citeproc{ref-bch2014}{2014}): a
BIC-tuned lasso of the outcome on the pool with the exposure
unpenalized, a BIC-tuned lasso of the exposure on the pool, the union of
the two selected sets, and post-selection OLS. The second is an
\emph{outcome-adaptive penalization analogue}: the pool columns are
rescaled by the squared OLS outcome coefficients, so that the lasso
penalty on the original scale is inversely proportional to them ---
covariates weakly related to the outcome are penalized more, as in the
outcome-adaptive lasso of Shortreed and Ertefaie
(\citeproc{ref-shortreed2017}{2017}) --- with a fixed penalty of
\(n^{-3/4}\) on a Gaussian regression of the exposure, the union with
the outcome-selected set of the first arm, and post-selection OLS. The
published procedure has a binary treatment, a penalized logistic
propensity, an imbalance-based tuning rule and inverse-probability
weighting (\citeproc{ref-shortreed2017}{Shortreed and Ertefaie 2017,
1112--14}); none of that procedure's inferential guarantees is claimed
for the analogue, and nothing below is a finding about the published
estimator. Both arms retain the mediator in 100 percent of replications
--- it is, after all, the best predictor of the outcome in the pool ---
and deliver a bias of \(-0.42\) on a true effect of \(0.5\) (MCSE
0.0015) with \textbf{zero percent} coverage of post-selection OLS
intervals for the true total effect. This is the consequence of feeding
an ineligible pool to a prediction-driven selector, not a defect of the
selectors: prediction-optimal selection is causal-role-blind by
construction, so letting the data choose the controls cannot repair the
pooled multiverse's problem. (Selection throughout uses the Gaussian BIC
\(n \log(\mathrm{RSS}/n) + \mathrm{df}
\log n\) along a 30-value penalty path, except for the fixed-penalty
stage of the analogue; 300 replications, no fit failures.) Applied
\emph{within} each candidate world --- the world's required controls
forced into both selection regressions and the OLS, double selection
over that world's optional set --- the same post-double-selection
procedure is unbiased with nominal coverage in the true world (bias
\(0.001\), coverage 95.0 percent) and reproduces each wrong world's
characteristic bias exactly; the structural share computed across the
four within-world arms is 97.6 percent, indistinguishable from the
enumerated multiverse's 97.5. The conclusion is narrow but useful:
data-driven selection and the disciplined multiverse are complements at
different levels --- selection can replace \emph{enumeration within} a
world (and brings efficiency when the optional pool is large), but no
selector can adjudicate \emph{between} worlds, because the worlds
disagree about which variables are eligible, not about which predict
(\texttt{sim/sim\_selectors.R}).

\section{Empirical Applications}\label{sec-reanalysis}

We first reanalyze both applications in Muñoz and Young
(\citeproc{ref-munozyoung2018}{2018}) with candidate DAGs, then take the
framework beyond reanalysis to a question the contested-control problem
has structured for fifty years: the union wage premium. All three
applications use public data; all results are reproducible from the
replication archive. For hurricanes we build a Muñoz-Young-style
multidimensional space (control sets, functional form, outliers,
estimation command; 144 cells against their 1,152); for job training we
use the 256 subsets of the Dehejia--Wahba covariate pool and extend this
space with functional-form choices --- the pool is not identical to
Muñoz and Young's; for the union premium we build the full space over
the entire control pool in two independent samples.

\subsection{Female-Named Hurricanes: Fragility Survives Every Causal
Reading}\label{female-named-hurricanes-fragility-survives-every-causal-reading}

\begin{figure}

\centering{

\includegraphics[width=0.88\linewidth,height=\textheight,keepaspectratio]{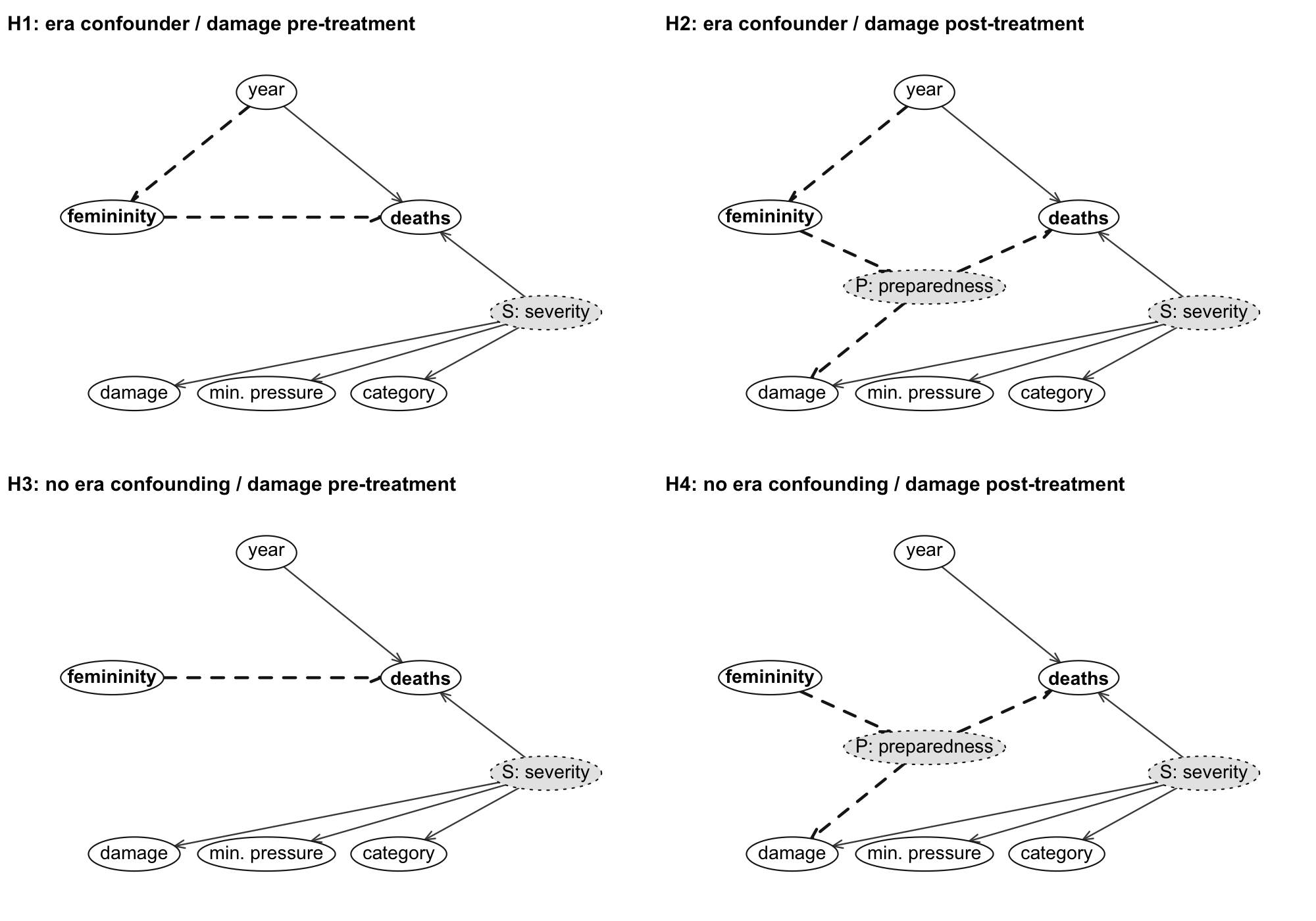}

}

\caption{\label{fig-daghur}Candidate graphs for the hurricane
application. Solid edges are common to all four worlds; dashed edges are
contested (the era edge year to femininity in H1 and H2; the
preparedness path in H2 and H4, which replaces the direct edge of H1 and
H3); dotted gray nodes are latent.}

\end{figure}

The first application asks whether feminine-named hurricanes kill more
people (\citeproc{ref-jung2014}{Jung et al. 2014}; 92 Atlantic storms,
1950--2012, as reanalyzed by \citeproc{ref-munozyoung2018}{Muñoz and
Young 2018}). We estimate negative binomial models of deaths on name
femininity (1--11 scale) across all \(2^4 = 16\) subsets of the
candidate controls: normalized damage, minimum pressure, category, and
year. The estimand, stated once for the record: the treatment contrast
is a one-point increase in the femininity index; the outcome scale is
the log conditional mean of deaths (the negative-binomial coefficient),
with log-OLS cells reported as a separate family below; the target
population is the 92 analyzed storms (the 94 landfalling Atlantic
hurricanes of 1950--2012 less the two that Jung et al.
(\citeproc{ref-jung2014}{2014}) exclude); and the estimating functional
is the femininity coefficient at mean severity under each licensed
control set. Because the moderators are centered at the sample means,
that coefficient is the average over the 92 storms of the storm-specific
log-linear slope in femininity --- a conditional contrast averaged over
the population, not a contrast of population-averaged fitted means; we
return to the difference below. Two structural questions are genuinely
contested. First, \emph{era}: before 1979 all storms received female
names while mortality declined secularly, making year a plausible
confounder. Second, \emph{mechanism}: the proposed causal channel is
that feminine names lower perceived threat and hence preparedness ---
but if so, property damage is partly \emph{downstream} of the name, and
controlling for it is overcontrol. In the graphs where damage is
pre-treatment (H1, H3) it is a proxy for latent severity --- an optional
precision control, not a confounder of name and deaths, because in those
graphs severity affects deaths but does not cause the assignment of a
name; in the graphs where it is post-treatment (H2, H4) it is a
descendant of the latent mediator preparedness, not itself a node on the
causal path, and the adjustment criterion forbids it as a descendant of
a causal node. Crossing the two contests yields four candidate DAGs:
year is required under the era-confounding graphs (H1, H2) and damage is
forbidden under the preparedness-mechanism graphs (H2, H4).
Figure~\ref{fig-daghur} draws the four graphs, with the contested edges
dashed and the latent severity and preparedness nodes marked, and
Table~\ref{tbl-roles-hurricane} gives the roles the adjustment criterion
assigns to each control under each graph, the number of admissible
control sets, and the number of licensed cells. In H2 and H4 the name
acts on deaths only through preparedness, so every licensed cell still
targets the total effect. H3, in which neither contested edge is
present, licenses all 16 control sets, and 12 of the 16 are licensed by
more than one world, so the licensed multiverse overlaps heavily across
worlds.

\begin{longtable}[]{@{}lllll@{}}
\caption{Roles of the candidate controls under the four hurricane
worlds, from
\texttt{mv\_classify()}.}\label{tbl-roles-hurricane}\tabularnewline
\toprule\noalign{}
Control & H1 & H2 & H3 & H4 \\
\midrule\noalign{}
\endfirsthead
\toprule\noalign{}
Control & H1 & H2 & H3 & H4 \\
\midrule\noalign{}
\endhead
\bottomrule\noalign{}
\endlastfoot
damage\_norm & optional & forbidden & optional & forbidden \\
min\_pressure & optional & optional & optional & optional \\
category & optional & optional & optional & optional \\
year & required & required & optional & optional \\
Admissible control sets (of 16) & 8 & 4 & 16 & 8 \\
Licensed cells (of 144) & 72 & 24 & 144 & 48 \\
\end{longtable}

\begin{figure}

\centering{

\includegraphics[width=0.85\linewidth,height=\textheight,keepaspectratio]{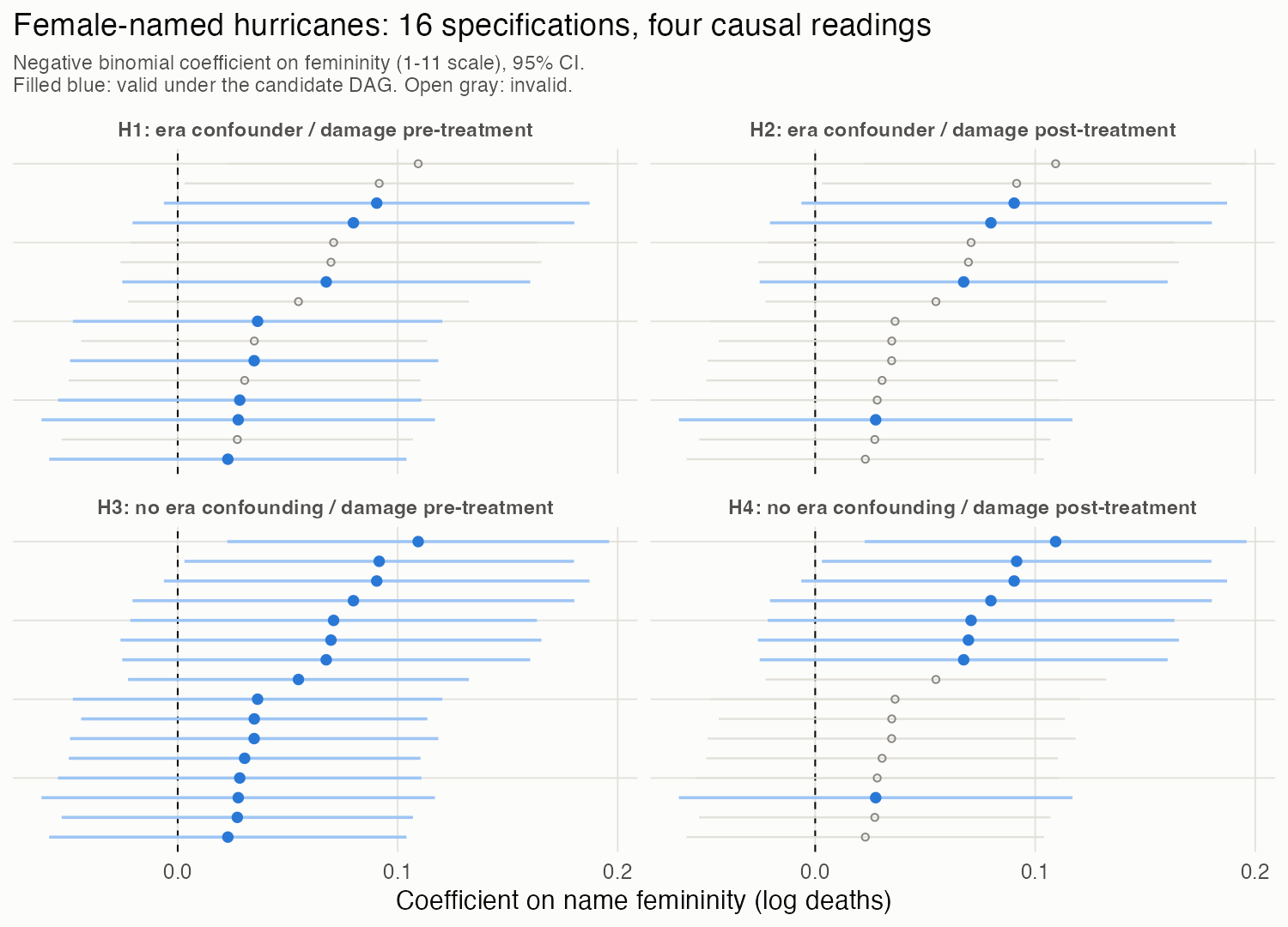}

}

\caption{\label{fig-hurricane}Female-named hurricanes: sixteen
specifications under four causal readings. Filled points are valid under
the candidate DAG.}

\end{figure}

The result (Figure~\ref{fig-hurricane}) is the mirror image of the
simulations in Section~\ref{sec-simulation}. On the control-set
dimension alone (16 specifications), the naive multiverse is fragile
(significance rate 12.5\%, robustness ratio 1.04) --- and so is every
DAG-conditional multiverse: per-graph mean coefficients range only from
0.048 to 0.076, significance rates from 0\% to 25\%, and the structural
share is 15.5\%.

\begin{figure}

\centering{

\pandocbounded{\includegraphics[keepaspectratio]{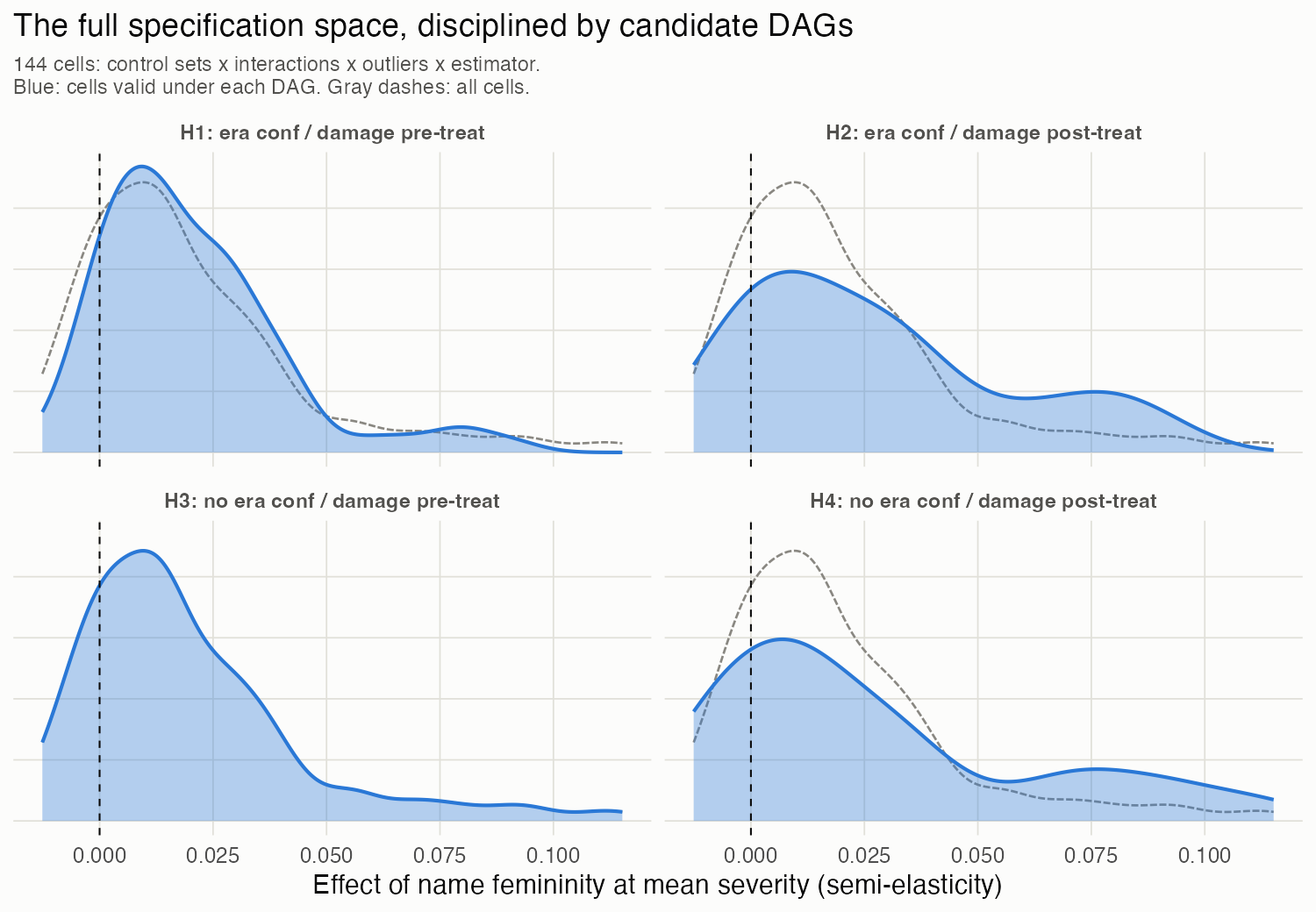}}

}

\caption{\label{fig-hurricanefull}The full 144-cell specification space
by candidate DAG.}

\end{figure}

Extending the space to the full Muñoz-Young-style grid sharpens the
verdict. We cross the 16 control sets with functional form (interactions
of femininity with damage and with pressure, where the moderator is
included, with moderators mean-centered so that the femininity
coefficient is the effect at mean severity), outlier handling (full
sample versus dropping the two deadliest storms), and estimation command
(negative binomial versus OLS on log deaths) --- 144 cells in total. One
of these dimensions is not a modeling choice on a par with the others:
dropping the two deadliest storms selects the sample on the outcome, and
a treatment contrast estimated on the selected sample targets a
different population from the full-sample contrast even under a valid
adjustment set. We therefore keep the trimmed cells in the grid, because
the original space has them, but report them as a
\emph{selection-sensitivity} family rather than pooling them with the
full-sample cells in any decomposition read as causal. Concretely, the
16-cell pool and the standardized 12-cell family introduced below are
full-sample families and carry the structural reading; the 72-cell
negative-binomial family, the 144-cell pooled grid and the high-severity
pressure-interaction family mix full-sample and trimmed cells and are
reported as descriptive summaries of the Muñoz--Young grid. The naive
significance rate falls to 2.8 percent (Muñoz and Young report under 5
percent on their 1,152-model space); the Jung-style interaction cells
themselves are significant in only 2.5 percent of cases; and every
DAG-conditional robustness ratio sits below 1. Within the 72-cell
negative-binomial family --- reported descriptively because it mixes
full and trimmed samples --- the structural share is \textbf{11.3
percent}; within the 72 log-OLS cells, whose coefficient targets the
conditional mean of a transformed outcome and is therefore a different
functional, it is 5.2 percent. Pooling the two families, as Muñoz and
Young's space does, gives 1.9 percent; we report that figure for
comparability with the original space but do not treat it as a
decomposition of one estimand, because a negative-binomial coefficient
and a log-OLS coefficient are not commensurable merely because both are
read as semi-elasticities (Figure~\ref{fig-hurricanefull}). For the
negative-binomial family the Dirichlet sweep of
Section~\ref{sec-framework} gives 5th, 50th, and 95th percentiles of
4.1, 9.6, and 12.6 percent, and over every weighting that gives each
world at least 5 percent the share lies between 3.8 and 13.8 percent
(6.6 to 13.2 percent at 10 percent); for the pooled 144 cells the
supremum on the same restricted domain is 2.0 percent. The paired
bootstrap adds the sampling side. Negative-binomial fits on resampled
storms do not always converge, and the bootstrap treats non-convergence
as a failure rather than retaining the last iterate: a fit counts as
failed when it errors, ends with a non-converged final iteration, or
carries a dispersion-iteration warning, and two prespecified retries (a
higher iteration limit, then a moment-based starting value for the
dispersion) are attempted before a cell is given up. Of the 36,000
negative-binomial fits, 32,035 converged at the first attempt, 3,111
after a retry, and 854 (2.4 percent, in 161 of the 500 resamples)
failed. The declared world set is fixed before resampling: a world whose
licensed cells include a failed fit is left out of that resample rather
than averaged over the surviving cells, and the share is left out with
it rather than recomputed over the surviving worlds, so that neither the
within-world nor the between-world mixture changes from resample to
resample; the number of contributing resamples is reported beside each
statistic (372 to 396 of 500 for the 16-cell world means and their
share, 339 for the 72-cell family, 364 for the standardized family;
Table~\ref{tbl-boot}). Fitting success is not random: the probability
that a resample contains a failed fit rises from 3 percent when none of
the three deadliest storms is drawn to over 80 percent when six copies
of them are (replication archive,
\texttt{phaseB\_nb\_failure\_pattern.txt}), so the percentiles below are
conditional numerical diagnostics --- conditional on fitting success ---
and not validated confidence limits. Under this algorithm the
negative-binomial share has a bootstrap standard error of 10 percentage
points (2.5th to 97.5th percentiles 0.3 to 34.0 percent), and each world
mean in the 16-cell pool has a standard error of 0.044 to 0.053 with a
percentile interval that includes zero. The same statistics computed
with unchecked fits and renormalized world means, the algorithm of an
earlier version of this analysis, differ by at most 0.003 in the
standard error of any average-severity world mean (0.007 for the
high-severity means) and by at most 0.017 in any percentile of a world
mean, and recomputing the shares over surviving worlds instead of fixing
the world set moves their 2.5th and 97.5th percentiles by at most 0.3
percentage points (replication archive,
\texttt{phaseB\_bootstrap\_hurricane\_algorithm\_comparison.csv} and
\texttt{phaseB\_bootstrap\_hurricane\_share\_v2\_vs\_v3.csv}), so
neither the convergence audit nor the fixed-world convention changes a
reading. Ninety-two storms do not pin down the share or any world mean
with much precision, and nothing in the resampling distribution favors
one world over another. The interpretation is now sharper than the
original could be: at the average-severity contrast, the spread of
estimates is almost entirely ordinary specification noise, and the
finding is weak \emph{within every plausible causal world}. Causal
discipline does not rescue the female-hurricanes effect at that
contrast; it confirms its fragility there.

Two further checks ask whether the between-world differences that do
appear --- the H4 mean exceeding the H1 mean by 0.027 in the 16-cell
pool and by 0.022 in the negative-binomial family --- are differences in
a common quantity or artifacts of comparing coefficients that target
different contrasts. The first is a standardization. For every
negative-binomial cell we evaluate the fitted conditional mean at each
storm's own covariates with its femininity at \(f_i\) and at \(f_i +
1\), restrict to the storms for which \(f_i + 1\) stays inside the
observed range of the index (73 of 92), and take the log of the ratio of
the two population sums: a population-averaged log rate ratio for a
one-point increase, evaluated over a common population, which equals the
coefficient exactly in a cell without interactions and differs from it
only through the interaction terms. The second is a calibration
diagnostic for the same cells: the ratio of the fitted total deaths over
that population to the observed total. The diagnostic sorts the 72
negative-binomial cells into two groups. Every cell whose control set
excludes normalized damage predicts between 0.75 and 1.0 of the observed
deaths; every cell that includes damage as a linear term in the log link
predicts 3.9 to 41 times the observed deaths in the full sample (1.8 to
4.6 times in the trimmed sample), because a single storm with normalized
damage near \$75 billion is assigned a fitted mean in the tens of
thousands. In the twelve full-sample cells that interact femininity with
damage, the standardized contrast is 0.57 to 0.92 against coefficients
of 0.026 to 0.055, and one storm accounts for more than 90 percent of
the summed predicted increase in deaths. Those cells' fitted means
cannot be standardized credibly, and no adjustment-set argument repairs
an estimation model that is this far from the data; they illustrate the
distinction of Section~\ref{sec-framework} between graph eligibility and
estimation-model adequacy. The standardized decomposition is therefore
defined on the fixed family of twelve full-sample negative-binomial
cells whose control set excludes damage. That family is licensed
identically by H1 and H2 (year required) and by H3 and H4 (year
optional), so within it the damage contest disappears by construction
and only the era contest remains: the standardized world means are 0.044
for H1 and H2 and 0.058 for H3 and H4, a gap of 0.014, and the
structural share is 3.7 percent. The reading is unambiguous. The part of
the between-world spread that the coefficient family attributes to the
damage contest --- whether damage is a pre-treatment severity proxy or a
post-treatment descendant of preparedness --- rests entirely on cells
whose fitted means over-predict deaths by an order of magnitude; it is
coefficient sensitivity to a poorly calibrated term, not a measured
structural disagreement. What survives standardization is a small era
contrast, well inside its sampling uncertainty (Table~\ref{tbl-boot}).
In this application the decomposition therefore isolates nothing that
deserves the name structural uncertainty, and we do not claim that it
does; its contribution is to show that the pooled fragility verdict
holds within every candidate world and that the between-world
differences the coefficient family displays are not causal quantities. A
grid with damage entered on the log scale would likely restore
calibration; we keep the original grid because the exercise is a
reanalysis of it, and note the alternative as the obvious next step.

Because the original claim centered on the most damaging storms, the
verdict at average severity needs a high-severity companion, and the
candidate graphs constrain how that companion may be defined. In the
worlds where damage is a descendant of the mediator preparedness (H2,
H4), restricting the analysis to storms with high \emph{realized} damage
defines a post-treatment subgroup, so the severity contrast must be
defined on a pre-treatment intensity measure such as minimum pressure.
We compute it from the cells that already interact femininity with
pressure: the femininity coefficient evaluated at a minimum pressure of
942 millibars, the 10th percentile of the 92 storms (the sample mean is
965), with a standard error from the coefficient covariance. Each world
contributes the pressure-interaction cells whose control set it
licenses; in H2 and H4 the cells that also interact femininity with
realized damage drop out, because damage is forbidden there. In the
negative-binomial family the high-severity coefficient is negative in
every world (world means of \(-0.036\), \(-0.032\), \(-0.030\), and
\(-0.019\) over 12, 4, 24, and 8 cells, with a mean standard error of
about 0.065) and is significant in none of the 24 cells; in the log-OLS
family it is slightly positive (0.010 to 0.015) and again never
significant. The paired bootstrap places each negative-binomial world
mean between about \(-0.17\) and \(+0.11\) (Table~\ref{tbl-boot}). At
the intense end of a pre-treatment severity scale, then, the
negative-binomial world means are negative, and none of the licensed
high-severity estimates is statistically significant (five of the 24
negative-binomial cells have positive point estimates, the largest about
0.023); the conditional bootstrap intervals remain wide. The fragility
verdict at the average-severity contrast extends to the high-severity
contrast defined on pressure. A contrast defined on realized damage, the
original paper's framing, is admissible only in H1 and H3, where damage
is pre-treatment.

\subsection{Job Training: An External Benchmark Discriminates among the
DAGs}\label{job-training-an-external-benchmark-discriminates-among-the-dags}

The second application revisits Muñoz and Young's within-study
comparison: NSW experimental treated units {[}Dehejia-Wahba sample,
\(n=185\){]} combined with 15,992 CPS-1 controls, estimating the effect
of training on 1978 earnings (in 1982 dollars, the scale of the
distributed data; Dehejia and Wahba
(\citeproc{ref-dehejiawahba1999}{1999}), Table 2) across all
\(2^8 = 256\) subsets of eight candidate controls (six demographics plus
1974 and 1975 earnings). Muñoz and Young describe their ingredients for
this application as past wages and unemployment status, age, race,
marital status and education, also 256 models, and report a mean of
\(-\$815\), a modeling standard error of \(\$2{,}639\) and sign
stability of 63 percent (\citeproc{ref-munozyoung2018}{Muñoz and Young
2018}, Table 6); our pool is the Dehejia--Wahba covariate set, so the
two spaces are not identical, and ours reproduces their pattern
qualitatively rather than their figures. Because the control-set space
leaves the defended worlds with one or two licensed specifications each,
we cross it with four within-world functional-form dimensions that do
not touch admissibility, each a transformation of a control already in
the set: a quadratic in age, a quadratic in education, and the
zero-earnings indicators \(u_{74} = 1(\text{re74} = 0)\) and
\(u_{75} = 1(\text{re75} =
0)\), each available only when its parent control is included. The
result is a space of 1,296 cells that contains the least-squares
comparison specification of Dehejia and Wahba
(\citeproc{ref-dehejiawahba1999}{1999}, Table 3, note (a)) --- which
omits marital status and is therefore licensed by none of the worlds
below --- estimated by OLS throughout. The estimand card: the treatment
contrast is NSW participation versus non-participation; the outcome is
1978 earnings (1982 dollars); the target population is the NSW treated,
the population of the experimental benchmark; and the estimating
functional is, in the first pass, the OLS coefficient on the treatment
indicator under each licensed cell --- which equals the average effect
on the treated only under effect homogeneity across the adjustment
strata, and can differ from it under valid adjustment (a binary
confounder with equal stratum probabilities, propensities 0.2 and 0.8
and stratum effects 0 and 1 gives a common-slope coefficient of 0.5
against an ATT of 0.8) --- and, in the second pass, an ATT-targeted
doubly robust estimator on the same cells. The experimental benchmark,
\(+\$1{,}794\), is itself an estimate, with a standard error of
\(\$633\) in the experimental sample
(\citeproc{ref-dehejiawahba1999}{Dehejia and Wahba 1999}), and it shares
the 185 treated units with every observational cell.

\begin{figure}

\centering{

\includegraphics[width=0.95\linewidth,height=\textheight,keepaspectratio]{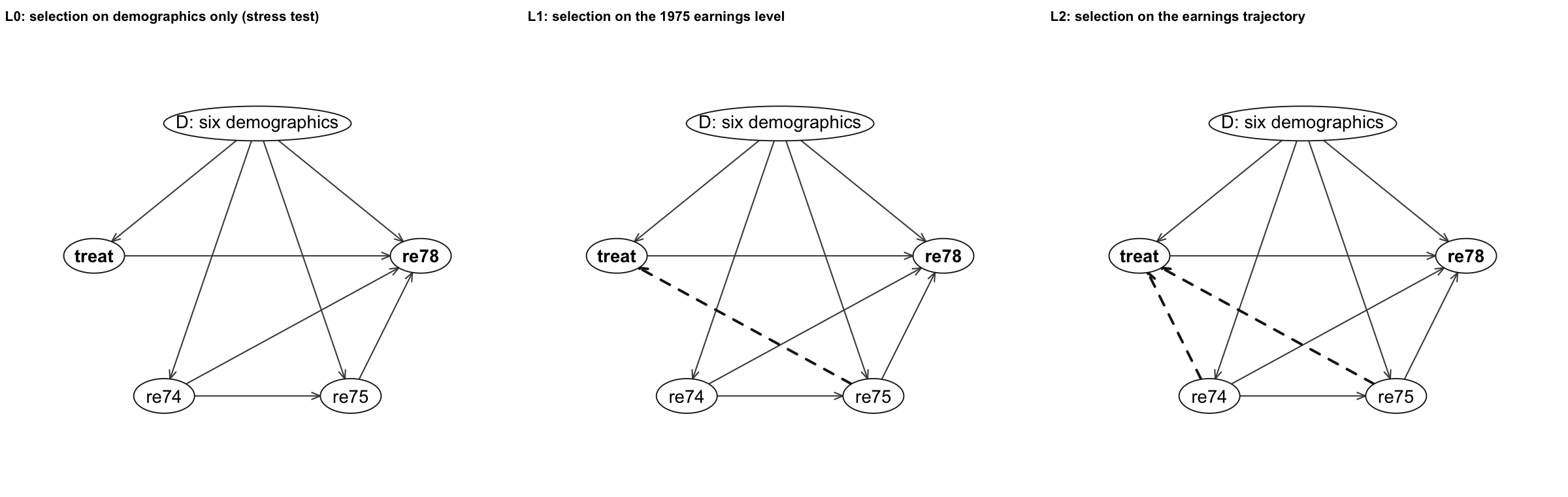}

}

\caption{\label{fig-daglal}Candidate graphs for the job-training
application. D stands for the six demographic controls, which every
world treats as confounders; the contested edges (dashed) are the
selection edges from lagged earnings into treatment.}

\end{figure}

The naive multiverse on the 256 control sets reproduces the original
pathology: mean \(-\$1{,}153\), modeling SD \(\$2{,}442\), and sign
stability of 49.2 percent, a coin flip (50.8 percent of specifications
are positive; the mean is negative). On the 1,296 cells the naive mean
is \(-\$124\) with a modeling SD of \(\$1{,}880\), and 73 percent of
cells are positive while the mean is negative (mean-sign sign stability
27 percent). The candidate DAGs (Figure~\ref{fig-daglal},
Table~\ref{tbl-roles-lalonde}) contest a single substantive question:
\emph{what drives selection into training?} Under L1 (selection on the
1975 earnings level) and L2 (selection on the earnings
\emph{trajectory}, the Ashenfelter dip), lagged earnings are
confounders. L0 (selection on demographics only, lagged earnings mere
precision variables) is a different kind of entry: it is the world
implicit in the early observational literature and in every
naive-multiverse cell that omits lagged earnings, and we can cite no
account of program selection under which it holds. We therefore carry it
as a \emph{historical stress-test world outside the defended candidate
set}, reported because it is where the naive multiverse's instability
lives, not because we defend it. The licensed counts must be read first.
On control sets alone, L0 admits four (the six demographics plus any
subset of the two earnings variables), L1 two (1975 earnings required,
1974 optional), and L2 one (all eight controls); with the
functional-form dimensions the worlds license 36, 24, and 16 cells,
nested in that order, so 24 of the 36 licensed cells are licensed by
more than one world.

\begin{longtable}[]{@{}
  >{\raggedright\arraybackslash}p{(\linewidth - 6\tabcolsep) * \real{0.2500}}
  >{\raggedright\arraybackslash}p{(\linewidth - 6\tabcolsep) * \real{0.2500}}
  >{\raggedright\arraybackslash}p{(\linewidth - 6\tabcolsep) * \real{0.2500}}
  >{\raggedright\arraybackslash}p{(\linewidth - 6\tabcolsep) * \real{0.2500}}@{}}
\caption{Roles of the candidate controls under the three job-training
worlds, from \texttt{mv\_classify()}; L0 is the stress-test world
outside the defended set.}\label{tbl-roles-lalonde}\tabularnewline
\toprule\noalign{}
\begin{minipage}[b]{\linewidth}\raggedright
Control
\end{minipage} & \begin{minipage}[b]{\linewidth}\raggedright
L0
\end{minipage} & \begin{minipage}[b]{\linewidth}\raggedright
L1
\end{minipage} & \begin{minipage}[b]{\linewidth}\raggedright
L2
\end{minipage} \\
\midrule\noalign{}
\endfirsthead
\toprule\noalign{}
\begin{minipage}[b]{\linewidth}\raggedright
Control
\end{minipage} & \begin{minipage}[b]{\linewidth}\raggedright
L0
\end{minipage} & \begin{minipage}[b]{\linewidth}\raggedright
L1
\end{minipage} & \begin{minipage}[b]{\linewidth}\raggedright
L2
\end{minipage} \\
\midrule\noalign{}
\endhead
\bottomrule\noalign{}
\endlastfoot
age, education, black, hispanic, married, nodegree & required & required
& required \\
re74 & optional & optional & required \\
re75 & optional & required & required \\
Admissible control sets (of 256) & 4 & 2 & 1 \\
Licensed cells (of 1,296) & 36 & 24 & 16 \\
\end{longtable}

\begin{figure}

\centering{

\includegraphics[width=0.9\linewidth,height=\textheight,keepaspectratio]{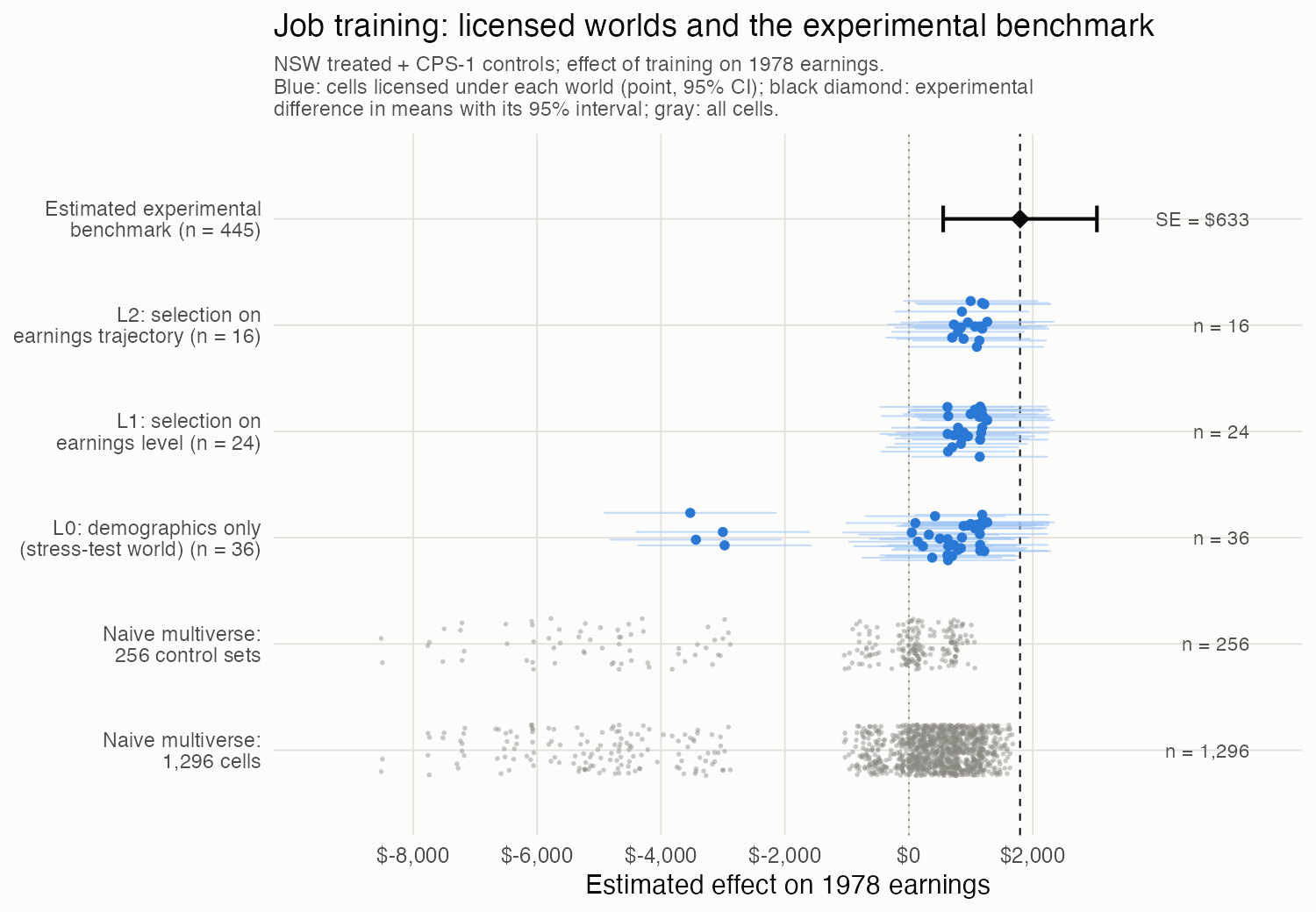}

}

\caption{\label{fig-lalonde}Job training: the naive multiverse (gray),
the cells licensed under each world (blue, with 95 percent intervals),
and the estimated experimental benchmark (black diamond with its 95
percent interval, SE \$633). The number of cells is printed on each row;
L0 is the stress-test world outside the defended set.}

\end{figure}

Under L0 the 36 licensed cells still swing widely (within-world SD
\(\$1{,}333\), range \(-\$3{,}528\) to \(+\$1{,}264\); 89 percent
positive). Under L1 the 24 licensed cells average \(+\$967\) (SD
\(\$222\)) and under L2 the 16 licensed cells average \(+\$1{,}003\) (SD
\(\$184\)) (Figure~\ref{fig-lalonde}). The zero-earnings indicators,
which the control-set space could not represent, move the defended
worlds' estimates toward the benchmark: within L1, cells without either
indicator average \(\$697\) and cells with one or both average \(\$922\)
to \(\$1{,}181\), in line with the matching literature
(\citeproc{ref-dehejiawahba1999}{Dehejia and Wahba 1999}). The
structural share across the three worlds is 13.2 percent at equal
weights (Dirichlet 5th, 50th, and 95th percentiles 4.6, 12.8, and 15.7
percent; between 2.3 and 16.2 percent over weightings that give each
world at least 5 percent), with a bootstrap standard error of 0.8
percentage points; restricted to the two defended worlds it is 0.8
percent, and L1 minus L2 is \(-\$36\) (bootstrap 2.5th to 97.5th
percentiles \(-\$97\) to \(+\$30\)), substantively nil.

What the benchmark does is discriminate among combinations of
identifying assumptions and estimators under stated assumptions; it does
not verify a graph. Every L1 and L2 cell lies inside the benchmark's 95
percent interval (\(\$554\) to \(\$3{,}035\)). Of L0's 36 cells, 24 are
L1's; the other 12, which omit 1975 earnings, all lie below that
interval (mean \(-\$900\), range \(-\$3{,}528\) to \(+\$500\)). Because
the observational cells and the benchmark share the treated units, the
paired bootstrap resamples the treated, the CPS controls, and the
experimental controls separately and recomputes both sides on the same
draw: the L1 mean falls short of the benchmark by \(\$828\) (bootstrap
SE \(\$416\), percentile interval \(-\$1{,}625\) to \(-\$7\)) and the L2
mean by \(\$791\) (SE \(\$424\), interval \(-\$1{,}606\) to \(+\$45\)),
while the L0 world mean, over all 36 of its cells, falls short by
\(\$1{,}450\) (SE \(\$409\), interval \(-\$2{,}219\) to \(-\$668\)), and
the 12 cells that only L0 licenses fall short by about \(\$2{,}700\) on
average.

The second pass replaces the coefficient by an estimator that targets
the population the benchmark describes. In every cell we fit a logistic
propensity model and a linear control-outcome regression on the cell's
own columns and form the augmented inverse-probability weighted estimate
of the effect on the treated (AIPW-ATT; fitted propensities truncated at
0.9, influence-function standard errors, the estimator of the companion
paper (\citeproc{ref-okubo2026att}{Okubo 2026})), so that the exposure
contrast, the outcome scale and the target population are the same in
every cell and the same as the benchmark's; the target remains fixed
while the adjustment set and nuisance-model specification vary across
cells. Overlap is adequate in the licensed cells: the largest control
propensity in any licensed cell is 0.89 in the 1,296-cell space (below
the truncation point, which binds only in eight unlicensed cells of the
full space) and 0.49 in the 256-cell space, and the effective sample
size of the control weights is at least 103 in every licensed cell
(median 228 to 256 in the extended space). The ATT-targeted estimates
move the defended worlds toward the benchmark and leave the structure of
the verdict unchanged. Over the 1,296 cells the L1 world averages
\(\$1{,}331\) (within-world SD \(\$133\)) and L2 \(\$1{,}405\) (SD
\(\$93\)), against \(\$967\) and \(\$1{,}003\) for the coefficients; the
cell with all eight controls and both zero-earnings indicators, the
ten-covariate specification of the propensity-score literature, gives
\(\$1{,}495\) (standard error \(\$678\)). The gaps to the benchmark
shrink from about \(\$800\) to \(\$463\) for L1 and \(\$390\) for L2 ---
on the 256 control sets, which the paired bootstrap refits, from
\(\$1{,}129\) and \(\$1{,}095\) to \(\$575\) (bootstrap SE \(\$453\),
percentile interval \(-\$1{,}390\) to \(+\$354\)) and \(\$523\) (SE
\(\$458\), interval \(-\$1{,}363\) to \(+\$430\)) --- so that both
defended worlds are compatible with the benchmark once the estimand is
matched to it. L0's 36 cells still span \(-\$3{,}622\) to \(+\$1{,}539\)
(SD \(\$1{,}440\)), and the 12 cells that only L0 licenses remain far
below the benchmark. L1 minus L2 is \(-\$73\) (256 sets: \(-\$52\),
interval \(-\$140\) to \(+\$35\)), and the structural share over the
three worlds is 14.6 percent on the 1,296 cells and 25.6 percent on the
256 sets (bootstrap SE 1.5 percentage points), the latter driven
entirely by L0. The ATT-targeted reanalysis thus narrows the defended
worlds' gaps to the experimental benchmark by roughly \(\$350\) to
\(\$400\). This change cannot be attributed uniquely to estimand
mismatch, because the estimating procedure and the nuisance models
change as well as the target; the constructed example above shows that
the targets \emph{can} differ under valid adjustment, and the data show
that the two procedures \emph{do} differ, without isolating the
mechanism. The qualitative reading --- the instability lives in
unlicensed cells, the defended worlds agree with each other, and the
benchmark discriminates among worlds --- is the same under both
functionals. Three lessons follow. First, the notorious instability of
the observational multiverse is \emph{produced by specifications that no
defended world licenses}: once lagged earnings are required, the
modeling SD falls from \(\$2{,}442\) to about \(\$200\) (coefficients)
or \(\$90\) to \(\$130\) (AIPW-ATT) across 24 and 16 cells. Second,
within-study comparisons can be read as \emph{external checks on
candidate worlds}: the benchmark rejects the cells that only L0 licenses
and, once the estimand is matched, is compatible with both L1 and L2
(with the coefficient functional it was borderline for L1, whose paired
interval excluded zero by seven dollars), which is discrimination among
worlds, not adjudication of one. Third, the residual gap to the
benchmark, about \(\$400\) to \(\$600\) with a paired standard error of
about \(\$453\), cannot be attributed from these data to unmeasured
confounding as against limited overlap, functional form, a population
mismatch, or sampling variation; specification search alone cannot
determine which of these mechanisms explains the gap, which is precisely
why robustness and validity must be kept distinct
(\citeproc{ref-oster2019}{Oster 2019}; \citeproc{ref-keele2020}{Keele,
Stevenson, and Elwert 2020}).

\subsection{The Union Wage Premium: A Fragility Verdict
Reversed}\label{the-union-wage-premium-a-fragility-verdict-reversed}

\begin{figure}

\centering{

\includegraphics[width=0.95\linewidth,height=\textheight,keepaspectratio]{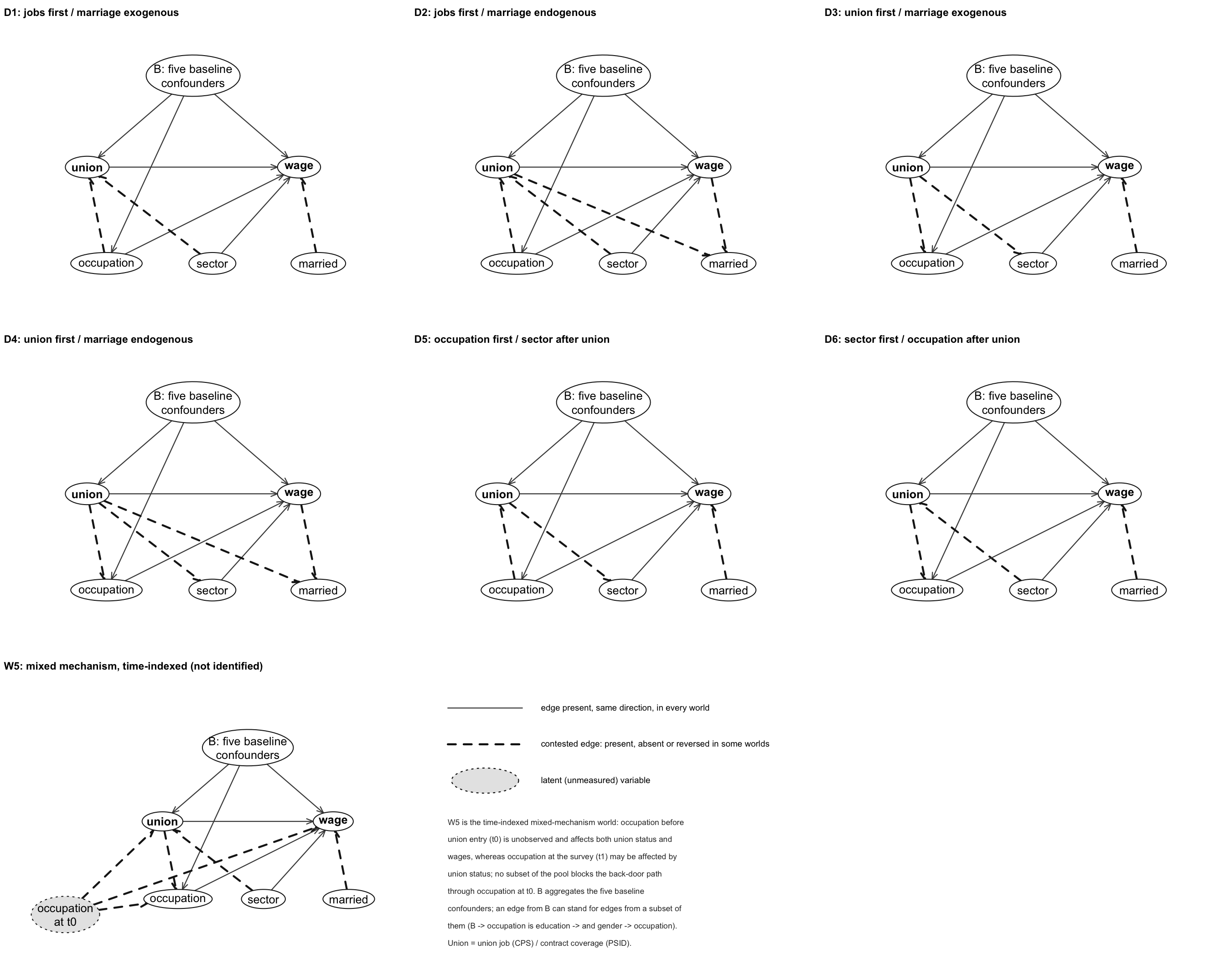}

}

\caption{\label{fig-dagunion}Candidate graphs for the union-premium
application (CPS variable names; the PSID graphs replace occupation and
sector with blue-collar status and industry). B stands for the five
baseline confounders, and an edge from B can represent edges from a
subset of them. A solid edge is present, with the same direction, in
every world; a dashed edge is contested --- present, absent or reversed
in some worlds (the marriage--wage edge is reversed in D2 and D4). Line
styles encode edge membership only: the occupation--wage and
sector--wage edges are present in every world and therefore solid,
although the causal role of occupation and sector --- pre- or
post-treatment --- is exactly what the worlds dispute, through the
dashed edges that connect them to union status. W5 is the time-indexed
mixed-mechanism world: occupation before union entry is unobserved and
affects both union status and wages, whereas occupation at the survey
may be affected by union status.}

\end{figure}

The first two applications reanalyze Muñoz and Young's own cases. The
third takes the framework to a quantity sociology has estimated for half
a century --- the union wage premium
(\citeproc{ref-freemanmedoff1984}{Freeman and Medoff 1984};
\citeproc{ref-card1996}{Card 1996};
\citeproc{ref-westernrosenfeld2011}{Western and Rosenfeld 2011};
\citeproc{ref-farber2021}{Farber et al. 2021a}) --- because it is the
discipline's clearest example of a \emph{structurally contested
control}. Every premium study must decide what to do with occupation and
industry. Under a \textbf{jobs-first} account, workers sort into
occupations and industries, some of which are organized; job location
affects both union status and pay and must therefore be adjusted for
under this account. Under a \textbf{union-first} account, union
attachment shapes subsequent job placement; occupation and industry are
then mediators of the very effect being estimated, and adjusting for
them is overcontrol. The union-first scenario is an author-specified
hypothesis, stated here because it is the natural rival to the
jobs-first reading; we do not attribute it to a particular study, and
the candidate set includes both readings so that the analysis does not
privilege either causal ordering. A second, quieter contest concerns
marital status: a precision covariate under one reading, a
\emph{descendant} of both union status and wages under another (stable,
well-paid employment raises marriage rates), in which case conditioning
on it biases the estimate. Crossing the two contests yields four
candidate DAGs, D1 to D4; education, experience, gender, race, and
region are confounders in all four (Figure~\ref{fig-dagunion},
Table~\ref{tbl-roles-union}). Because nothing forces occupation and
industry to share a causal role, we add two mixed-role candidates: D5,
in which occupation is settled before union entry (skills and
credentials sort workers into occupations, some of them organized) while
sector placement follows attachment, and D6, its mirror, in which sector
is chosen first and occupation follows (apprenticeship and job-ladder
placement); marriage is exogenous in both. A seventh world, W5, is added
below because it is not identified by adjustment. In the terms of the
protocol of Section~\ref{sec-framework}, every contested edge in this
candidate set is supported as a substantively justified hypothesis, not
by a mechanism citation: the edges are the author's hypotheses,
motivated by the literature rather than demonstrated by it. The
two-sided sorting model of Card (\citeproc{ref-card1996}{1996, 977--78})
is a model of selection on skill, not of occupation or industry as such,
and the occupation controls of Farber et al.
(\citeproc{ref-farber2021}{2021a}) enter as a robustness check on a
household-level baseline (\citeproc{ref-farber2021wp}{Farber et al.
2021b, 15--19}, the working-paper version whose pagination we cite);
both are cited here as the sources of the \emph{jobs-first} intuition
that workers are sorted into union jobs, not as evidence for the
specific edges drawn. The union-first edges likewise express the
hypothesis that union attachment causes later placement, and the
marriage edges follow the reading of stable, well-paid employment as a
cause of marriage.

\begin{longtable}[]{@{}
  >{\raggedright\arraybackslash}p{(\linewidth - 14\tabcolsep) * \real{0.1250}}
  >{\raggedright\arraybackslash}p{(\linewidth - 14\tabcolsep) * \real{0.1250}}
  >{\raggedright\arraybackslash}p{(\linewidth - 14\tabcolsep) * \real{0.1250}}
  >{\raggedright\arraybackslash}p{(\linewidth - 14\tabcolsep) * \real{0.1250}}
  >{\raggedright\arraybackslash}p{(\linewidth - 14\tabcolsep) * \real{0.1250}}
  >{\raggedright\arraybackslash}p{(\linewidth - 14\tabcolsep) * \real{0.1250}}
  >{\raggedright\arraybackslash}p{(\linewidth - 14\tabcolsep) * \real{0.1250}}
  >{\raggedright\arraybackslash}p{(\linewidth - 14\tabcolsep) * \real{0.1250}}@{}}
\caption{Roles of the candidate controls under the union-premium worlds
(CPS pool), from \texttt{mv\_classify()}. D1, D2: jobs first; D3, D4:
union first; D2, D4: marriage a descendant of union status and wages;
D5: occupation before union entry, sector after; D6: sector before,
occupation after; W5: time-indexed mixed mechanism, not identified by
adjustment (n.i.). The PSID pool (with SMSA added to the baseline set)
gives the same pattern with 4, 2, 4, 2, 4, 4, and 0 licensed
cells.}\label{tbl-roles-union}\tabularnewline
\toprule\noalign{}
\begin{minipage}[b]{\linewidth}\raggedright
Control
\end{minipage} & \begin{minipage}[b]{\linewidth}\raggedright
D1
\end{minipage} & \begin{minipage}[b]{\linewidth}\raggedright
D2
\end{minipage} & \begin{minipage}[b]{\linewidth}\raggedright
D3
\end{minipage} & \begin{minipage}[b]{\linewidth}\raggedright
D4
\end{minipage} & \begin{minipage}[b]{\linewidth}\raggedright
D5
\end{minipage} & \begin{minipage}[b]{\linewidth}\raggedright
D6
\end{minipage} & \begin{minipage}[b]{\linewidth}\raggedright
W5
\end{minipage} \\
\midrule\noalign{}
\endfirsthead
\toprule\noalign{}
\begin{minipage}[b]{\linewidth}\raggedright
Control
\end{minipage} & \begin{minipage}[b]{\linewidth}\raggedright
D1
\end{minipage} & \begin{minipage}[b]{\linewidth}\raggedright
D2
\end{minipage} & \begin{minipage}[b]{\linewidth}\raggedright
D3
\end{minipage} & \begin{minipage}[b]{\linewidth}\raggedright
D4
\end{minipage} & \begin{minipage}[b]{\linewidth}\raggedright
D5
\end{minipage} & \begin{minipage}[b]{\linewidth}\raggedright
D6
\end{minipage} & \begin{minipage}[b]{\linewidth}\raggedright
W5
\end{minipage} \\
\midrule\noalign{}
\endhead
\bottomrule\noalign{}
\endlastfoot
education, experience, gender, ethnicity, region & required & required &
required & required & required & required & n.i. \\
occupation & required & required & forbidden & forbidden & required &
forbidden & n.i. \\
sector & required & required & forbidden & forbidden & forbidden &
required & n.i. \\
married & optional & forbidden & optional & forbidden & optional &
optional & n.i. \\
Admissible control sets (of 256) & 2 & 1 & 2 & 1 & 2 & 2 & 0 \\
Licensed cells (of 768) & 8 & 4 & 8 & 4 & 8 & 8 & 0 \\
\end{longtable}

\begin{figure}

\centering{

\pandocbounded{\includegraphics[keepaspectratio]{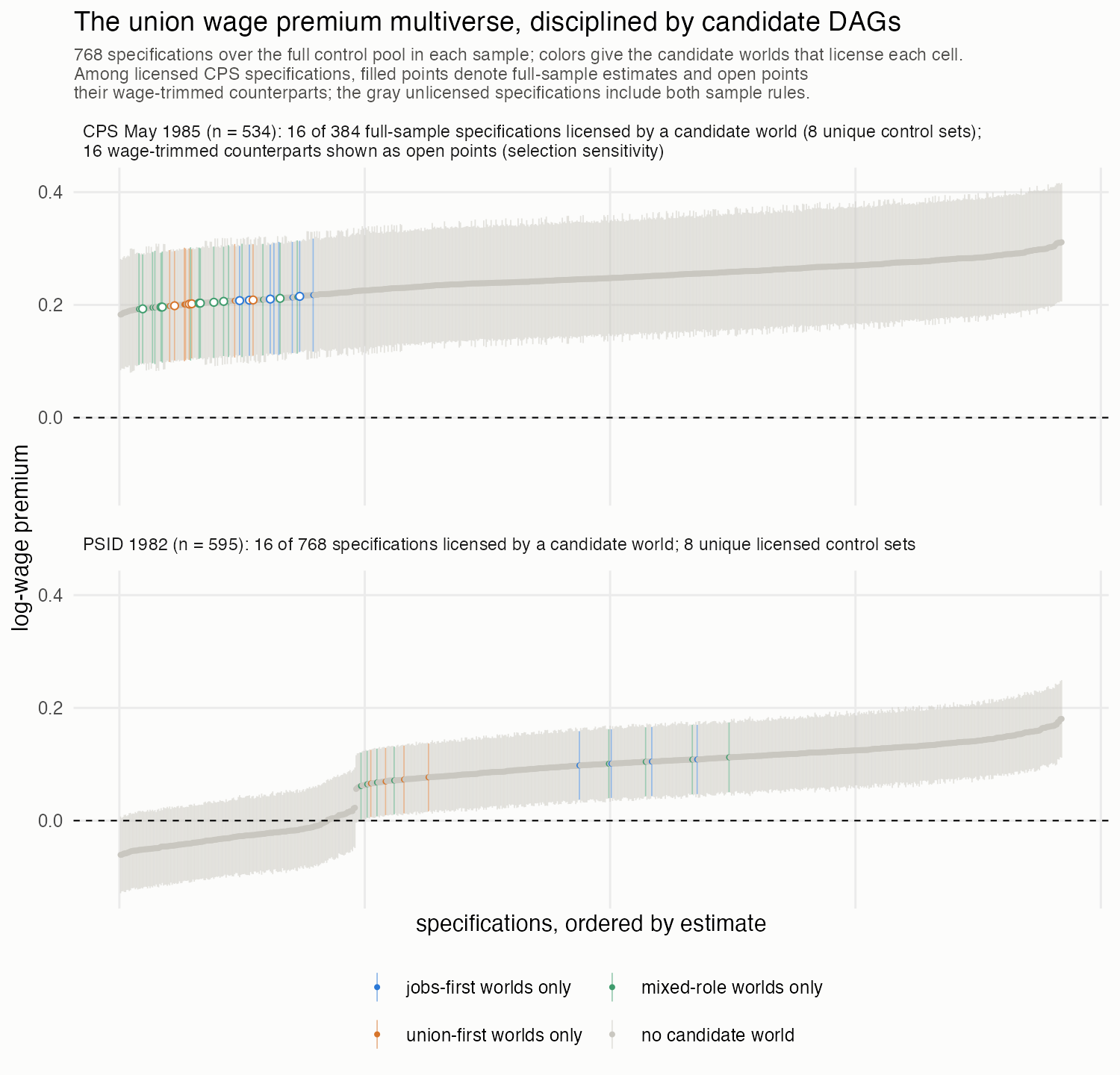}}

}

\caption{\label{fig-union}The union wage premium multiverse in two
samples (CPS May 1985, top; PSID 1982, bottom), 768 specifications each,
ordered by estimate. Colors give the candidate worlds that license each
cell --- jobs-first worlds only (D1, D2), union-first worlds only (D3,
D4), mixed-role worlds only (D5, D6), or none. No cell is licensed by
both a jobs-first and a union-first world, so that combination does not
appear in the legend. Among licensed CPS specifications, filled points
denote full-sample estimates (the structural family; 16 cells) and open
points their wage-trimmed counterparts (selection sensitivity; 16
cells). The gray unlicensed specifications include both sample rules.
The PSID space has no trimming dimension (16 of 768 cells licensed). The
panel headers print the counts; eight distinct control sets are licensed
in each sample. The naive cloud (gray) fails the Young-Holsteen
robustness standard in the PSID even though every licensed cell in every
identified world is positive and significant.}

\end{figure}

We estimate the premium in two independent samples: the canonical CPS
May 1985 extract {[}\(n=534\); Berndt
(\citeproc{ref-berndt1991}{1991}){]} and the 1982 wave of the
Cornwell--Rupert PSID panel {[}\(n=595\); Cornwell and Rupert
(\citeproc{ref-cornwellrupert1988}{1988}){]}. The two exposures are not
the same variable, and neither is personal union membership. In the CPS
extract the indicator records whether the individual ``works on a union
job'' (\citeproc{ref-berndt1991}{Berndt 1991, 193}); in the PSID sample
it records whether the individual's wage is set by a union contract
(\citeproc{ref-cornwellrupert1988}{Cornwell and Rupert 1988, 152}) ---
coverage, which includes non-members in covered jobs. Membership,
coverage and placement in a union job are distinct interventions, and we
make no claim about the first; the contrast estimated below is union job
(CPS) or union contract coverage (PSID) against neither. The PSID sample
is also a selected one: 595 heads of household aged 18 to 65 in 1976 who
report positive wages in private non-farm employment in every year from
1976 to 1982, so the target population is that continuously employed
cohort --- household heads aged 18 to 65 in 1976 with positive
private-sector wages in each of the seven years --- not the 1982 labor
force. The estimand card, with those labels: the treatment contrast is
union job or coverage versus neither; the outcome scale is the log wage,
so the coefficient is read as an approximate proportional premium in log
points; the target population is each analyzed sample as just described;
and the estimating functional is the OLS coefficient on the union
indicator under each licensed control set --- a graph-conditioned
coefficient, which equals the average effect in the sample only under
effect homogeneity across the adjustment strata, an assumption we carry
explicitly and do not test here. The two samples are kept in separate
displays throughout. In each we build the full Muñoz-Young-style space
--- every subset of the entire control pool, crossed with experience
functional form and (for the CPS) outlier trimming: 768 specifications
per sample. Enumerating the \emph{full} pool, including subsets that
omit schooling or experience, is deliberate: it is exactly what
mrobust-style software does by default, so the naive rows of
Figure~\ref{fig-union} show the multiverse a practitioner would actually
run. The CPS trimming dimension, however, restricts the sample to hourly
wages between 1 and 40 dollars --- here it drops one worker at 44.5 ---
and a restriction on the realized outcome changes the target and can
bias a treatment contrast even under a constant effect, exactly as the
dropped-storm cells did in the hurricane grid. By the
target-compatibility rule of Section~\ref{sec-framework} the CPS
decomposition read as structural is therefore defined on the 384
full-sample cells, and the 384 wage-trimmed cells are reported beside it
as a selection-sensitivity family; the PSID space has no trimming
dimension.

The two samples yield opposite pooled robustness assessments, both
qualified by the graph-conditioned analysis. In the CPS, the naive
multiverse over all 768 cells meets every reported robustness criterion:
mean 0.246, every specification positive and significant, robustness
ratio 4.1. But 95.8 percent of those specifications are licensed by
\emph{no} candidate world --- most omit schooling or experience, which
every DAG requires --- and the licensed full-sample worlds sit tightly
below the naive mean: 0.213--0.215 under the jobs-first graphs,
0.202--0.204 under the union-first graphs, and 0.210 and 0.196 under the
two mixed-role graphs (within-world modeling SD 0.003--0.004). The
wage-trimmed counterparts of the same cells give 0.210--0.213,
0.203--0.205, 0.206 and 0.197: the exclusion of one worker moves every
licensed world mean by less than 0.004 log points (the jobs-first worlds
and D5 down, the union-first worlds and D6 up), which is selection
sensitivity of that size, not additional graph uncertainty. The naive
curve is ``robust'' around a value no causal account endorses. In the
PSID, the naive multiverse instead looks \emph{fragile}: mean 0.076,
modeling SD 0.065, sign stability 78 percent, significance rate 75
percent, robustness ratio 1.05 --- decisively below the conventional
threshold of 2 (\citeproc{ref-youngholsteen2017}{Young and Holsteen
2017}). Yet every licensed specification --- sixteen distinct cells,
twenty world--cell pairs --- is positive and significant, with per-world
robustness ratios of 2.2--3.4: 0.103--0.105 in the jobs-first worlds,
0.071--0.073 in the union-first worlds, 0.106 in D5 and 0.066 in D6. The
fragility is produced by specifications (controlling for job attributes
while omitting schooling) that no candidate world licenses. Because D2's
cells are D1's with marriage excluded, and D4's are D3's, four of the
licensed cells in each sample's structural family are licensed by two
worlds; no cell is licensed by both a jobs-first and a union-first
world, and the mixed-role worlds' cells are their own
(Figure~\ref{fig-union}). The structural share is 84 percent in the CPS
full-sample family and 95 percent in the PSID at equal weights over the
six identified worlds (80 and 94 percent over D1 to D4 alone, so the
mixed-role candidates raise it slightly rather than dissolving it; in
the CPS trimmed family it is 71 percent, the difference reflecting how
the one excluded worker shifts the tightly clustered world means). The
Dirichlet sweep gives 5th, 50th, and 95th percentiles of 68, 81, and 88
percent in the CPS and 91, 95, and 96 percent in the PSID; over
weightings that give each world at least 5 percent the share lies
between 61 and 91 percent in the CPS and between 90 and 96 percent in
the PSID (76 to 89 and 94 to 96 percent at 10 percent). Outside those
domains no bound holds: as weight concentrates on any single world,
whose within-world variance is positive here, \(\rho\) falls toward zero
(Section~\ref{sec-framework}), and the paired bootstrap, which resamples
workers and refits all 768 cells, shows how much the share itself moves
with the sample (CPS 2.5th to 97.5th percentiles 11 to 99 percent; PSID
66 to 100 percent; Table~\ref{tbl-boot}). The reading does not rest on
the share, however, but on the world-by-world check and on the contrasts
in effect units: every licensed specification in every world is positive
and significant, and the jobs-first minus union-first gap is 0.032 log
points in the PSID (bootstrap SE 0.011, percentile interval 0.011 to
0.054) and 0.011 in the CPS full-sample family (SE 0.019, interval
\(-0.026\) to \(0.050\)). The PSID gap is the one that matters
substantively, and it is estimated with enough precision to be read as a
real disagreement between the worlds rather than as noise; the CPS gap
is not. The mixed-role worlds sharpen the PSID reading: D5, which holds
occupation fixed and lets industry follow union status, sits with the
jobs-first worlds (0.106), and D6, which holds industry fixed and lets
occupation follow, sits below the union-first worlds (0.066; D5 minus D6
0.040 log points, bootstrap SE 0.011, interval 0.019 to 0.063), so the
disagreement is about the role of occupation, not of industry. These are
statements about six adjustment-identifiable candidates; the seventh
candidate, W5, supplies no identified estimate, and the reversal of the
fragility verdict is conditional on the candidate set.

\subsubsection*{A world not identified by adjustment: the time-indexed
mixed
mechanism}\label{a-world-not-identified-by-adjustment-the-time-indexed-mixed-mechanism}
\addcontentsline{toc}{subsubsection}{A world not identified by
adjustment: the time-indexed mixed mechanism}

The jobs-first and union-first worlds are pure types, and a single
timeless role switch cannot express the mechanism a labor sociologist
would most likely propose: occupation before union entry sorts workers
into union jobs and sets their pay, and occupation after entry is
reshaped by union status. World W5 (Figure~\ref{fig-dagunion}) writes
this down with a time index, occupation at \(t_0\) as a confounder and
occupation at \(t_1\) as a mediator, with the earlier occupation causing
the later one. A cross-section observes only the \(t_1\) measure. Under
W5 the back-door path from union status (a union job or contract
coverage, the exposure actually measured) through occupation at \(t_0\)
to wages is open, the observed occupation is a mediator and a descendant
of the latent confounder, and conditioning on it neither blocks that
path nor is admissible; \texttt{mv\_classify()} returns the role
\texttt{not\ identified} for every control with the attribute
\texttt{identified\ =\ FALSE} and zero admissible subsets, in both
pools. W5 is therefore carried in Table~\ref{tbl-roles-union} with
\(|\mathcal{S}_g| = 0\), excluded from \(\rho\) with the remaining
weights renormalized; this exclusion is reported explicitly. It is the
reason the question taken up below --- a premium of 7 or of 10 percent?
--- cannot be closed with these data, and the reason no claim in this
section is robust to every structure we considered: under W5 the premium
is not identified by any control set in a cross-section. A panel that
observes occupation before entry supplies the missing measurement; it
does not by itself supply identification, which under W5 still requires
that the pre-entry occupation, together with the baseline confounders,
closes every back-door path (Section~\ref{sec-panel}).

Two substantive payoffs follow. First, the framework recovers a
direction of confounding the pooled curve conceals: occupation is a
\emph{suppressor}, not an absorber, of the union premium. In both
samples union jobs concentrate in lower-paid blue-collar occupations (55
against 18 percent of blue- and white-collar workers covered in the
PSID; 28 percent of production workers against 3 to 8 percent of sales,
office and management employees on union jobs in the CPS), so holding
job location fixed \emph{raises} the estimated premium (0.103 against
0.071 log points in the PSID, that is, 10.8 against 7.4 percent), the
opposite of the human-capital intuition that job controls soak up the
effect; the mixed-role worlds locate the suppression in occupation
rather than industry. Second, the residual question --- 7 percent or 10
percent? --- is not a robustness question at all. It asks whether union
coverage raises wages partly by relocating workers across occupations or
only within them, a question about how unions work that no specification
curve on these data can settle; the decomposition's contribution is to
say so explicitly, to quantify the difference between world means as
0.03 to 0.04 log points, and to record that one candidate mechanism (W5)
leaves the premium unidentified altogether.

\section{Panel Extension}\label{sec-panel}

Panel data add two dimensions, which a companion paper (in preparation)
develops with the Japanese Life Course Panel Surveys (waves 1--19;
microdata available under SSJDA terms). First, a \emph{timing} axis: a
variable can be a confounder for the wave-\(t\) treatment and a mediator
for the wave-\((t{-}1)\) treatment, so a candidate graph is a graph over
the panel and each role carries a wave index; W5 of
Section~\ref{sec-reanalysis} is the cross-sectional shadow of this
structure. A panel supplies the measurement W5 lacks --- occupation
before union entry --- and with it an adjustment set can exist under W5;
whether that set identifies the premium is a further assumption (no
unmeasured common cause of entry and wages beyond the measured history),
which the panel does not test. The Cornwell--Rupert data themselves
illustrate the stakes: their seven-wave within (fixed-effects)
regression, which also conditions on the time-varying occupation
(blue-collar) and industry (manufacturing) indicators, weeks worked, and
the other covariates of their Table I, gives an estimated union
coefficient of about 0.014 that is not statistically significant
(\citeproc{ref-cornwellrupert1988}{Cornwell and Rupert 1988}, Table I),
against 0.07 to 0.11 in the 1982 cross-section under every identified
world here. Individual fixed effects remove time-invariant differences
between workers; the additional adjustment for time-varying job
attributes can also block mediated pathways of the union-first kind. The
panel and cross-sectional estimates should therefore not be treated as
directly comparable without aligning their controls, populations and
assumptions, and we do not read either as a check on the other. Second,
an \emph{estimator} axis: pooled OLS, unit fixed effects, and
heterogeneity-robust difference-in-differences estimators answer
different questions under staggered adoption and effect heterogeneity
(\citeproc{ref-imaikim2019}{Imai and Kim 2019};
\citeproc{ref-callawaysantanna2021}{Callaway and Sant'Anna 2021};
\citeproc{ref-dechaisemartin2020}{de Chaisemartin and D'Haultfœuille
2020}). Estimand-preserving estimator variation enters within worlds;
estimand-changing variation belongs in separate displays, one
decomposition per estimand, never in the between-world component.
Pre-trends and placebo waves test implications of the identifying
assumptions of each world, and within-person contrasts change the
estimand; none of them is a substitute for the external benchmark of
Section~\ref{sec-reanalysis}, but each can reject a candidate world
whose implications fail.

\section{The dagmv R Package}\label{sec-software}

The framework ships as an R package, dagmv. Its design follows three
principles. First, a \textbf{dependency-free core}: DAG parsing (a
dagitty-compatible syntax), d-separation (via moralization), and the
generalized adjustment criterion are implemented in base R, so the
package installs anywhere R runs; dagitty and fixest are optional
interoperability layers. Second, an \textbf{estimand-first API} that
mirrors the framework: \texttt{dag\_parse()} encodes candidate graphs;
\texttt{mv\_classify()} returns the required/forbidden/optional table
per graph; \texttt{mv\_run()} fits each unique specification once and
maps admissibility per graph afterwards (so adding candidate graphs
costs no additional estimation), with engines for OLS, negative
binomial, and fixed-effects panels; \texttt{mv\_decompose()} returns the
DAG-conditional Young-Holsteen metrics and the within/between
decomposition; \texttt{mv\_plot()} draws the disciplined multiverse.
Third, \textbf{auditable correctness}: the admissibility engine
evaluates the generalized adjustment criterion directly for every
specification--world pair (it never enumerates from role labels), and a
world without an admissible set is returned by \texttt{mv\_classify()}
with the role \texttt{not\ identified}, kept by \texttt{mv\_run()} and
\texttt{mv\_decompose()} with zero specifications, and named in a
message and in the printed decomposition rather than dropped. Version
0.1.2 (GitHub tag \texttt{v0.1.2}) ships a \texttt{testthat} suite of
568 expectations under testthat 3.2.1 with the dagitty cross-check
described below skipped --- its two expectations run only where dagitty
is installed --- covering back-door adjustment, M-bias and butterfly
bias, mediators and their descendants, instruments, colliders and their
descendants, a descendant of the exposure off the causal path
(admissible under the criterion as stated), latent-variable
restrictions, the eighteen numbered models of Cinelli, Forney, and Pearl
(\citeproc{ref-cinelli2024}{2024}), the role-label counterexample of
Section~\ref{sec-framework}, the parser, the reporting of non-identified
worlds, and the decomposition arithmetic against hand-computed values.
Version 0.1.3 adds three expectations to that suite; the test logs of
both versions are in the replication archive. Two reporting changes
deserve a note: the within-world component is printed as ``specification
dispersion within worlds'' from version 0.1.3 onward (0.1.2 prints
``sampling + specification noise,'' the pooled-simulation reading of
Section~\ref{sec-simulation} rather than the fixed-data reading of the
applications), and from the same version \texttt{mv\_run()} warns when a
fixed adjustment term named by the user is absent from the candidate
graphs, since the criterion cannot vouch for a variable it has not seen.
The convergence safeguards and the ATT-targeted estimator used in the
applications are implemented in the replication archive's
\texttt{spec\_builders.R}, not in the package's generic \texttt{lm},
\texttt{glm.nb} and \texttt{feols} engines; transferring the
negative-binomial safeguards into the engine is a natural next release.
Two randomized cross-checks need no additional package and run on every
installation: \(d\)-separation against explicit path enumeration (100
random DAGs, 800 queries) and admissibility against exact population OLS
bias in random linear structural equation models (120 DAGs, 1,226
graph--set pairs), both with zero discrepancies. A third cross-check
compares \texttt{adjustment\_valid()} with
\texttt{dagitty::isAdjustmentSet()} over all control subsets of 100
random DAGs of five to seven nodes; it is part of the suite and runs
whenever dagitty is installed (it is skipped, and reported as skipped,
otherwise), so that any user with dagitty can reproduce the comparison
with \texttt{testthat::test\_local()}.

Table~\ref{tbl-software} positions dagmv against the maintained
alternatives: mrobust and MULTIVRS
(\citeproc{ref-youngholsteen2017}{Young and Holsteen 2017},
\citeproc{ref-multivrs2021}{2021}), specr 1.0.0
(\citeproc{ref-specr2020}{Masur and Scharkow 2023}), multiverse 0.6.2
(\citeproc{ref-multiverse2024}{Sarma and Kay 2024}), DAGassist 0.3.0
(\citeproc{ref-dagassist2026}{Goff and Denly 2026}), and RobustiPy
(arXiv v4) (\citeproc{ref-robustipy2025}{Valdenegro Ibarra et al.
2026}). The rows compare like with like: what each tool fits, whether it
accepts a user-curated specification space, whether it screens
specifications against a graph, how it treats more than one graph, and
what it reports. Two rows deserve a gloss. specr already decomposes the
variance of the estimates by analytic choice with a multilevel model
(\texttt{icc\_specs()}); what dagmv adds is a decomposition by
\emph{graph membership}, where membership is tied to validity screening
rather than to a declared choice. And DAGassist's
\texttt{pdag\_robustness()} enumerates every acyclic orientation of a
set of uncertain edges and reports whether the covariate roles and
minimal adjustment sets change; it does not fit the model under each
orientation or name the orientations as rival worlds.

\begin{longtable}[]{@{}
  >{\raggedright\arraybackslash}p{(\linewidth - 12\tabcolsep) * \real{0.1500}}
  >{\raggedright\arraybackslash}p{(\linewidth - 12\tabcolsep) * \real{0.1300}}
  >{\raggedright\arraybackslash}p{(\linewidth - 12\tabcolsep) * \real{0.1200}}
  >{\raggedright\arraybackslash}p{(\linewidth - 12\tabcolsep) * \real{0.1300}}
  >{\raggedright\arraybackslash}p{(\linewidth - 12\tabcolsep) * \real{0.1500}}
  >{\raggedright\arraybackslash}p{(\linewidth - 12\tabcolsep) * \real{0.1600}}
  >{\raggedright\arraybackslash}p{(\linewidth - 12\tabcolsep) * \real{0.1600}}@{}}
\caption{Feature comparison with existing software (versions as stated
in the text; capabilities as documented in each tool's manual or paper
at the time of writing).}\label{tbl-software}\tabularnewline
\toprule\noalign{}
\begin{minipage}[b]{\linewidth}\raggedright
Capability
\end{minipage} & \begin{minipage}[b]{\linewidth}\raggedright
mrobust / MULTIVRS
\end{minipage} & \begin{minipage}[b]{\linewidth}\raggedright
specr
\end{minipage} & \begin{minipage}[b]{\linewidth}\raggedright
multiverse
\end{minipage} & \begin{minipage}[b]{\linewidth}\raggedright
DAGassist
\end{minipage} & \begin{minipage}[b]{\linewidth}\raggedright
RobustiPy
\end{minipage} & \begin{minipage}[b]{\linewidth}\raggedright
\textbf{dagmv}
\end{minipage} \\
\midrule\noalign{}
\endfirsthead
\toprule\noalign{}
\begin{minipage}[b]{\linewidth}\raggedright
Capability
\end{minipage} & \begin{minipage}[b]{\linewidth}\raggedright
mrobust / MULTIVRS
\end{minipage} & \begin{minipage}[b]{\linewidth}\raggedright
specr
\end{minipage} & \begin{minipage}[b]{\linewidth}\raggedright
multiverse
\end{minipage} & \begin{minipage}[b]{\linewidth}\raggedright
DAGassist
\end{minipage} & \begin{minipage}[b]{\linewidth}\raggedright
RobustiPy
\end{minipage} & \begin{minipage}[b]{\linewidth}\raggedright
\textbf{dagmv}
\end{minipage} \\
\midrule\noalign{}
\endhead
\bottomrule\noalign{}
\endlastfoot
Estimation backend & Stata estimation commands & any R model function &
arbitrary R code inside branches & the user's model call (R,
incl.~fixest) & OLS, logit, fixed-effects OLS (Python) & OLS, negative
binomial, fixed effects (R) \\
User-specified specification space & yes & yes & yes, with conditional
branch exclusions and a per-universe table of assignments and code & no
(single model call; original vs.~DAG-derived) & yes & yes \\
Graph adjudication (causal-validity screening) & no & no & no &
user-supplied DAG & no & per candidate DAG \\
Multiple candidate graphs & no & no & no & orientation enumeration over
uncertain edges (up to 1,024) and added-edge branches; roles and sets
compared, models not refitted, no named rival graphs & no &
\textbf{named worlds, each fitted} \\
Integrated reporting & pooled robustness metrics, model influence &
specification curve, variance components by analytic choice & multiverse
table, export to Milliways & role table, adjustment sets, original
vs.~DAG-derived comparison & curve, bootstrap and joint inference,
out-of-sample metrics, BMA over controls, predictive feature attribution
(SHAP) for the full model & world-conditional metrics, within/between
decomposition by graph membership, weight sweep and restricted-simplex
bounds \\
\end{longtable}

A complete analysis is a dozen lines. The example below runs as printed
on the CPS May 1985 extract distributed with the AER package, with a
reduced pool of four controls so that it fits on the page: it declares
two worlds that disagree about occupation, fits all 16 control subsets
once, and prints the per-world metrics and the decomposition.

\begin{Shaded}
\begin{Highlighting}[]
\FunctionTok{library}\NormalTok{(dagmv); }\FunctionTok{data}\NormalTok{(}\StringTok{"CPS1985"}\NormalTok{, }\AttributeTok{package =} \StringTok{"AER"}\NormalTok{)}
\NormalTok{d }\OtherTok{\textless{}{-}} \FunctionTok{transform}\NormalTok{(CPS1985, }\AttributeTok{lwage =} \FunctionTok{log}\NormalTok{(wage),}
               \AttributeTok{union =} \FunctionTok{as.numeric}\NormalTok{(union }\SpecialCharTok{==} \StringTok{"yes"}\NormalTok{))}
\NormalTok{base }\OtherTok{\textless{}{-}} \FunctionTok{paste}\NormalTok{(}\StringTok{"education {-}\textgreater{} union ; education {-}\textgreater{} lwage ;"}\NormalTok{,}
              \StringTok{"experience {-}\textgreater{} union ; experience {-}\textgreater{} lwage ;"}\NormalTok{,}
              \StringTok{"married {-}\textgreater{} lwage ; union {-}\textgreater{} lwage"}\NormalTok{)}
\NormalTok{dags }\OtherTok{\textless{}{-}} \FunctionTok{list}\NormalTok{(}
  \AttributeTok{jobs\_first  =} \FunctionTok{paste}\NormalTok{(}\StringTok{"dag \{"}\NormalTok{, base,}
                      \StringTok{"; occupation {-}\textgreater{} union ; occupation {-}\textgreater{} lwage \}"}\NormalTok{),}
  \AttributeTok{union\_first =} \FunctionTok{paste}\NormalTok{(}\StringTok{"dag \{"}\NormalTok{, base,}
                      \StringTok{"; union {-}\textgreater{} occupation ; occupation {-}\textgreater{} lwage \}"}\NormalTok{))}
\NormalTok{pool }\OtherTok{\textless{}{-}} \FunctionTok{c}\NormalTok{(}\StringTok{"education"}\NormalTok{, }\StringTok{"experience"}\NormalTok{, }\StringTok{"occupation"}\NormalTok{, }\StringTok{"married"}\NormalTok{)}
\FunctionTok{mv\_decompose}\NormalTok{(}\FunctionTok{mv\_run}\NormalTok{(d, }\StringTok{"lwage"}\NormalTok{, }\StringTok{"union"}\NormalTok{, pool, dags))}
\end{Highlighting}
\end{Shaded}

The package, together with the accompanying application scripts and
public data, supports reproduction of the reported analyses;
\texttt{mv\_plot()} draws the disciplined multiverse, and a separate
software paper (in preparation) documents implementation details.

\section{Discussion}\label{sec-discussion}

\subsection{What Changes in Practice}\label{what-changes-in-practice}

For authors, the framework replaces one summary (a specification curve
with pooled metrics) with three: the role-classification table (which
controls are contested, and what each side implies), the DAG-conditional
robustness table, and the structural share \(\rho\) read beside the
between-world range. For readers and reviewers, that display answers the
question a pooled curve cannot: is this spread evidence \emph{against}
the finding, or evidence that the discipline has not settled a causal
question? Our simulation and empirical results --- 97 percent (simulated
contested structure); 11 percent within the negative-binomial
coefficient family and 4 percent once the fitted means are standardized
on a common population (hurricanes, where the standardization shows the
between-world spread to be coefficient sensitivity rather than
structural disagreement); 13 to 26 percent across three worlds of which
two are defended and under 1 percent between the two defended worlds
(job training, with an external benchmark discriminating among them);
and 84--95 percent (union premium, on the full-sample families, where
the between-world gap \emph{is} the substantive question, conditional on
a candidate set with one non-identified member) --- show that the answer
is not predictable in advance, which is precisely why it must be
reported.

For reviewers and editors, the framework yields checkable submission
standards. A multiverse analysis should arrive with (i) its
candidate-graph set with, for each contested edge, the mechanism
citation that supports it or the substantive justification of the
stress-test scenario it represents, labelled as one or the other, (ii)
the role-classification table with the number of licensed specifications
per world, (iii) the structural share at equal weights with its
Dirichlet quantiles, restricted-simplex bounds, and the between-world
range with a paired-bootstrap interval, and (iv) the dimension inventory
that discloses how much within-world spread the design builds in. Two
failure modes then become visible on sight: a ``robust'' verdict earned
by pooling specifications no candidate world licenses (the CPS union
premium), and a ``fragile'' verdict produced the same way (its PSID
mirror image) --- neither is detectable from a pooled specification
curve. Much of this reporting can be automated once the candidate graphs
and the specification space have been justified and specified: the
graphs are a figure, the classification is computed mechanically, and
the sweep is one line of code.

For the ongoing debate about multiverse analysis, the framework reframes
the dichotomy between ``billions of regressions'' and ``a few thoughtful
models'' (\citeproc{ref-auspurg2025}{Auspurg 2025};
\citeproc{ref-ganslmeier2025reply}{Ganslmeier and Vlandas 2025b}): the
statistical rules for a justified model set stay where Auspurg puts
them, the causal rule becomes a small set of explicit graphs, brute
force enters within each graph, and neither side's discretion is hidden.
The framework does not adjudicate the other half of that debate ---
whether sample and measurement choices dominate conditioning choices ---
and nothing in our applications bears on it.

\subsection{Limitations and
Extensions}\label{limitations-and-extensions}

Five limitations bound our claims. First, the decomposition is
conditional on the candidate set: it cannot see structures no one
proposed (Scenario B) or sufficiency violations shared by all candidates
(Scenario C); the remedies lie outside the specification distribution
--- external benchmarks that discriminate among worlds
(Section~\ref{sec-reanalysis}), tests of each world's implications, and
assumption-based sensitivity analysis for the confounding no candidate
represents (\citeproc{ref-cinellihazlett2020}{Cinelli and Hazlett 2020};
\citeproc{ref-oster2019}{Oster 2019}), which we view as complements, not
competitors. Second, our core theory covers a point exposure and the
total effect; time-varying treatments, where a control can be confounder
and mediator at once, require the panel extension sketched in
Section~\ref{sec-panel} and developed in companion work (in
preparation). Third, the mixture identity holds for any set of numbers,
but its \emph{reading} as a decomposition of one estimand requires cells
that are compatible in target, sample and estimation model, not merely
licensed by a graph. Graph eligibility validates an adjustment
functional; it does not validate a regression coefficient as an estimate
of that functional's contrast, an interaction evaluated at a reference
point, or a sample selected on the outcome. The applications treat this
differently and say so: the job-training comparison uses an ATT-targeted
estimator on the treated population, the hurricane comparison
standardizes fitted means on a common population and finds that the
cells separating the worlds are exactly the cells whose fitted means are
not credible, and the union comparison remains a comparison of
graph-conditioned coefficients under an effect-homogeneity assumption.
Where the applications claim that the decomposition isolates structural
uncertainty, therefore, the claim is made for the job-training and union
comparisons under those stated conditions and withdrawn for the
hurricane coefficient family. Fourth, \(\rho\) is a descriptive share on
the fixed sample, and the paired bootstrap of
Section~\ref{sec-reanalysis} shows that its sampling distribution can be
very wide even where the reading is clear: in the CPS union sample the
2.5th to 97.5th percentiles run from 11 to 99 percent although every
licensed cell is positive and significant, and in the hurricane data
every world mean's interval includes zero; Scenario D shows that a high
fixed-data share can arise from sampling noise alone. The share should
therefore be read beside the bootstrap intervals of the world contrasts
in effect units (Table~\ref{tbl-boot}), which is where the sampling
evidence about structural disagreement lives; with 92 storms or 534
workers those intervals are wide, and a share computed on such a sample
says little on its own. Fifth, candidate graphs could be complemented by
graphs learned from data --- orientations of an estimated skeleton, as
in Hu and van der Pas (\citeproc{ref-huvdpas2025}{2025}), whose list of
adjustment sets with validity frequencies is the learned counterpart of
our role table; we see this as the natural next step, with the present
framework as its transparent, elicitation-based special case. The graphs
in our applications are literature-informed retrospective stress tests,
not pre-registered elicitations; an independent elicitation exercise is
the natural test of the protocol.

\subsection{Conclusion}\label{conclusion}

Multiverse analysis taught social science to see the model space; the
graphical literature taught it to judge specifications. The two lessons
are compatible, and jointly they say: run the multiverse, but only
within causal worlds someone is prepared to defend --- and report how
much of the disagreement is about the world rather than the model. The
empirical payoff of doing so is not subtle. Across the three canonical
multiverses of this paper, causal discipline left one fragility verdict
standing at the contrast examined, traced one notorious instability to
unlicensed specifications, and reversed one fragility verdict as a
reading of the same estimates; in each case the disciplined display
changed what the pooled curve should be taken to mean. The decomposition
summarizes how estimates vary across the stated graph--specification
mixture. World-specific contrasts and their uncertainty identify the
disagreements that remain substantively consequential; resolving them
requires evidence beyond the specification distribution itself.

\section*{Supplemental Materials}\label{supplemental-materials}
\addcontentsline{toc}{section}{Supplemental Materials}

\emph{For online publication alongside the article.}

\begin{longtable}[]{@{}llll@{}}
\caption{Design III: sensitivity of the
decomposition.}\label{tbl-designiii}\tabularnewline
\toprule\noalign{}
\(n\) & signal scale \(s\) & structural share \(\rho\) & between-DAG
range \\
\midrule\noalign{}
\endfirsthead
\toprule\noalign{}
\(n\) & signal scale \(s\) & structural share \(\rho\) & between-DAG
range \\
\midrule\noalign{}
\endhead
\bottomrule\noalign{}
\endlastfoot
500 & 1.0 & 89.1\% & 0.41 \\
2,000 & 1.0 & 97.1\% & 0.41 \\
500 & 0.5 & 39.7\% & 0.09 \\
2,000 & 0.5 & 72.4\% & 0.10 \\
\end{longtable}

\begin{longtable}[]{@{}
  >{\raggedright\arraybackslash}p{(\linewidth - 12\tabcolsep) * \real{0.1600}}
  >{\raggedright\arraybackslash}p{(\linewidth - 12\tabcolsep) * \real{0.3467}}
  >{\raggedright\arraybackslash}p{(\linewidth - 12\tabcolsep) * \real{0.1067}}
  >{\raggedright\arraybackslash}p{(\linewidth - 12\tabcolsep) * \real{0.1067}}
  >{\raggedright\arraybackslash}p{(\linewidth - 12\tabcolsep) * \real{0.1067}}
  >{\raggedright\arraybackslash}p{(\linewidth - 12\tabcolsep) * \real{0.1067}}
  >{\raggedright\arraybackslash}p{(\linewidth - 12\tabcolsep) * \real{0.0667}}@{}}
\caption{Paired bootstrap (B = 500 resamples of units, seed 20260906;
every cell refitted on each resample; n = resamples contributing to the
statistic). Hurricane coefficients are on the log-deaths scale; the
standardized means are the population-averaged one-point log rate ratio
over the fixed 12-cell family (negative binomial, full sample, no damage
term), identical for H1 and H2 and for H3 and H4; the high-severity
means are the femininity coefficient at a minimum pressure of 942 mb
(10th percentile) in the negative-binomial pressure-interaction cells;
job-training quantities are 1978 earnings in 1982 dollars (the
benchmark's own bootstrap resamples the experimental controls jointly
with the treated units shared with the observational cells; AIPW-ATT
rows refer to the 256 control-set cells); union quantities are log
points (six identified worlds; CPS rows refer to the full-sample family
unless labelled trimmed; \(\rho\) over D1 to D4 also shown); \(\rho\) in
percent; the share is missing on a resample whenever any declared
world's mean is missing (fixed world set), and n counts the contributing
resamples. Negative-binomial fits use a convergence-aware algorithm
(failure = error, non-converged final IRLS, or a dispersion-iteration
warning; prespecified retries with a higher iteration limit and a
moment-based initial dispersion): of 36,000 fits in the hurricane
bootstrap, 32,035 converged at the first attempt, 3111 after a retry,
and 854 failed, in 161 resamples; a world with a failed licensed cell is
left out of that resample rather than averaged over the surviving cells,
and the share is left out with it rather than recomputed over the
surviving worlds. Failures are associated with resampling the deadliest
storms, so the percentiles are conditional numerical diagnostics, not
validated confidence limits. No fit failed in any other
application.}\label{tbl-boot}\tabularnewline
\toprule\noalign{}
\begin{minipage}[b]{\linewidth}\raggedright
Application
\end{minipage} & \begin{minipage}[b]{\linewidth}\raggedright
Statistic
\end{minipage} & \begin{minipage}[b]{\linewidth}\raggedright
Point
\end{minipage} & \begin{minipage}[b]{\linewidth}\raggedright
Boot. SE
\end{minipage} & \begin{minipage}[b]{\linewidth}\raggedright
2.5th
\end{minipage} & \begin{minipage}[b]{\linewidth}\raggedright
97.5th
\end{minipage} & \begin{minipage}[b]{\linewidth}\raggedright
n
\end{minipage} \\
\midrule\noalign{}
\endfirsthead
\toprule\noalign{}
\begin{minipage}[b]{\linewidth}\raggedright
Application
\end{minipage} & \begin{minipage}[b]{\linewidth}\raggedright
Statistic
\end{minipage} & \begin{minipage}[b]{\linewidth}\raggedright
Point
\end{minipage} & \begin{minipage}[b]{\linewidth}\raggedright
Boot. SE
\end{minipage} & \begin{minipage}[b]{\linewidth}\raggedright
2.5th
\end{minipage} & \begin{minipage}[b]{\linewidth}\raggedright
97.5th
\end{minipage} & \begin{minipage}[b]{\linewidth}\raggedright
n
\end{minipage} \\
\midrule\noalign{}
\endhead
\bottomrule\noalign{}
\endlastfoot
hurricanes & H1 world mean (16-cell pool) & 0.048 & 0.044 & -0.056 &
0.112 & 373 \\
hurricanes & H2 world mean & 0.066 & 0.053 & -0.061 & 0.145 & 396 \\
hurricanes & H3 world mean & 0.055 & 0.044 & -0.049 & 0.117 & 372 \\
hurricanes & H4 world mean & 0.076 & 0.052 & -0.047 & 0.146 & 395 \\
hurricanes & H4 minus H1 & 0.027 & 0.023 & -0.020 & 0.068 & 373 \\
hurricanes & \(\rho\), 16-cell pool & 15.5 & 8.6 & 0.5 & 28.6 & 372 \\
hurricanes & \(\rho\), NB family (72 cells) & 11.3 & 10.1 & 0.3 & 34.0 &
339 \\
hurricanes & \(\rho\), pooled 144 cells & 1.9 & 6.3 & 0.2 & 24.3 &
339 \\
hurricanes & H1/H2 standardized mean (12-cell family) & 0.044 & 0.050 &
-0.077 & 0.123 & 377 \\
hurricanes & H3/H4 standardized mean (12-cell family) & 0.058 & 0.048 &
-0.056 & 0.130 & 364 \\
hurricanes & H3 minus H1, standardized & 0.014 & 0.018 & -0.016 & 0.052
& 364 \\
hurricanes & \(\rho\), standardized 12-cell family & 3.7 & 7.1 & 0.0 &
25.8 & 364 \\
hurricanes & H1 high-severity mean (NB) & -0.036 & 0.068 & -0.165 &
0.105 & 405 \\
hurricanes & H2 high-severity mean (NB) & -0.032 & 0.065 & -0.172 &
0.081 & 435 \\
hurricanes & H3 high-severity mean (NB) & -0.030 & 0.062 & -0.141 &
0.107 & 386 \\
hurricanes & H4 high-severity mean (NB) & -0.019 & 0.056 & -0.135 &
0.083 & 415 \\
job training & L0 world mean (36 cells) & 345 & 607 & -711 & 1,615 &
500 \\
job training & L1 world mean (24 cells) & 967 & 616 & -100 & 2,208 &
500 \\
job training & L2 world mean (16 cells) & 1,003 & 619 & -76 & 2,268 &
500 \\
job training & experimental benchmark & 1,794 & 678 & 537 & 3,168 &
500 \\
job training & L1 minus benchmark & -828 & 416 & -1,625 & -7 & 500 \\
job training & L2 minus benchmark & -791 & 424 & -1,606 & 45 & 500 \\
job training & L0 minus benchmark & -1,450 & 409 & -2,219 & -668 &
500 \\
job training & L1 minus L2 & -36 & 33 & -97 & 30 & 500 \\
job training & \(\rho\) & 13.2 & 0.8 & 11.7 & 14.7 & 500 \\
job training & \(\rho\), L1 and L2 only & 0.8 & 1.4 & 0.0 & 4.6 & 500 \\
job training & L0 AIPW-ATT mean (256 sets) & 29 & 605 & -1,056 & 1,263 &
500 \\
job training & L1 AIPW-ATT mean & 1,219 & 624 & 109 & 2,503 & 500 \\
job training & L2 AIPW-ATT mean & 1,271 & 624 & 147 & 2,561 & 500 \\
job training & L1 AIPW-ATT minus benchmark & -575 & 453 & -1,390 & 354 &
500 \\
job training & L2 AIPW-ATT minus benchmark & -523 & 458 & -1,363 & 430 &
500 \\
job training & L0 AIPW-ATT minus benchmark & -1,765 & 430 & -2,569 &
-900 & 500 \\
job training & L1 minus L2, AIPW-ATT & -52 & 45 & -140 & 35 & 500 \\
job training & \(\rho\), AIPW-ATT (256 sets) & 25.6 & 1.5 & 22.7 & 28.5
& 500 \\
union CPS 1985 & D1 world mean & 0.213 & 0.051 & 0.120 & 0.312 & 500 \\
union CPS 1985 & D2 world mean & 0.215 & 0.050 & 0.121 & 0.313 & 500 \\
union CPS 1985 & D3 world mean & 0.202 & 0.047 & 0.111 & 0.293 & 500 \\
union CPS 1985 & D4 world mean & 0.204 & 0.047 & 0.111 & 0.295 & 500 \\
union CPS 1985 & D5 world mean (mixed role) & 0.210 & 0.050 & 0.113 &
0.302 & 500 \\
union CPS 1985 & D6 world mean (mixed role) & 0.196 & 0.048 & 0.104 &
0.287 & 500 \\
union CPS 1985 & jobs-first minus union-first & 0.011 & 0.019 & -0.026 &
0.050 & 500 \\
union CPS 1985 & \(\rho\) & 83.8 & 26.2 & 11.0 & 99.3 & 500 \\
union CPS 1985 & \(\rho\), D1 to D4 only & 80.3 & 32.3 & 2.2 & 99.3 &
500 \\
union CPS 1985 & \(\rho\), wage-trimmed family & 71.2 & 26.5 & 10.8 &
99.0 & 500 \\
union CPS 1985 & jobs-first minus union-first, trimmed family & 0.008 &
0.019 & -0.029 & 0.046 & 500 \\
union CPS 1985 & trimmed minus full, mean over worlds & -0.001 & 0.002 &
-0.006 & 0.002 & 500 \\
union PSID 1982 & D1 world mean & 0.103 & 0.028 & 0.052 & 0.158 & 500 \\
union PSID 1982 & D2 world mean & 0.105 & 0.028 & 0.053 & 0.160 & 500 \\
union PSID 1982 & D3 world mean & 0.071 & 0.026 & 0.024 & 0.121 & 500 \\
union PSID 1982 & D4 world mean & 0.073 & 0.026 & 0.025 & 0.124 & 500 \\
union PSID 1982 & D5 world mean (mixed role) & 0.106 & 0.028 & 0.054 &
0.160 & 500 \\
union PSID 1982 & D6 world mean (mixed role) & 0.066 & 0.026 & 0.019 &
0.117 & 500 \\
union PSID 1982 & jobs-first minus union-first & 0.032 & 0.011 & 0.011 &
0.054 & 500 \\
union PSID 1982 & \(\rho\) & 95.3 & 9.0 & 66.0 & 99.8 & 500 \\
union PSID 1982 & \(\rho\), D1 to D4 only & 94.5 & 12.5 & 54.6 & 99.8 &
500 \\
\end{longtable}

\begin{longtable}[]{@{}
  >{\raggedright\arraybackslash}p{(\linewidth - 18\tabcolsep) * \real{0.3544}}
  >{\raggedright\arraybackslash}p{(\linewidth - 18\tabcolsep) * \real{0.0380}}
  >{\raggedright\arraybackslash}p{(\linewidth - 18\tabcolsep) * \real{0.0759}}
  >{\raggedright\arraybackslash}p{(\linewidth - 18\tabcolsep) * \real{0.0759}}
  >{\raggedright\arraybackslash}p{(\linewidth - 18\tabcolsep) * \real{0.0759}}
  >{\raggedright\arraybackslash}p{(\linewidth - 18\tabcolsep) * \real{0.0759}}
  >{\raggedright\arraybackslash}p{(\linewidth - 18\tabcolsep) * \real{0.0759}}
  >{\raggedright\arraybackslash}p{(\linewidth - 18\tabcolsep) * \real{0.0759}}
  >{\raggedright\arraybackslash}p{(\linewidth - 18\tabcolsep) * \real{0.0759}}
  >{\raggedright\arraybackslash}p{(\linewidth - 18\tabcolsep) * \real{0.0759}}@{}}
\caption{Structural share \(\rho(w)\) in percent: at equal weights;
infimum and supremum over the restricted simplices that give every world
at least 5 percent (inf .05, sup .05) and at least 10 percent (inf .10,
sup .10) of the weight (numerical optimization from the per-world means
and variances; for decompositions with at most five worlds verified
against a fine grid on the restricted simplex with local refinement,
maximum discrepancy 5e-07; the six-world union rows rest on the
optimization alone); and the 5th, 50th, and 95th percentiles of a
symmetric Dirichlet sweep (\(\alpha = 1\), \(N = 10{,}000\) draws, seed
20260906). Simulation rows pool estimates across Monte Carlo
replications.}\label{tbl-simplex}\tabularnewline
\toprule\noalign{}
\begin{minipage}[b]{\linewidth}\raggedright
Decomposition
\end{minipage} & \begin{minipage}[b]{\linewidth}\raggedright
G
\end{minipage} & \begin{minipage}[b]{\linewidth}\raggedright
Equal
\end{minipage} & \begin{minipage}[b]{\linewidth}\raggedright
inf .05
\end{minipage} & \begin{minipage}[b]{\linewidth}\raggedright
sup .05
\end{minipage} & \begin{minipage}[b]{\linewidth}\raggedright
inf .10
\end{minipage} & \begin{minipage}[b]{\linewidth}\raggedright
sup .10
\end{minipage} & \begin{minipage}[b]{\linewidth}\raggedright
Dir. 5th
\end{minipage} & \begin{minipage}[b]{\linewidth}\raggedright
50th
\end{minipage} & \begin{minipage}[b]{\linewidth}\raggedright
95th
\end{minipage} \\
\midrule\noalign{}
\endfirsthead
\toprule\noalign{}
\begin{minipage}[b]{\linewidth}\raggedright
Decomposition
\end{minipage} & \begin{minipage}[b]{\linewidth}\raggedright
G
\end{minipage} & \begin{minipage}[b]{\linewidth}\raggedright
Equal
\end{minipage} & \begin{minipage}[b]{\linewidth}\raggedright
inf .05
\end{minipage} & \begin{minipage}[b]{\linewidth}\raggedright
sup .05
\end{minipage} & \begin{minipage}[b]{\linewidth}\raggedright
inf .10
\end{minipage} & \begin{minipage}[b]{\linewidth}\raggedright
sup .10
\end{minipage} & \begin{minipage}[b]{\linewidth}\raggedright
Dir. 5th
\end{minipage} & \begin{minipage}[b]{\linewidth}\raggedright
50th
\end{minipage} & \begin{minipage}[b]{\linewidth}\raggedright
95th
\end{minipage} \\
\midrule\noalign{}
\endhead
\bottomrule\noalign{}
\endlastfoot
hurricanes, 16-cell pool & 4 & 15.5 & 3.9 & 23.2 & 7.4 & 21.4 & 5.3 &
12.6 & 20.4 \\
hurricanes, NB family (72 cells) & 4 & 11.3 & 3.8 & 13.8 & 6.6 & 13.2 &
4.1 & 9.6 & 12.6 \\
hurricanes, log-OLS family (72 cells) & 4 & 5.2 & 1.0 & 9.0 & 2.0 & 8.1
& 1.2 & 4.1 & 7.6 \\
hurricanes, pooled (144 cells) & 4 & 1.9 & 0.5 & 2.0 & 0.9 & 2.0 & 0.6 &
1.7 & 2.0 \\
job training, 256 control sets & 3 & 25.5 & 4.9 & 33.4 & 9.3 & 32.2 &
10.7 & 26.6 & 33.3 \\
job training, 1,296 cells & 3 & 13.2 & 2.3 & 16.2 & 4.5 & 16.1 & 4.6 &
12.8 & 15.7 \\
job training, L1 and L2 only & 2 & 0.8 & 0.1 & 0.8 & 0.3 & 0.8 & 0.1 &
0.6 & 0.8 \\
union, CPS 1985, full-sample family (D1 to D6) & 6 & 83.8 & 60.6 & 90.7
& 75.5 & 88.5 & 68.1 & 81.3 & 88.1 \\
union, CPS 1985, full-sample family (D1 to D4) & 4 & 80.3 & 50.6 & 84.8
& 66.1 & 83.9 & 58.8 & 77.3 & 83.1 \\
union, CPS 1985, wage-trimmed family (D1 to D6) & 6 & 71.2 & 41.5 & 84.0
& 59.1 & 80.0 & 46.6 & 67.0 & 79.2 \\
union, PSID 1982 (D1 to D6) & 6 & 95.3 & 90.1 & 96.2 & 94.0 & 95.9 &
91.0 & 94.7 & 95.8 \\
union, PSID 1982 (D1 to D4) & 4 & 94.5 & 84.8 & 94.9 & 91.0 & 94.8 &
85.8 & 93.8 & 94.7 \\
Design I & 4 & 97.5 & 90.6 & 98.4 & 94.8 & 98.2 & 92.9 & 96.9 & 98.1 \\
Design II, scenario A & 4 & 97.1 & 88.0 & 98.2 & 93.5 & 98.0 & 91.8 &
96.4 & 97.8 \\
Design II, scenario B & 3 & 95.0 & 72.6 & 96.6 & 84.3 & 96.4 & 78.6 &
93.6 & 96.4 \\
Design II, scenario C & 4 & 96.5 & 87.8 & 97.4 & 93.1 & 97.2 & 90.6 &
95.8 & 97.0 \\
Design III, n = 500, s = 1 & 4 & 89.1 & 66.2 & 92.6 & 79.0 & 91.9 & 73.0
& 86.7 & 91.4 \\
Design III, n = 2,000, s = 1 & 4 & 97.1 & 88.0 & 98.1 & 93.5 & 97.9 &
91.6 & 96.4 & 97.8 \\
Design III, n = 500, s = 0.5 & 4 & 39.7 & 15.5 & 48.0 & 25.5 & 46.1 &
18.4 & 34.9 & 44.7 \\
Design III, n = 2,000, s = 0.5 & 4 & 72.4 & 41.5 & 78.9 & 57.0 & 77.6 &
47.3 & 68.1 & 76.5 \\
Scenario D & 4 & 0.1 & 0.0 & 0.2 & 0.1 & 0.2 & 0.0 & 0.1 & 0.2 \\
\end{longtable}

\begin{longtable}[]{@{}
  >{\raggedright\arraybackslash}p{(\linewidth - 12\tabcolsep) * \real{0.2400}}
  >{\raggedright\arraybackslash}p{(\linewidth - 12\tabcolsep) * \real{0.1300}}
  >{\raggedright\arraybackslash}p{(\linewidth - 12\tabcolsep) * \real{0.2000}}
  >{\raggedright\arraybackslash}p{(\linewidth - 12\tabcolsep) * \real{0.1600}}
  >{\raggedright\arraybackslash}p{(\linewidth - 12\tabcolsep) * \real{0.1300}}
  >{\raggedright\arraybackslash}p{(\linewidth - 12\tabcolsep) * \real{0.0700}}
  >{\raggedright\arraybackslash}p{(\linewidth - 12\tabcolsep) * \real{0.0700}}@{}}
\caption{Monte Carlo standard errors and fit-failure counts by
simulation design (shares, coverage, and their MCSEs in percent). The
replication is the Monte Carlo unit: the pooled share carries a
batch-means MCSE (10 batches); the per-replication share is the share
computed on one replication's multiverse, as in the applications; bias
refers to the true world's valid-specification mean (Design II: G1,
scenario A and C; Design V: the post-selection OLS estimate) and
coverage to the average coverage of the individual licensed cells' 95
percent intervals (Design V: of the post-selection OLS interval) with
MCSE = SD of the per-replication value over \(\sqrt{R}\) (Design V
coverage: \(\sqrt{c(1-c)/R}\)). Design IV summarizes 95 contested
structures (its uncertainty is a bootstrap standard error over
structures); the world-blind arms of Design V are the out-of-domain
stress test; in Designs IV and V a failed fit aborts the script, so a
completed run implies none (0*). Scenario D sets \(Z_4 \to Y\) and
\(U_2 \to Y\) to zero (\(n = 2{,}000\), true effect
0.2).}\label{tbl-simmcse}\tabularnewline
\toprule\noalign{}
\begin{minipage}[b]{\linewidth}\raggedright
Design / scenario
\end{minipage} & \begin{minipage}[b]{\linewidth}\raggedright
Pooled share (MCSE)
\end{minipage} & \begin{minipage}[b]{\linewidth}\raggedright
Per-replication share: median {[}5th, 95th{]}
\end{minipage} & \begin{minipage}[b]{\linewidth}\raggedright
Bias, true world (MCSE)
\end{minipage} & \begin{minipage}[b]{\linewidth}\raggedright
Coverage, true world (MCSE)
\end{minipage} & \begin{minipage}[b]{\linewidth}\raggedright
Reps
\end{minipage} & \begin{minipage}[b]{\linewidth}\raggedright
Fit failures
\end{minipage} \\
\midrule\noalign{}
\endfirsthead
\toprule\noalign{}
\begin{minipage}[b]{\linewidth}\raggedright
Design / scenario
\end{minipage} & \begin{minipage}[b]{\linewidth}\raggedright
Pooled share (MCSE)
\end{minipage} & \begin{minipage}[b]{\linewidth}\raggedright
Per-replication share: median {[}5th, 95th{]}
\end{minipage} & \begin{minipage}[b]{\linewidth}\raggedright
Bias, true world (MCSE)
\end{minipage} & \begin{minipage}[b]{\linewidth}\raggedright
Coverage, true world (MCSE)
\end{minipage} & \begin{minipage}[b]{\linewidth}\raggedright
Reps
\end{minipage} & \begin{minipage}[b]{\linewidth}\raggedright
Fit failures
\end{minipage} \\
\midrule\noalign{}
\endhead
\bottomrule\noalign{}
\endlastfoot
I (illustrative) & 97.5 (0.1) & 99.6 {[}98.9, 99.9{]} & & & 300 & 0 \\
II, A: truth among candidates & 97.1 (0.3) & 99.6 {[}99.0, 99.9{]} &
0.0008 (0.0017) & 95.2 (1.3) & 200 & 0 \\
II, B: truth excluded & 95.0 (0.5) & 99.3 {[}98.1, 99.8{]} & & & 200 &
0 \\
II, C: unmeasured confounder & 96.5 (0.3) & 99.4 {[}98.3, 99.8{]} &
0.1963 (0.0017) & 0.0 (0.0) & 200 & 0 \\
III, n = 500, s = 1.0 & 89.1 (0.6) & 99.1 {[}96.4, 99.8{]} & 0.0010
(0.0036) & & 200 & 0 \\
III, n = 2000, s = 1.0 & 97.1 (0.3) & 99.7 {[}99.0, 99.9{]} & 0.0005
(0.0017) & & 200 & 0 \\
III, n = 500, s = 0.5 & 39.7 (2.8) & 96.2 {[}81.4, 99.4{]} & 0.0009
(0.0033) & & 200 & 0 \\
III, n = 2000, s = 0.5 & 72.4 (1.8) & 99.1 {[}96.4, 99.8{]} & -0.0007
(0.0017) & & 200 & 0 \\
D (contested coefficients zero) & 0.1 (0.1) & 49.1 {[}4.8, 93.1{]} &
0.0010 (0.0014) & & 200 & 0 \\
IV (100 random structures) & median 96.5 (0.2) & & median abs. bias
0.0034 (0.0005) & & 20 per structure & 0* \\
V, post-double-selection, world-blind & & & -0.4176 (0.0015) & 0.0 (0.0)
& 300 & 0* \\
V, outcome-adaptive penalization analogue, world-blind & & & -0.4161
(0.0015) & 0.0 (0.0) & 300 & 0* \\
V, post-double-selection within the true world (required controls
forced) & & & 0.0005 (0.0014) & 95.0 (1.3) & 300 & 0* \\
\end{longtable}

\section*{Reproducibility}\label{reproducibility}
\addcontentsline{toc}{section}{Reproducibility}

All results derive from public data and are reproducible end-to-end from
the replication archive: the dagmv package (version 0.1.3, which
produced the shipped outputs, installed from the tagged release
\texttt{v0.1.3}; version 0.1.2, with which the numbers were first
produced, uses the earlier printed labels and does not emit the message
about fixed terms outside the candidate graphs described in
Section~\ref{sec-software}, but reproduces the same values, and the test
logs of both versions are in the archive's \texttt{pkg\_logs/} folder),
\texttt{poc\_simulation.R}, \texttt{sim\_main.R},
\texttt{sim\_variants.R} (Designs I to III and Scenario D, with Monte
Carlo standard errors), \texttt{sim\_random\_graphs.R},
\texttt{sim\_selectors.R}, and \texttt{phaseB\_mcse.R}
(Section~\ref{sec-simulation}); \texttt{spec\_builders.R} (the
specification spaces as functions of the data),
\texttt{reanalysis\_hurricane.R},
\texttt{reanalysis\_hurricane\_full.R}, \texttt{reanalysis\_lalonde.R},
\texttt{reanalysis\_union.R}, \texttt{phaseB\_bootstrap.R} with
\texttt{phaseB\_wrappers.R} (the paired bootstrap with the
convergence-aware negative-binomial fits and their per-fit log, the
AIPW-ATT cells, the overlap counts, and the high-severity contrast; the
wrappers hold the per-application statistics and apply the fixed-mixture
convention in every branch), \texttt{phaseB\_simplex.R}
(restricted-simplex bounds and Dirichlet sweeps),
\texttt{phaseB\_roles.R} and \texttt{phaseB\_dag\_figures.R} (role
tables and graph figures), \texttt{phaseB\_regression\_checks.R} (the
finite-mixture checks of Section~\ref{sec-framework} and the
failure-convention check through every wrapper),
\texttt{phaseB\_nb\_failure\_pattern.R},
\texttt{phaseB\_history\_comparison.R} (the comparisons with the earlier
bootstrap algorithms, computed from their archived summaries in
\texttt{baselines/}; the earlier algorithms are not refitted), and
\texttt{phaseB\_tables.R} (Section~\ref{sec-reanalysis} and the
Supplemental Materials). The simulation and application scripts
regenerate their results from the data; \texttt{phaseB\_mcse.R},
\texttt{phaseB\_simplex.R}, \texttt{phaseB\_nb\_failure\_pattern.R},
\texttt{phaseB\_history\_comparison.R} and \texttt{phaseB\_tables.R}
read the outputs written by the scripts before them in the order given
in the archive's README, so a clean run must follow that order.
\texttt{stage\_figures.sh} (run with \texttt{bash}) copies the ten
figure files from the outputs to the folder the manuscript source reads,
and the PDF is built with Quarto; the archive claims numerical
reproduction of the tables and figures, not a byte-identical PDF. Data
sources: Jung et al.~hurricane data (92 storms); NSW-Dehejia-Wahba and
CPS-1 files; the CPS May 1985 extract (\citeproc{ref-berndt1991}{Berndt
1991}) and the PSID 1976--1982 panel
(\citeproc{ref-cornwellrupert1988}{Cornwell and Rupert 1988}), both
distributed with standard R packages. Archive location:
https://github.com/sokubo/paper-multiverse-dag-replication, whose README
names the commit checked against this manuscript and which carries a
release-check record: the snapshot was downloaded without author
credentials, its contents agreed with its file manifest, and the
documented sequence was re-run from a clean copy with the shipped
outputs set aside, reproducing all ten figures byte for byte and every
result file but an elapsed-time line; the record ships with the archive
together with the run log and session information, and a tag matching
the posted version is added at posting. The R package is at
https://github.com/sokubo/dagmv. Every shipped output was produced by
one run of the documented sequence under R 4.6.0 on macOS. The numerical
results agree with the earlier R 4.3.3/Linux outputs at the precision
reported in the paper, and the archive includes the comparison record.
The archive's \texttt{analysis/README.md} pins the package and
dependency versions, records the environment in which the shipped
outputs and runtimes were produced, gives the execution order and states
which steps regenerate results and which read archived ones, and maps
every reported number to a script and an output file; the
application-specific fitting functions (the convergence-aware
negative-binomial fit and the AIPW-ATT estimator in
\texttt{spec\_builders.R}) are part of the archive, not of the package
engine. The archive contains the code and materials; the data are the
public files named above.

\section*{References}\label{references}
\addcontentsline{toc}{section}{References}

\phantomsection\label{refs}


\begin{CSLReferences}{1}{0}
\bibitem[\citeproctext]{ref-auspurg2025}
Auspurg, Katrin. 2025. {``Robustness Is Better Assessed with a Few
Thoughtful Models Than with Billions of Regressions.''}
\emph{Proceedings of the National Academy of Sciences} 122 (43):
e2521917122.

\bibitem[\citeproctext]{ref-bch2014}
Belloni, Alexandre, Victor Chernozhukov, and Christian Hansen. 2014.
{``Inference on Treatment Effects After Selection Among High-Dimensional
Controls.''} \emph{Review of Economic Studies} 81 (2): 608--50.

\bibitem[\citeproctext]{ref-berndt1991}
Berndt, Ernst R. 1991. \emph{The Practice of Econometrics: Classic and
Contemporary}. Reading, MA: Addison-Wesley.

\bibitem[\citeproctext]{ref-breznau2022}
Breznau, Nate, Eike Mark Rinke, Alexander Wuttke, Hung H. V. Nguyen,
Muna Adem, Jule Adriaans, Amalia Alvarez-Benjumea, et al. 2022.
{``Observing Many Researchers Using the Same Data and Hypothesis Reveals
a Hidden Universe of Uncertainty.''} \emph{Proceedings of the National
Academy of Sciences} 119 (44): e2203150119.

\bibitem[\citeproctext]{ref-callawaysantanna2021}
Callaway, Brantly, and Pedro H. C. Sant'Anna. 2021.
{``Difference-in-Differences with Multiple Time Periods.''}
\emph{Journal of Econometrics} 225 (2): 200--230.

\bibitem[\citeproctext]{ref-card1996}
Card, David. 1996. {``The Effect of Unions on the Structure of Wages: A
Longitudinal Analysis.''} \emph{Econometrica} 64 (4): 957--79.

\bibitem[\citeproctext]{ref-cinelli2022preprint}
Cinelli, Carlos, Andrew Forney, and Judea Pearl. 2022. {``A Crash Course
in Good and Bad Controls.''} Author manuscript, March 21, 2022;
subsequently published in revised form in Sociological Methods \&
Research (2024).
\url{https://carloscinelli.com/files/Cinelli\%20et\%20al\%20-\%20A\%20Crash\%20Course\%20in\%20Good\%20and\%20Bad\%20Controls.pdf}.

\bibitem[\citeproctext]{ref-cinelli2024}
---------. 2024. {``A Crash Course in Good and Bad Controls.''}
\emph{Sociological Methods \& Research} 53 (3): 1071--1104.

\bibitem[\citeproctext]{ref-cinellihazlett2020}
Cinelli, Carlos, and Chad Hazlett. 2020. {``Making Sense of Sensitivity:
Extending Omitted Variable Bias.''} \emph{Journal of the Royal
Statistical Society Series B} 82 (1): 39--67.

\bibitem[\citeproctext]{ref-cornwellrupert1988}
Cornwell, Christopher, and Peter Rupert. 1988. {``Efficient Estimation
with Panel Data: An Empirical Comparison of Instrumental Variables
Estimators.''} \emph{Journal of Applied Econometrics} 3 (2): 149--55.

\bibitem[\citeproctext]{ref-dechaisemartin2020}
de Chaisemartin, Clément, and Xavier D'Haultfœuille. 2020. {``Two-Way
Fixed Effects Estimators with Heterogeneous Treatment Effects.''}
\emph{American Economic Review} 110 (9): 2964--96.

\bibitem[\citeproctext]{ref-dehejiawahba1999}
Dehejia, Rajeev H., and Sadek Wahba. 1999. {``Causal Effects in
Nonexperimental Studies: Reevaluating the Evaluation of Training
Programs.''} \emph{Journal of the American Statistical Association} 94
(448): 1053--62.

\bibitem[\citeproctext]{ref-delgiudice2021}
Del Giudice, Marco, and Steven W. Gangestad. 2021. {``A Traveler's Guide
to the Multiverse: Promises, Pitfalls, and a Framework for the
Evaluation of Analytic Decisions.''} \emph{Advances in Methods and
Practices in Psychological Science} 4 (1): 2515245920954925.

\bibitem[\citeproctext]{ref-elwertwinship2014}
Elwert, Felix, and Christopher Winship. 2014. {``Endogenous Selection
Bias: The Problem of Conditioning on a Collider Variable.''}
\emph{Annual Review of Sociology} 40: 31--53.

\bibitem[\citeproctext]{ref-farber2021wp}
Farber, Henry S., Daniel Herbst, Ilyana Kuziemko, and Suresh Naidu.
2021b. {``Unions and Inequality over the Twentieth Century: New Evidence
from Survey Data.''} Working Paper 24587, revised April 2021. National
Bureau of Economic Research.

\bibitem[\citeproctext]{ref-farber2021}
---------. 2021a. {``Unions and Inequality over the Twentieth Century:
New Evidence from Survey Data.''} \emph{Quarterly Journal of Economics}
136 (3): 1325--85.

\bibitem[\citeproctext]{ref-freemanmedoff1984}
Freeman, Richard B., and James L. Medoff. 1984. \emph{What Do Unions
Do?} New York: Basic Books.

\bibitem[\citeproctext]{ref-ganslmeier2025}
Ganslmeier, Michael, and Tim Vlandas. 2025a. {``Estimating the Extent
and Sources of Model Uncertainty in Political Science.''}
\emph{Proceedings of the National Academy of Sciences} 122 (25):
e2414926122.

\bibitem[\citeproctext]{ref-ganslmeier2025reply}
---------. 2025b. {``Reply to {Auspurg}: On the Limits of {`Justified'}
Model Spaces.''} \emph{Proceedings of the National Academy of Sciences}
122 (43): e2523374122.

\bibitem[\citeproctext]{ref-dagassist2026}
Goff, Graham, and Michael Denly. 2026. {``{DAGassist}: Test Robustness
with Directed Acyclic Graphs.''} R package version 0.3.0, CRAN.

\bibitem[\citeproctext]{ref-henckel2022}
Henckel, Leonard, Emilija Perković, and Marloes H. Maathuis. 2022.
{``Graphical Criteria for Efficient Total Effect Estimation via
Adjustment in Causal Linear Models.''} \emph{Journal of the Royal
Statistical Society Series B} 84 (2): 579--99.

\bibitem[\citeproctext]{ref-huvdpas2025}
Hu, Zhongyi, and Stéphanie van der Pas. 2025. {``Selecting Valid
Adjustment Sets with Uncertain Causal Graphs.''} arXiv preprint
arXiv:2511.01662.

\bibitem[\citeproctext]{ref-imaikim2019}
Imai, Kosuke, and In Song Kim. 2019. {``When Should We Use Unit Fixed
Effects Regression Models for Causal Inference with Longitudinal
Data?''} \emph{American Journal of Political Science} 63 (2): 467--90.

\bibitem[\citeproctext]{ref-jung2014}
Jung, Kiju, Sharon Shavitt, Madhu Viswanathan, and Joseph M. Hilbe.
2014. {``Female Hurricanes Are Deadlier Than Male Hurricanes.''}
\emph{Proceedings of the National Academy of Sciences} 111 (24):
8782--87.

\bibitem[\citeproctext]{ref-keele2020}
Keele, Luke, Randolph T. Stevenson, and Felix Elwert. 2020. {``The
Causal Interpretation of Estimated Associations in Regression Models.''}
\emph{Political Science Research and Methods} 8 (1): 1--13.

\bibitem[\citeproctext]{ref-lundberg2021}
Lundberg, Ian, Rebecca Johnson, and Brandon M. Stewart. 2021. {``What Is
Your Estimand? Defining the Target Quantity Connects Statistical
Evidence to Theory.''} \emph{American Sociological Review} 86 (3):
532--65.

\bibitem[\citeproctext]{ref-specr2020}
Masur, Philipp K., and Michael Scharkow. 2023. {``{specr}: Conducting
and Visualizing Specification Curve Analyses.''} R package version
1.0.0, CRAN.

\bibitem[\citeproctext]{ref-montgomery2018}
Montgomery, Jacob M., Brendan Nyhan, and Michelle Torres. 2018. {``How
Conditioning on Posttreatment Variables Can Ruin Your Experiment and
What to Do about It.''} \emph{American Journal of Political Science} 62
(3): 760--75.

\bibitem[\citeproctext]{ref-munozyoung2018}
Muñoz, John, and Cristobal Young. 2018. {``We Ran 9 Billion Regressions:
Eliminating False Positives Through Computational Model Robustness.''}
\emph{Sociological Methodology} 48 (1): 1--33.

\bibitem[\citeproctext]{ref-okubo2026att}
Okubo, Shoki. 2026. {``Optimal Covariate Adjustment Beyond the Average
Treatment Effect: Treated-Population and Overlap-Weighted Estimands.''}
arXiv preprint arXiv:2609.11222.

\bibitem[\citeproctext]{ref-oster2019}
Oster, Emily. 2019. {``Unobservable Selection and Coefficient Stability:
Theory and Evidence.''} \emph{Journal of Business \& Economic
Statistics} 37 (2): 187--204.

\bibitem[\citeproctext]{ref-pearl2009}
Pearl, Judea. 2009. \emph{Causality: Models, Reasoning, and Inference}.
2nd ed. Cambridge: Cambridge University Press.

\bibitem[\citeproctext]{ref-perkovic2018}
Perković, Emilija, Johannes Textor, Markus Kalisch, and Marloes H.
Maathuis. 2018. {``Complete Graphical Characterization and Construction
of Adjustment Sets in {Markov} Equivalence Classes of Ancestral
Graphs.''} \emph{Journal of Machine Learning Research} 18 (220): 1--62.

\bibitem[\citeproctext]{ref-rotnitzkysmucler2020}
Rotnitzky, Andrea, and Ezequiel Smucler. 2020. {``Efficient Adjustment
Sets for Population Average Causal Treatment Effect Estimation in
Graphical Models.''} \emph{Journal of Machine Learning Research} 21
(188): 1--86.

\bibitem[\citeproctext]{ref-multiverse2024}
Sarma, Abhraneel, and Matthew Kay. 2024. {``{multiverse}: Create
{`Multiverse Analysis'} in {R}.''} R package version 0.6.2, CRAN.

\bibitem[\citeproctext]{ref-short2026}
Short, Cassie Ann, Nate Breznau, Maria Bruntsch, Micha Burkhardt, Niko
A. Busch, Elena Cesnaite, Maximilian Frank, et al. 2026.
{``Multicurious: A Multidisciplinary Guide to Multiverse Analysis.''}
\emph{Advances in Methods and Practices in Psychological Science} 9 (2):
1--21. \url{https://doi.org/10.1177/25152459261434881}.

\bibitem[\citeproctext]{ref-shortreed2017}
Shortreed, Susan M., and Ashkan Ertefaie. 2017. {``Outcome-Adaptive
Lasso: Variable Selection for Causal Inference.''} \emph{Biometrics} 73
(4): 1111--22.

\bibitem[\citeproctext]{ref-shpitser2010}
Shpitser, Ilya, Tyler J. VanderWeele, and James M. Robins. 2010. {``On
the Validity of Covariate Adjustment for Estimating Causal Effects.''}
In \emph{Proceedings of the 26th Conference on Uncertainty in Artificial
Intelligence}, 527--36.

\bibitem[\citeproctext]{ref-simonsohn2020}
Simonsohn, Uri, Joseph P. Simmons, and Leif D. Nelson. 2020.
{``Specification Curve Analysis.''} \emph{Nature Human Behaviour} 4
(11): 1208--14.

\bibitem[\citeproctext]{ref-slez2019}
Slez, Adam. 2019. {``The Difference Between Instability and Uncertainty:
Comment on {Young} and {Holsteen} (2017).''} \emph{Sociological Methods
\& Research} 48 (2): 400--430.

\bibitem[\citeproctext]{ref-steegen2016}
Steegen, Sara, Francis Tuerlinckx, Andrew Gelman, and Wolf Vanpaemel.
2016. {``Increasing Transparency Through a Multiverse Analysis.''}
\emph{Perspectives on Psychological Science} 11 (5): 702--12.

\bibitem[\citeproctext]{ref-robustipy2025}
Valdenegro Ibarra, Daniel, Jiani Yan, Duiyi Dai, and Charles Rahal.
2026. {``{RobustiPy}: An Efficient Next-Generation Multiversal Library
with Model Selection, Averaging, Resampling, and Explainable {AI}.''}
arXiv preprint arXiv:2506.19958v4 (version 4, 18 May 2026; first version
June 2025).

\bibitem[\citeproctext]{ref-vanderweele2019}
VanderWeele, Tyler J. 2019. {``Principles of Confounder Selection.''}
\emph{European Journal of Epidemiology} 34 (3): 211--19.

\bibitem[\citeproctext]{ref-westernrosenfeld2011}
Western, Bruce, and Jake Rosenfeld. 2011. {``Unions, Norms, and the Rise
in {U.S.} Wage Inequality.''} \emph{American Sociological Review} 76
(4): 513--37.

\bibitem[\citeproctext]{ref-westreich2013}
Westreich, Daniel, and Sander Greenland. 2013. {``The {Table 2} Fallacy:
Presenting and Interpreting Confounder and Modifier Coefficients.''}
\emph{American Journal of Epidemiology} 177 (4): 292--98.

\bibitem[\citeproctext]{ref-whitelu2011}
White, Halbert, and Xun Lu. 2011. {``Causal Diagrams for Treatment
Effect Estimation with Application to Efficient Covariate Selection.''}
\emph{Review of Economics and Statistics} 93 (4): 1453--59.

\bibitem[\citeproctext]{ref-wysocki2022}
Wysocki, Anna C., Katherine M. Lawson, and Mijke Rhemtulla. 2022.
{``Statistical Control Requires Causal Justification.''} \emph{Advances
in Methods and Practices in Psychological Science} 5 (2):
25152459221095823.

\bibitem[\citeproctext]{ref-young2009}
Young, Cristobal. 2009. {``Model Uncertainty in Sociological Research:
An Application to Religion and Economic Growth.''} \emph{American
Sociological Review} 74 (3): 380--97.

\bibitem[\citeproctext]{ref-young2019}
---------. 2019. {``The Difference Between Causal Analysis and
Predictive Models: Response to {`Comment on {Young} and {Holsteen}
(2017)'}.''} \emph{Sociological Methods \& Research} 48 (2): 431--47.

\bibitem[\citeproctext]{ref-youngcumberworth2025}
Young, Cristobal, and Erin Cumberworth. 2025. \emph{Multiverse Analysis:
Computational Methods for Robust Results}. Cambridge: Cambridge
University Press.

\bibitem[\citeproctext]{ref-youngholsteen2017}
Young, Cristobal, and Katherine Holsteen. 2017. {``Model Uncertainty and
Robustness: A Computational Framework for Multimodel Analysis.''}
\emph{Sociological Methods \& Research} 46 (1): 3--40.

\bibitem[\citeproctext]{ref-multivrs2021}
---------. 2021. {``{MULTIVRS}: Stata Module to Conduct Multiverse
Analysis.''} Statistical Software Components S458927, Boston College
Department of Economics.

\end{CSLReferences}
\end{document}